\documentclass[a4paper]{article}

\usepackage{amsmath,amssymb,amsthm,bbm,natbib,color,nccmath}
\usepackage[margin=1in]{geometry}
\usepackage{caption}
\usepackage{graphicx}
\usepackage{subcaption}
\usepackage{afterpage}
\usepackage[hang,flushmargin]{footmisc}
\usepackage{float}
\usepackage{enumitem,kantlipsum}
\usepackage{multirow}
\usepackage[ruled,vlined]{algorithm2e}
\usepackage[
  colorlinks,
  citecolor=blue,
  linkcolor=black,
  anchorcolor=red,
  urlcolor=blue
]{hyperref}    
\usepackage{booktabs}
\usepackage{tikz}
\usepackage{pgfplots}
\usepackage{stmaryrd}
\pgfplotsset{compat=1.17}

\newtheorem{theorem}{Theorem}
\newtheorem{lemma}{Lemma}
\newtheorem{proposition}{Proposition}
\newtheorem{assumption}{Assumption}
\newtheorem{definition}{Definition}
\newtheorem{corollary}{Corollary}

\theoremstyle{definition}
\newtheorem{remark}{Remark}
\allowdisplaybreaks

\newcommand{\R}{\mathbb{R}}
\newcommand{\N}{\mathbb{N}}
\newcommand{\D}{\mathcal{D}}
\newcommand{\eps}{\epsilon}
\newcommand{\EE}[1]{\mathbb{E}\left[{#1}\right]}
\newcommand{\EEst}[2]{\mathbb{E}\left[{#1}\  \middle| \ {#2}\right]}

\newcommand{\PP}[1]{\mathbb{P}\left\{{#1}\right\}}
\newcommand{\PPst}[2]{\mathbb{P}\left\{{#1}\  \middle| \ {#2}\right\}}

\newcommand{\Pp}[2]{\mathbb{P}_{{#1}}\left\{{#2}\right\}}
\newcommand{\One}[1]{{\mathbbm{1}}\left\{{#1}\right\}}

\newcommand{\iidsim}{\stackrel{\textnormal{iid}}{\sim}}

\newcommand{\X}{\mathcal{X}}
\newcommand{\Y}{\mathcal{Y}}

\newcommand{\cR}{\mathcal{R}}
\newcommand{\FDR}{\textnormal{FDR}}
\newcommand{\cFDR}{\textnormal{cFDR}}
\newcommand{\FDP}{\textnormal{FDP}}
\renewcommand{\hat}{\widehat}
\newcommand{\ebh}{\textnormal{e-BH}}
\newcommand{\pbh}{\textnormal{p-BH}}
\newcommand{\sub}{\textnormal{sub}}
\newcommand{\hier}{\textnormal{hier}}
\definecolor{ForestGreen}{RGB}{34,139,34}

\usepackage{authblk}

\allowdisplaybreaks

\title{Selection from Hierarchical Data with Conformal e-values}
\author{Yonghoon Lee}
\author{Zhimei Ren}
\affil{Department of Statistics and Data Science, the Wharton School,\\ University of Pennsylvania}

\date{\today}

\begin{document}

\maketitle

\begin{abstract}

Distribution-free predictive inference beyond the construction of prediction sets 
has gained a lot of interest in recent applications. 
One such application is the selection task, where the objective is to 
design a reliable selection rule to pick out individuals with desired unobserved outcomes 
while controlling the error rate. 
In this work, we address the selection problem in the context of hierarchical data, 
where groups of observations may exhibit distinct within-group distributions. 
This generalizes existing techniques beyond the standard i.i.d./exchangeable data settings. 
For hierarchical data, we introduce methods to construct valid 
conformal e-values, enabling control of the false discovery rate (FDR) 
through the e-BH procedure. 
In particular, we introduce and compare two approaches---{\em subsampling conformal e-values}
and {\em hierarchical conformal e-values}. 
Empirical results demonstrate that both approaches achieve valid FDR control while highlighting a tradeoff between stability and power. The subsampling-based method, though random, typically offers higher power, whereas the hierarchical approach, being deterministic, tends to be slightly less powerful. The effectiveness of the proposed methods is illustrated in two real-world applications.


\end{abstract}


\section{Introduction}
In recent years, artificial intelligence has demonstrated 
remarkable predictive power across a wide range of fields, 
showing great promise in assisting humans with decision-making. 
Consider the task of selecting units from a large pool of candidates
whose unobserved outcomes meet specific desirable criteria.
Leveraging machine-predicted outcomes to identify and shortlist a subset of promising candidates can significantly streamline the decision-making process, reducing the need for extensive experiments or manual investigations.
Indeed, machine learning algorithms have been used to find protein structures with desired 
functions~\citep{watson2023novo}, to identify promising drug candidates~\citep{dara2022machine}, 
or to select a cohort of patients for clinical trials~\citep{lehman2012risk,xiong2019cohort}. 

In many of these applications, making a wrong selection can be costly---whether in terms of time, money, or even human welfare. When relying on machine learning algorithms to make predictions,
it is crucial to have a reliable method for assessing the uncertainty of the predictions 
and for controlling the error rate of the selection procedure. 
To this end,~\citet{jin2023selection} 
proposed {\em conformal selection (CS)}, a prediction-assisted selection framework that  
controls the false discovery rate (FDR) of the selected set in a model-free fashion. 
Formally, suppose we have a set of labeled data $(X_i,Y_i)_{1 \leq i \leq n}$, 
and then given multiple new inputs $X_{n+1}, \cdots, X_{n+m}$ (the candidate pool); 
our goal is to select a subset of them whose unobserved outcomes $Y_{n+1}, \cdots, Y_{n+m}$ 
meet a certain condition, e.g., $Y_{n+j}$ that exceeds some threshold $c$.
The FDR of a selected set is defined as 
\[\text{FDR} = \EE{\frac{\sum_{j=1}^m\One{\text{$Y_{n+j}$ does not satisfy the condition, but unit $n+j$  selected}}}{\sum_{j=1}^m \One{\text{unit $n+j$  selected}}}} \leq \alpha,\]
where we follow the convention that $0/0 = 0$ and $\alpha \in (0,1)$ is a pre-defined level. 
At a high level, CS formulates the selection task as a multiple hypotheses testing problem, where each 
unit in the test set is associated with a hypothesis 
and rejecting a null hypothesis means selecting 
the corresponding unit. It constructs a p-value for 
each hypothesis, and obtains the selection set by 
applying the Benjamini Hochberg~\citep{benjamini1995controlling} (BH) 
procedure to the p-values.
It is proved in~\citet{jin2023selection} that 
CS achieves FDR control as long as the training and the test data are {\em jointly exchangeable}. 

\subsection{Selection with hierarchical data}
In reality, data often exhibit more complex structures, 
requiring procedures that function under assumptions weaker than joint exchangeability. 
In this work, we examine a setting in which the data have a hierarchical structure---data points 
are organized into groups and exchangeability only holds within and across groups. 
Below, we present several examples of hierarchical data-generating processes.

\paragraph{Multi-environment data.}
In many modern applications, data are collected from multiple environments, 
where the data-generating distributions vary across environments~\citep{meinshausen2015maximin,rothenhausler2021anchor,guo2024statistical}.
For example, researchers may collect data from multiple sites 
to test a scientific hypothesis~\citep{higgins2009re}; 
electronic health record data often aggregate data from various hospitals~\citep{singh2022generalizability}; and many genetic studies 
include cohorts representing diverse populations~\citep{keys2020cross}.
The data-generating process of such multi-environment data
is often modeled hierarchically: the distribution of each 
distribution is considered a random draw from a prior distribution, and 
observations within each environment are sampled from that environment-specific 
distribution~\citep{duchi2024predictive,jeong2022calibrated,jeong2024out}.

\paragraph{Cluster-randomized trials.}
Cluster randomized trials (CRTs) are a class of experimental 
methods widely used in empirical studies. 
In CRTs, treatments are randomly assigned at the cluster level, 
where clusters can represent villages, cities, schools, or similar groups~\citep{murray1998design,donner2000design,10.1214/18-AOS1765,su2021model,jin2023toward}.
In these applications, the data-generating process  
often follows a hierarchical structure, where the clusters are drawn from 
a (prior) distribution
and the units in each class from a cluster-specific distribution~\citep{wang2024model,wang2024conformal}.

\paragraph{Repeated observations.}
When the units in a study have multiple independent observations,
a natural hierarchical structure emerges: the units 
can be treated as clusters, with the repeated observations acting as
elements within those clusters. Such a setting arises when, for example, 
a medical sample has multiple doctor ratings~\citep{liu2020deep}, or when a data point receives 
multiple annotations~\citep{stutz2023conformal}.\\

Our work considers the selection problem in the context of hierarchical data: 
given multiple new clusters of test points, we aim to select test units/clusters whose unobserved
outcomes meet a certain condition with guarantees on the {\em selected} units/clusters.
For example, in job hiring, the candidates may be coming from different regions with different distributions of skills:
how can we select candidates whose (yet-to-be-observed) work performance exceeds a certain threshold, while controlling 
the error rate of the selection process? In cluster randomized trials, how to select 
the individuals from distinct populations with large individual treatment effects 
with a controlled error rate? 
When multiple annotations are available for each sample, it is 
of interest to pick out the samples with at least one high-quality
annotation and/or to identify which ones are of high quality.

In this paper, we offer a solution to these questions, building on conformal inference~\citep{vovk2005algorithmic} and e-values~\citep{vovk2021values,grunwald2024safe,shafer2021testing,howard2021time}.
Throughout, we assume the data satisfy \textit{hierarchical exchangeability}---a concept formally defined later---which informally means that 
the groups of observations are exchangeable, and the observations within each group are exchangeable as well. 
Notably, the exchangeability of the entire dataset considered in ~\citet{jin2023selection} is a special case under this framework. 
In the context of such hierarchical data, we study the selection task given new groups of feature inputs. 
We will first discuss the selection of individual test points, 
and then extend the procedure to the selection of both groups and individuals.

\subsection{Our contributions}
We provide a general recipe for model-free 
selection with hierarchical data, which allows practitioners 
to leverage the power of complex machine learning algorithms for selection 
while enjoying rigorous error control guarantees on the selected units.
We highlight our main contributions below.

\begin{itemize}
\item {\em Model-free selection for hierarchical data.}
We consider a suite of selection tasks under a hierarchical data-generating process. We formulate the selection problem as a (structured) 
multiple hypothesis testing problem, and provide a testing procedure 
with provable FDR control. The key idea is to construct an 
{\em e-value}---to be introduced shortly---for 
each hypothesis, and apply e-value-based multiple testing procedures to control the FDR. 

\item {\em Instantiation and extension of e-value-based multiple testing.} 
Our work provides valid and powerful instantiations of
the e-BH procedure~\citep{wang2022false} tailored to our setting, achieving power comparable to the (heuristic) p-value-based methods while 
ensuring rigorous FDR control.
Moreover, we extend the e-value-based multiple testing framework 
to accommodate settings with a random number of hypotheses 
and hierarchical data structure, 
under which we identify sufficient conditions for FDR control.
To show that the e-values satisfy the properties required for FDR control,
we adopt a careful treatment to address the challenge brought by the hierarchical structure, 
which can be of independent interest.

\item {\em Empirical evaluation.} 
We evaluate the validity and power of the proposed methods through extensive simulations and real data analysis\footnote{The data that support the findings of this work are openly available in \url{https://github.com/socialfoundations/folktables} and \url{https://datadryad.org/dataset/doi:10.5061/dryad.16c9m51}.}. 
The results show that the proposed methods achieve the desired FDR control with competitive power.

\end{itemize}

Our paper is organized as follows. In Section~\ref{sec:problem}, we introduce the 
problem setup and the relevant background. 
Section~\ref{sec:warmup} provides a review of existing methods and 
discusses why their direct generalization fails in our setting.
Section~\ref{sec:method} presents our main results, 
including the construction of e-values for hierarchical data and the FDR control procedure.
In the Appendix, we explore extensions to other selection tasks and settings.
Sections~\ref{sec:experiments} and~\ref{sec:real_data} present empirical results, 
and we conclude in Section~\ref{sec:discussion}.

\subsection{Notations}
We write $\R$ to denote the real space, $\R^d$ to denote the $d$-dimensional real space, 
and $\N$ to denote the set of positive integers.
For $n \in \N$, $[n]$ denotes the set $\{1,2,\cdots,n\}$, and $v_{1:n}$ denotes the vector $(v_1,v_2,\cdots,v_n)^\top$. $\mathcal{S}_n = \{\sigma : [n] \rightarrow [n],  \text{$\sigma$ is a bijection}\}$ denotes the set of all permutations of $[n]$. For any $a,b\in \R$, 
$a \vee b = \max(a,b)$ and $a \wedge b = \min(a,b)$. For a set $A$, $|A|$ refers to its cardinality. For a random variable $X$ that takes values in a space $\X$---i.e., $X$ is a measurable function with codomain $\X$---we write $X \in \X$ for simplicity.

\section{Problem setup}
\label{sec:problem}
\begin{figure}[h!]
    \centering
    \includegraphics[width=\textwidth]{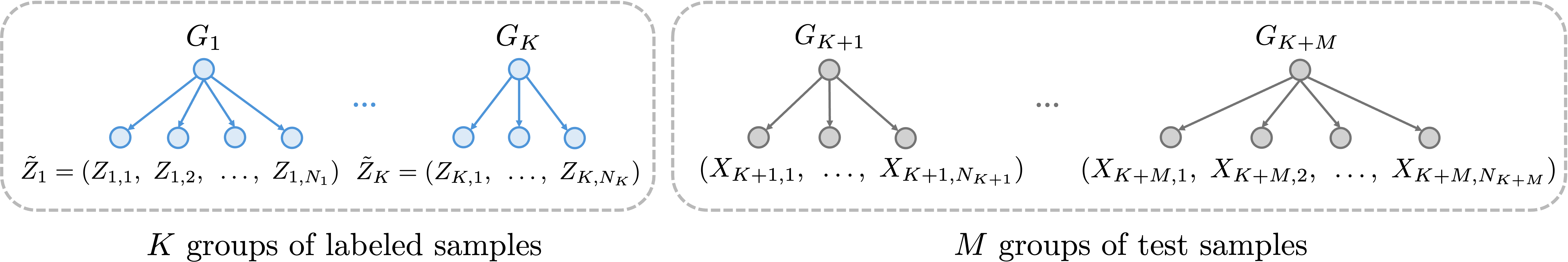}
    \caption{Illustration of the hierarchical structure.}
    \label{fig:dgp}
\end{figure}

Suppose we have labeled samples from $K \ge 1$ 
groups, with $N_k$ samples in group $k\in [K]$: 
$Z_{k,1},\ldots,Z_{k,N_k}$---we treat $K$ as fixed, while the group sizes $N_k$ are treated random. Each $Z_{k,i}$ denotes the tuple $(X_{k,i},Y_{k,i})$, where 
$X_{k,i} \in \X \subseteq \R^d$ is the feature vector and 
$Y_{k,i} \in \Y \subseteq \R$ is the outcome.
Denote by $G_1,\ldots,G_K \in \mathcal{G}$ ($\mathcal{G} \subset \R^{d_G}$) the group-specific feature observations 
and $\tilde{Z}_k = (Z_{k,1}, \cdots, Z_{k,N_k})$ the vector of samples in the 
$k$-th group.
Suppose we are given new inputs from $M$ groups: $(G_k,X_{k,i})_{K+1 \leq k \leq K+M, 1 \leq i \leq N_k}$,  
without their corresponding outcomes $(Y_{k,i})_{K+1 \leq k \leq K+M, 1 \leq i \leq N_k}$. See Figure~\ref{fig:dgp} for a graphical illustration.

We impose the following assumptions on the data-generating process.
\begin{assumption}[Hierarchical exchangeability]\label{asm:dist}
The dataset $(G_k, \tilde{Z}_k)_{1 \leq k \leq K+M}$ satisfies the following:
\begin{enumerate}
    \item The sequence of random vectors $\tilde{Z}_1, \cdots, \tilde{Z}_K, \tilde{Z}_{K+1}, \cdots, \tilde{Z}_{K+M}$ satisfies \textit{hierarchical exchangeability}, i.e., for any $\sigma \in \mathcal{S}_{K+M}$,
    \[(\tilde{Z}_1, \tilde{Z}_2, \cdots, \tilde{Z}_{K+M}) \stackrel{d}{=} (\tilde{Z}_{\sigma(1)}, \tilde{Z}_{\sigma(2)}, \cdots, \tilde{Z}_{\sigma(K+M)}),\]
    and furthermore, for any $m \geq 1$, $\sigma \in \mathcal{S}_m$ and $k \in [K+M]$,
    \[\scalebox{0.93}{$(\tilde{Z}_1, \cdots, \tilde{Z}_{K+M}) \mid \{N_k = m\} \stackrel{d}{=} (\tilde{Z}_1, \cdots \tilde{Z}_{k-1}, (Z_{k,\sigma(1)}, \cdots, Z_{k,\sigma(m)}), \cdots, \tilde{Z}_{K+M}) \mid \{N_k = m\}.$}
    \]
    \item The group size $N_k$ is independent of the individual observations in the $k$-th group, as well as the observations in other groups, i.e., for any  $k \in [K+M]$ and $m \geq 1$,
    \[(Z_{k,1},\cdots,Z_{k,m}), (N_l, \tilde{Z}_l)_{l \neq k} \mid \{N_k = m\} \;\;\stackrel{d}{=}\;\; (Z_{k,1},\cdots,Z_{k,m}), (N_l, \tilde{Z}_l)_{l \neq k}.\]
\end{enumerate}
\end{assumption}
The first condition states that the dataset has between-group exchangeability 
in the sense that the groups of observations $\tilde{Z}_1, \cdots, \tilde{Z}_{K+M}$ 
are exchangeable, as well as the within-group exchangeability, i.e., for each $k$, $\tilde{Z}_k = (Z_{k,1}, \cdots, Z_{k,N_k})$ is a vector of exchangeable variables
given the other groups.\footnote{This condition was previously discussed in~\citet{lee2023distribution}, which studies the construction of prediction sets under a hierarchical data structure.}
A special case satisfying Assumption~\ref{asm:dist} is the following model:
\begin{equation}\label{eqn:model}
\begin{split}
    &G_1, G_2, \cdots, G_{K+M} \iidsim P_G,\\
    &N_1, N_2, \cdots, N_{K+M} \iidsim P_{N},\\
    &Z_{k,1}, Z_{k,2}, \cdots, Z_{k,N_k} \mid G_k, N_k \iidsim P_{Z \mid G}, 
\end{split}
\end{equation}
where $P_G$ is some distribution over the distributions on $(X,Y)$, 
$P_N$ is the distribution on $\N$ of the group sizes, 
and $P_{Z \mid G}$ is the distribution of the observations within each group.

Note that the hierarchical exchangeability condition is weaker than, i.e., implied by, full exchangeability. Specifically, Assumption~\ref{asm:dist} essentially states that the within-group observations are ``symmetric'', and thus their joint distribution is invariant with respect to within-group permutations, and that the group-level observations are also symmetric. This hierarchical exchangeability, or symmetry, directly reflects the structure of the data, while imposing a weaker assumption on the joint distribution than the standard exchangeability.

Under the hierarchical exchangeable setting described above, 
we consider the task of selecting test points whose outcomes meet some 
desired property.
Without loss of generality, we focus on the selection of individuals
with outcomes surpassing a threshold, noting that the methods discussed 
in this paper can be generalized to accommodate other types of selection criteria (e.g., individuals with $Y \in \mathcal{C}$ 
for some set $\mathcal{C}$ characterizing the property of interest).
Following~\citet{jin2023selection}, we formulate the task 
as testing the hypotheses:
\begin{equation}\label{eqn:hypothesis}
H_{j,i} : Y_{K+j, i} \leq c(X_{K+j,i}),\qquad j\in[M], i\in [N_{K+j}].
\end{equation}
Above, $c : \X \rightarrow \R$ is a predefined threshold function---that is, 
we allow the thresholds to depend on the covariates through a fixed function.
For notational simplicity, we will write $C_{j,i} = c(X_{j,i})$ for each $(j,i)$ pair
hereafter.
Note that the $H_{j,i}$’s are ``random hypotheses''---they are technically events and differ from the classical notion of hypotheses, but we follow the convention in the related literature to facilitate formulating the problem as a multiple testing problem. The action of rejecting $H_{j,i}$ demonstrates evidence that the $i$-th individual in the $j$-th test group has an outcome exceeding the threshold $C_{j,i}$. Accordingly, 
we select the individuals whose corresponding null hypotheses are rejected.

We now aim to develop a multiple testing procedure that returns 
a rejection set $\mathcal{R}\subseteq \{(j,i) : j \in [M], i \in [N_{K+j}]\}$ 
with FDR control. 
With our notation, the FDR is defined as the expectation of 
false discovery (selection) proportion (FDP), given by
\[
\FDP = \frac{\sum_{j=1}^{M} \sum_{i=1}^{N_{K+j}} \One{(j,i) \in \cR, Y_{K+j,i} \leq C_{K+j,i}}}{|\cR| \vee 1}.\]

Recall that in our setup, the group size $N_k$ is random, and
consequently, the total number of hypotheses is also random. 
The expectation in the definition of FDR is therefore 
taken over the randomness of 
both the samples and the group sizes---this feature distinguishes our setting from 
standard multiple testing problems. Towards controlling FDR,  
we shall aim to control a slightly stronger notion of FDR, 
which conditions on the group sizes:
\begin{equation}\label{eqn:cond_fdr}
\cFDR = \EEst{\FDP}{N_{K+1:K+M}}.
\end{equation}
By the law of total expectation, controlling the above conditional FDR implies the control 
of the marginal FDR, i.e., $\EE{\FDP}$. 
The notion of $\cFDR$ provides better interpretability by effectively fixing the number of hypotheses, reflecting realistic scenarios in which the user already has information about the number of inputs when conducting inference.

Beyond the individual selection task formulated above, we will discuss 
extensions to hierarchical selection tasks (group nulls, hybrid nulls),
counterfactual selection, and the setting with distribution shift 
in Section~\ref{sec:extension} of the Appendix.

\subsection{Multiple testing with e-values}
The main statistical tool we use for testing the hypotheses in~\eqref{eqn:hypothesis} is the 
e-value~\citep{grunwald2024safe,shafer2021testing,vovk2021values,ramdas2024hypothesis}.
Similar to a p-value, an e-value is a measure of the strength of evidence against 
the null hypothesis. 
Formally, for a null hypothesis $H_0$ (deterministic or random), 
the e-value $e$ is defined as the realization of an e-variable $E$ s.t. 
$E \ge 0$ and $\mathbb{E}_{H_0}[E]\le 1$. 
When $H_0$ is random, 
we consider variables $E$ satisfying $E\ge 0$ and $\EE{E \cdot \One{H_0 \text{ is true}}} \le 1$.
Following the convention in the literature, 
we do not distinguish between the e-variable and its realization, using $e$ to denote both. 

Suppose there are $m$ hypotheses $H_1,\ldots,H_m$,
and each hypothesis $H_j$ is associated with an e-value $e_j$ for all $j \in [m]$.
\citet{wang2022false} proposed  the e-BH procedure, which takes 
the e-values as input, ranks them in a descending order 
$e_{(1)} \ge \cdots \ge e_{(m)}$, and rejects the hypotheses  
whose corresponding e-values exceed $m / (\alpha k^*)$. Here, 
$k^* = \max\{k \in [m]: e_{(k)} \ge m/(\alpha k)\}$, with the convention that 
$\max\{\varnothing\} = 0$. The e-BH procedure provably controls the FDR at level $\alpha$
as long as the e-values are valid. In fact, as pointed out by~\citet{wang2022false}, 
it suffices to impose the following conditions on the e-values to ensure FDR control:
\begin{align} 
\label{eq:compound_eval}
e_j \ge 0 \, \text{ and }  \sum_{j \in [m]} \EE{e_j \One{H_j \text{ true}}} \le m.
\end{align}
It can be easily seen that~\eqref{eq:compound_eval} is a relaxation of 
the ``bona fide'' e-values , which requires $\EE{e_j}\le 1$ for all $j\in[m]$. 
The (set of) e-values defined by~\eqref{eq:compound_eval}
are termed as the \textit{compound e-values} in~\citet{ignatiadis2024compound}.

Our work builds upon the e-BH procedure, extending it to the hierarchical data setting, 
where the number of hypotheses is random and the data has a hierarchical structure.
En route, we construct e-values for the hypotheses in~\eqref{eqn:hypothesis}, 
which are themselves new examples of efficient e-values for offline multiple 
testing problems.

\subsection{Related work}

Conformal prediction and distribution-free inference have gained significant attention in recent literature. An overview of this area is provided in~\citet{vovk2005algorithmic}, \citet{shafer2008tutorial},~\citet{angelopoulos2021gentle}, and~\citet{angelopoulos2024theoretical}. Conformal prediction and split conformal prediction—general frameworks for distribution-free predictive inference—are introduced in~\citet{vovk2005algorithmic} and~\citet{papadopoulos2008inductive}. Additionally, \citet{barber2021predictive} propose Jackknife+, a computationally feasible method for constructing distribution-free prediction sets without data splitting.

Several recent works have sought to extend the conformal prediction framework to structured or asymmetric data, 
where the i.i.d.~or exchangeability assumptions may not be reasonable. 
\citet{dunn2022distribution} and \citet{lee2023distribution} 
consider the hierarchical setting, providing methodologies that construct
distribution-free prediction sets for a test input in a new cluster;
\citet{liu2024multi} and~\citet{duchi2024predictive} consider the same setting,
but focus on cluster-level outcomes.  
\citet{dobriban2023symmpi} address the case of data with group symmetries. 
Other lines of work explore inference under distribution shift. \citet{tibshirani2019conformal} 
introduce {\em weighted conformal prediction}, which provides a distribution-free prediction set given knowledge of the likelihood ratio of feature distributions. Their method has been further developed in several subsequent works, 
such as \citet{lei2021conformal} and \citet{candes2023conformalized}.
In another line of work,~\citet{barber2023conformal,cauchois2024robust,ai2024not,gui2024distributionally}
consider prediction interval construction accounting for the worst-case distributional shifts within 
a class of distributions.

Predictive inference on multiple test points has also been a focus in a number of recent studies. 
For example, \citet{vovk2013transductive} discuss constructing a prediction region for the vector of multiple test outcomes, 
\citet{lee2024simultaneous} study the construction of simultaneous prediction sets for multiple outcomes under covariate shift, 
and \citet{lee2024batchpredictiveinference} propose a method for inference on a function of test points. 
Our work is closely related to \citet{jin2023selection, jin2023model}, which introduces a methodology for selecting test points 
under the i.i.d.~assumption or distribution shift. These works extend the results of \citet{10.1214/22-AOS2244}, 
which study the outlier detection problem; see also~\citet{marandon2024adaptive,bashari2024derandomized,liang2024integrative,gui2024conformal}.

Our work is also closely related to multiple hypothesis testing. 
Specifically, the control of the false discovery rate employs the e-BH procedure 
from~\citet{wang2022false}, which extends the Benjamini-Hochberg procedure~\citep{benjamini1995controlling} 
to work with e-values instead of p-values. E-values have attracted considerable attention in 
recent literature of multiple testing, with methods such as e-filter~\citep{gablenz2024catch} (which extends the p-filter~\citep{10.1214/18-AOS1765}) addressing multiple testing with a group structure. 
Further recent advancements in e-value-based multiple testing include works 
by~\citet{ignatiadis2024values, ren2024derandomised, lee2024boosting, fischer2024online,ignatiadis2024compound, xu2025bringing, goeman2025partitioning}, 
among others.

\section{Warm-up}
\label{sec:warmup}
To build intuition for our proposal, we begin with a review of the CS procedure~\citep{jin2023selection} 
and discuss why direct generalization of CS does not yield 
desirable procedures in our setting.

\subsection{Review of conformal selection} 
CS considers $K$ labeled samples $Z_1,\ldots,Z_K$ 
and $M$ test samples $X_{K+1},\ldots,X_{K+M}$ whose 
outcomes $Y_{K+1},\ldots, Y_{K+M}$ are unobserved. 
It assumes that the combined sequence $Z_1,\ldots$, $Z_{K+M}$ are jointly exchangeable.
This setup corresponds to a special case of our setting, 
where each group contains exactly one sample.

CS begins with splitting the labeled dataset into a training and a calibration fold;
on the training set, it constructs a score function $s : \X \times \Y \rightarrow \R$, 
where smaller values of $s(X,c)$ provide stronger 
evidence for rejecting $Y \leq c$. A simple example is $s(x,y) = y-\hat{\mu}(x)$, where $\hat{\mu}(\cdot)$ is an estimator of the mean function $\mu(x) = \EEst{Y}{X=x}$.
To streamline the exposition, we treat the score function as 
fixed---i.e., we condition on the training data and let $\{Z_{k}\}_{k \in [K]}$ denote the calibration set.

Next, for each sample $Z_{k,i}$, the procedure computes a score 
$\hat{V}_{k,i} = s(X_{k,i}, C_{k,i})$ for $k \in [K+M]$, $i \in [N_k]$. 
In particular, $\hat{V}_{K+j,i}$ serves as a test statistic for $H_{j,i}$, 
with smaller values indicating stronger evidence against the null. 
For each test point $j\in[M]$, CS constructs a p-value by comparing its
score to those of the null calibration points:
\begin{align}
\label{eq:cs_pval}
p_j = \frac{1+\sum_{k=1}^K \One{\hat V_{K+j} \ge \hat V_k, Y_k \le C_k}}{K+1}.
\end{align}
\citet{jin2023selection} proved that $p_j$ 
is a valid p-value in that 
$\PP{p_j \le t, Y_{K+j} \le C_{K+j}} \le t$, 
for any $t \in (0,1)$.
They further proved that applying BH (referred to as 
p-BH hereafter for distinction) to the $(p_j)_{j\in[M]}$ 
controls the FDR.


\subsection{Failure of direct generalization of CS}

A natural attempt to generalize the CS procedure to our setting 
is to construct a p-value for each hypothesis and apply the p-BH procedure.
As before, we assume without loss of generality---via sample splitting---that 
the score function $s(\cdot, \cdot)$ is fixed, and treat $\{Z_{k,i}\}_{k\in[K],i\in[N_k]}$
as the calibration points.

In our hierarchical exchangeable setting, to adapt the p-value defined in~\eqref{eq:cs_pval}, we construct a p-value for hypothesis $(j,i)$ by first drawing a single unit uniformly at random from each calibration group:
\begin{equation}\label{eqn:i_k_ast}
    i_k^* \mid N_k \sim \text{Unif}(\{1,2,\cdots,N_k\}), \text{ for } k\in[K].
\end{equation}
We then define the {\em subsampling conformal p-value} as:
\begin{align}\label{eq:sub_pval}
p^{\text{sub}}_{j,i} = \frac{1 + \sum_{k=1}^K 
\One{\hat V_{K+j,i} \ge \hat V_{k, i_k^*}, Y_{k,i_k^*} \le C_{k,i_k^*}}}{K+1}.
\end{align}
Hierarchical exchangeability ensures that the collection $Z_{1,i_1^*},\ldots,Z_{K,i_K^*},Z_{K+j,i}$ are jointly exchangeable,
so the validity of the $p^\sub_{j,i}$'s~\eqref{eq:sub_pval} follows directly 
from the same argument used for $p_j$'s~\eqref{eq:cs_pval}. 

However, the FDR control of $\cR_\pbh\{(p^{\text{sub}}_{j,i})_{j\in[M],i\in[N_{K+j}]}\}$---the rejection set obtained by applying p-BH to the $p_{j,i}$'s---does not follow, due to the potentially complex dependence structure among the p-values (note that the p-BH procedure requires certain conditions on the p-values, such as independence or positive dependence (PRDS condition); see \citep{benjamini2001control}).  We provide a counterexample in Section~\ref{appx:counter_sub} of the Appendix where this procedure results in a substantial FDR violation.

\subsection{Intuition behind our proposal}\label{sec:intuition}
Our proposal takes a different route---we will construct an e-value for 
each hypothesis such that applying the e-BH procedure to the e-values 
has `similar' empirical performance as applying p-BH to $p_{j,i}$'s in most cases,
while the e-value property ensures rigorous FDR control.

The construction builds on the observation that 
$\cR_\pbh\{(p^\sub_{j,i})_{j\in[M],i\in[N_{K+j}]}\}$
can be obtained by applying the e-BH to a set of heuristic ``e-values''.
The equivalence is formalized in the following proposition,
with its proof provided in Section~\ref{sec:prop_proof} of the Appendix.
\begin{proposition}\label{prop:equiv}
For any $j\in[M]$ and $i\in[N_{K+j}]$, define 
\begin{align}\label{eq:heuristic_eval}
& \tilde e_{j,i} = \frac{\One{\hat V_{K+j,i} < T^\pbh}}
{1 + \sum_{k=1}^K\One{\hat V_{k,i_k^*}< T^\pbh, Y_{k,i_k^*} \le C_{k,i_k^*}}}\cdot (K+1),
\end{align}
where $T^\pbh$ is a stopping time-type threshold defined as
\begin{equation}\label{eq:t_pbh}
\begin{split} 
& T^\pbh = \sup\big\{t \in \mathbb{R}: \hat{\FDP}^\pbh(t) \le \alpha\big\}, \text{ where} \\
& \hat{\FDP}^\pbh(t) = \frac{\sum_{k=1}^K \One{\hat V_{k,i_k^*}<t, Y_{k,i_k^*} \le C_{k,i_k^*}} +1}{1\vee \sum_{j=1}^M \sum_{i=1}^{N_{k+j}} \One{\hat V_{K+j,i}<t}} \cdot \frac{\sum_{j=1}^M N_{K+j}}{K+1}.
\end{split}
\end{equation}
Then there is 
$\cR_\pbh\{(p^\sub_{j,i})_{j\in[M],i\in[N_{K+j}]}\} 
= \cR_\ebh\{(\tilde e_{j,i})_{j\in[M],i\in[N_{K+j}]}\}$.
\end{proposition}
In the special case $N_k \equiv 1$, both 
rejection sets coincide with that of CS and 
FDR control is guaranteed. 
In the general case, if $\tilde e_{j,i}$'s {\em were} compound e-values, we could directly 
apply e-BH and complete the task. 
However, they are not valid (compound) e-values---otherwise 
$\cR_\pbh$ would already control the FDR
Nevertheless, we will show in the sequel 
that with appropriate modification of the threshold $T$, 
we can construct valid e-values whose empirical power closely matches that of 
their p-value counterpart.

\section{Main results}
\label{sec:method}
We present our main results in this section.  
We first extend previous results on e-values to the control of cFDR defined in~\eqref{eqn:cond_fdr} in Section~\ref{sec:cfdr}.
In Sections~\ref{sec:sub_eval} and~\ref{sec:hier_eval}, we present two e-value construction methods---the
subsampling  conformal e-values and the hierarchical conformal e-values. The two approaches present 
an interesting tradeoff between stability and statistical power, with the former being 
random but empirically more powerful while the latter stable yet less powerful. We 
leave the choice between these two approaches to the users depending on their practical 
desiderata.

\subsection{Controlling the cFDR}
\label{sec:cfdr}
Define the {\em conditional e-value} for $H_{j,i}$ to be a random variable $e_{j,i}$, which satisfies
\begin{equation}\label{eqn:e_value}
e_{j,i} \ge 0 \text{ and }    \EEst{e_{j,i} \cdot \One{H_{j,i}}}{N_{K+1:K+M}} \leq 1.
\end{equation}
Modifying the proof of~\citet{wang2022false}, we can show that applying the e-BH 
procedure to the above notion 
of e-values at level $\alpha$ controls the cFDR at level $\alpha$.

\begin{lemma}\label{lem:ebh}
    Suppose that random variables $(e_{j,i})_{1 \leq j \leq M, 1 \leq i \leq N_{K+j}}$ 
    satisfy the condition in~\eqref{eqn:e_value}. 
    Then the e-BH procedure applied to $(e_{j,i})_{1 \leq j \leq M, 1 \leq i \leq N_{K+j}}$ 
    at level $\alpha$ guarantees $\cFDR \leq \alpha$, where $\cFDR$ denotes the conditional false discovery rate~\eqref{eqn:cond_fdr}.
\end{lemma}

\begin{remark}
    The proof of Lemma~\ref{lem:ebh} in fact shows that the same result holds for compound e-values that satisfy the following weaker property:
    \[\EEst{\sum_{j=1}^M \sum_{i=1}^{N_{K+j}} e_{j,i} \One{H_{j,i}}}{N_{K+1:K+M}} \leq \sum_{j=1}^M N_{K+j}.\]
    Nonetheless, within the framework of conformal selection, the above condition does not represent a substantial relaxation, since most typical choices of e-values are inherently symmetric, reflecting the exchangeability assumption on which they are based.
\end{remark}

The complete proof of Lemma~\ref{lem:ebh} is deferred to the Appendix (Section~\ref{sec:proofs}). 
We note that even if our goal is solely to control the marginal FDR, the condition in~\eqref{eqn:e_value} on the group size-conditional expectation is still necessary to achieve the guarantee through the e-BH procedure, as the number of hypotheses is random. 


\subsection{Selection with subsampling conformal e-values}
\label{sec:sub_eval}



\begin{figure}[h!]
\centering
\includegraphics[width = 0.8\textwidth]{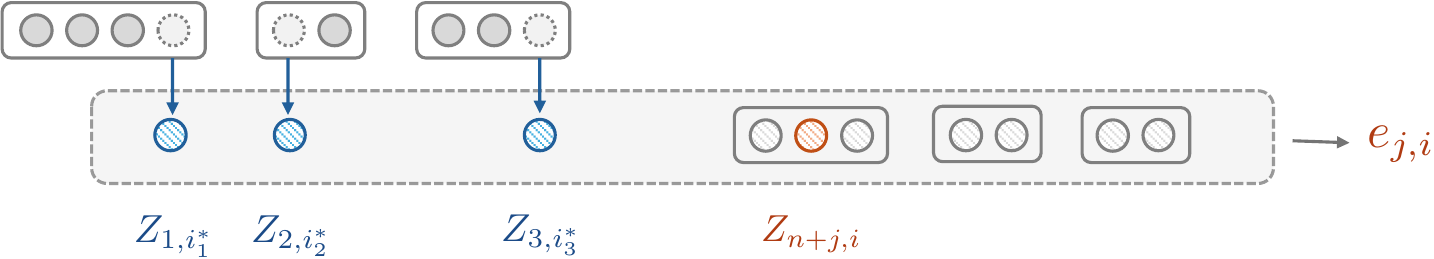}   
\caption{Illustration of the construction of subsampling conformal e-values. 
The subsampling procedure selects one unit from each group uniformly at random in the 
calibration data. The e-value $e_{j,i}$ is constructed based on the subsampled units 
and the test data.}
\label{fig:sub_eval}
\end{figure}

As alluded to earlier, we will construct valid e-values by modifying the $\tilde e_{j,i}$'s defined in~\eqref{eq:heuristic_eval}.
Recall that for each calibration group $k\in[K]$, 
we draw a single unit $i_k^*$ uniformly at random.
For the $i$-th unit in the $j$-th test group, 
we define the {\em subsampling conformal e-value} as
\begin{equation}\label{eqn:e_j_i}
    e_{j,i} = \frac{\One{\hat{V}_{K+j,i} < T_j}}{\sum_{k=1}^K \One{\hat{V}_{k,i_k^*} < T_j, Y_{k,i_k^*} \leq C_{k,i_k^*}} + 1} \cdot (K+1),
\end{equation}
where the threshold $T_j$ is group-specific and defined as
\begin{equation}\label{eqn:stopping}
\begin{split}
&T_j = \sup\left\{t \in \R : \widehat{\FDP}_j(t) \leq \tilde{\alpha}\right\}, \text{ with }\\
&\widehat{\FDP}_j(t) = \frac{\sum_{k=1}^K \One{\hat{V}_{k,i_k^*} < t, Y_{k,i_k^*} \leq C_{k,i_k^*}} + 1}{1 \vee \sum_{l \neq j} \sum_{i'=1}^{N_{K+l}} \One{\hat{V}_{K+l,i'} < t}} \cdot \frac{\sum_{l=1}^{M} N_{K+l}}{K+1}.
\end{split}
\end{equation}
Above, $\tilde \alpha \in (0,1)$ is a pre-defined threshold chosen by 
the user---it does not necessarily need to be $\alpha$, 
and its choice only affects the power;
see Figure~\ref{fig:sub_eval} for a schematic illustration of the  
e-value construction process.
Intuitively, the non-null scores tend to be small, so the numerator of~\eqref{eqn:e_j_i}  
tends to be positive when $H_{j,i}$ is not true; the null scores, on the other hand, tend to be large, leading to a relatively small denominator.
The resulting e-value is therefore expected to be large for non-nulls.

Compared with~\eqref{eq:heuristic_eval}, we can see that the 
only difference between $e_{j,i}$ and $\tilde e_{j,i}$ lies in the threshold used. 
In particular, $\widehat{\FDP}_j$ differs from $\widehat{\FDP}^{\textnormal{p-BH}}$ only in that the summation in the denominator 
excludes the $j$-th test group. Thus, the threshold $T_j$ in the construction of the e-value can 
also be viewed as a `correction' of $T^{\textnormal{p-BH}}$ that enables valid $\FDR$ control.
This seemingly minor modification is, in fact, crucial for establishing the e-value property—and, consequently, for ensuring FDR control of the e-BH procedure when applied to the
$e_{j,i}$'s.

From another perspective, roughly speaking, the method based on the above e-value~\eqref{eqn:e_j_i} can be viewed as a test with rejection rule $\widehat{V}_{K+j,i} < T_j$ that rejects the null when the corresponding score is less than a threshold $T_j$, which is designed to approximately control $\FDP$ through $\widehat{\FDP}_j$---but to ensure theoretical validity, the e-value~\eqref{eqn:e_j_i} with a valid normalizing constant, combined with the e-BH procedure, is followed.

The full procedure is summarized in Algorithm~\ref{alg:main}.
In the following theorem, we formally show 
that $e_{j,i}$'s are valid conditional e-values, and that applying the e-BH
procedure to $(e_{j,i})_{j\in[M],i\in[N_{K+j}]}$ controls cFDR.
Its proof is deferred to the Appendix.

\begin{theorem}\label{thm:individual}
Under Assumption~\ref{asm:dist}, for any choice of $\tilde{\alpha} \in (0,1)$,  
the statistic $e_{j,i}$ defined by~\eqref{eqn:e_j_i} is an e-value for $H_{j,i}$ 
conditional on $N_{K+1:K+M}$ (i.e., $e_{j,i}$ satisfies~\eqref{eqn:e_value}). 
Consequently,
the e-BH procedure applied to $(e_{j,i})_{1 \leq j \leq M, 1 \leq i \leq N_{K+j}}$ at level $\alpha$ guarantees $\cFDR \leq \alpha$.
\end{theorem}

\begin{algorithm}[h!]
\caption{Selection of hierarchical data with subsampling conformal e-values}
\label{alg:main}
\KwIn{Calibration data $(X_{k,i},Y_{k,i})_{1 \leq k \leq K, 1 \leq i \leq N_k}$, Score function $s : \X \times \Y \rightarrow \R$, Test inputs $(X_{k,i})_{K+1 \leq k \leq K+M, 1 \leq i \leq N_k}$, Cutoff function $c : \X \rightarrow \R$, Target level $\alpha$, Parameter for threshold $\tilde{\alpha}$.}

\vskip+2ex

{\bf Step 1:} Draw $i_k^* \mid N_k \sim \text{Unif}(\{1,2,\cdots,N_k\})$ for $k \in [K]$.
\vskip 1ex

{\bf Step 2:} Compute $C_{k,i_k^*} = c(X_{k,i_k^*})$ and $\hat{V}_{k,i_k^*} = s(X_{k,i_k^*},C_{k,i_k^*})$ for $k\in[K]$, and $\hat{V}_{K+j,i} = s(X_{K+j,i}, c(X_{K+j,i}))$ for $j \in [M]$, $i \in [N_{k+j}]$.
\vskip 1ex

{\bf Step 3:} Compute the threshold $T_j$ according to~\eqref{eqn:stopping}. 
\vskip 1ex

{\bf Step 4:} Compute the e-value $e_{j,i}$ according to~\eqref{eqn:e_j_i} for $j\in[M]$ and $i \in [N_{K+j}]$.
\vskip 1ex

{\bf Step 5:} Sort the e-values: $e_{(1)} \geq e_{(2)} \geq \cdots \geq e_{(n_\text{test})}$, where $n_\text{test} = \sum_{k=1}^{M} N_{k+K}$.
\vskip 1ex

{\bf Step 6:} (e-BH procedure): Compute $l^* = \max\Big\{l \in [n_\text{test}]: \frac{l e_{(l)}}{n_\text{test}} \geq \frac{1}{\alpha}\Big\}$. 
\vskip +2ex
\KwOut{Selection set $\mathcal{R} = \{(j,i) : e_{j,i} \geq e_{(l^*)}\}$.}
\end{algorithm}


\begin{remark}[Choice of $\tilde \alpha$]
\label{rmk:alpha}

By Theorem~\ref{thm:individual}, any choice of $\tilde \alpha \in(0,1)$ 
guarantees FDR control; however, it does influence the power of the resulting procedure.  
For improved statistical power, a recommended choice for $\tilde{\alpha}$ is a 
value slightly smaller than $\alpha$, e.g.,
$\tilde{\alpha} = 0.9\alpha$~\citep{ren2024derandomised}.
The intuition behind this choice is that $\tilde{e}_{j,i}$'s are 
{\em sharp} with respect to e-BH in the sense that the last non-zero 
$\tilde e_{j,i}$ lies (almost) exactly on the threshold (see panel (a) 
in Figure~\ref{fig:alpha_choice}); after modifying the threshold 
$T^\pbh$ to be $T_j$, the resulting e-values become {\em perturbed}: if 
the last non-zero e-value falls slightly below the threshold, 
no rejections are made (see panel (b) in Figure~\ref{fig:alpha_choice}); 
By choosing $\tilde \alpha$ to be slightly smaller than $\alpha$, 
the resulting e-values become less sharp, introducing slack that helps avoid the no-rejection case under such perturbations (see panel (c) in Figure~\ref{fig:alpha_choice}). Observe that, in the definition of the e-value in~\eqref{eqn:e_j_i}, a smaller $T_j$ increases the value of the e-value whenever it is nonzero. Thus, a smaller value of $\tilde{\alpha}$, which leads to a smaller value of $T_j$, in turn leads to overall larger values of the e-values.
Note also that the rejection threshold remains $m/(\alpha k)$, while the curve $m/(\tilde{\alpha} k)$ illustrates the region in which the nonzero e-values can be located.

\begin{figure}[h!]
\centering
\begin{minipage}{0.3\textwidth}
\centering
\begin{tikzpicture}
  \begin{axis}[
    width=6cm,         
    height=5cm,        
    axis lines = middle,
    xlabel = $k$,
    ylabel = $\tilde e_{(k)}$,
    ylabel style={yshift=3ex},  
    domain=1:10,
    samples=100,
    ymin=0, ymax=1,
    xmin=0, xmax=10,
    restrict y to domain=0:1,
    enlargelimits=true,
    ytick=\empty,
    legend pos=north east
  ]
    \addplot[gray, thick, dashed] {1/x};
    \addlegendentry{$y = \frac{m}{\alpha k}$}
    \addplot[
      only marks,
      mark=*,
      mark size=2.5pt,
      thick,
      draw = black!60,
      fill = black!10
    ] coordinates {
      (1, 1/6)
      (2, 1/6)
      (3, 1/6)
      (4, 1/6)
      (5, 1/6)
      (6, 1/6)
      (7, 0)
      (8, 0)
      (9, 0)
      (10, 0)
    };
  \end{axis}
\end{tikzpicture}
{(a) $\cR_\ebh(\tilde e_{j,i})$}
\end{minipage}
\hfill
\begin{minipage}{0.3\textwidth}
\centering
\begin{tikzpicture}
  \begin{axis}[
    width=6cm,         
    height=5cm,        
    axis lines = middle,
    xlabel = $k$,
    ylabel = $e^\sub_{(k)}(\alpha)$,
    ylabel style={yshift=3ex},  
    domain=1:10,
    samples=100,
    ymin=0, ymax=1,
    xmin=0, xmax=10,
    restrict y to domain=0:1,
    enlargelimits=true,
    ytick=\empty,
    legend pos=north east
  ]
    \addplot[gray, thick, dashed] {1/x};
    \addlegendentry{$y = \frac{m}{\alpha k}$}
    \addplot[
      only marks,
      mark=*,
      mark size=2.5pt,
      thick,
      draw = black!60,
      fill = black!10
    ] coordinates {
      (1, 1/6)
      (2, 1/6)
      (3, 1/6)
      (4, 1/7)
      (5, 3/24)
      (6, 3/24)
      (7, 0)
      (8, 0)
      (9, 0)
      (10, 0)
    };
  \end{axis}
\end{tikzpicture}
{(b) $\cR_\ebh(e^\sub_{j,i}(\alpha))$}
\end{minipage}
\hfill
\begin{minipage}{0.3\textwidth}
\centering
\begin{tikzpicture}
  \begin{axis}[
    width=6cm,         
    height=5cm,        
    axis lines = middle,
    xlabel = $k$,
    ylabel = $e^\sub_{(k)}(\tilde \alpha)$,
    ylabel style={yshift=3ex},  
    domain=1:10,
    samples=100,
    ymin=0, ymax=1,
    xmin=0, xmax=10,
    restrict y to domain=0:1,
    enlargelimits=true,
    ytick=\empty,
    legend pos=north east
  ]
    \addplot[gray, thick, dashed] {1/x};
    \addlegendentry{$y = \frac{m}{\alpha k}$};
    \addplot[red!70!white, thick, dashed] {2/x};
    \addlegendentry{$y = \frac{m}{\tilde \alpha k}$}
    \addplot[
      only marks,
      mark=*,
      mark size=2.5pt,
      thick,
      draw = black!60,
      fill = black!10
    ] coordinates {
      (1, 2/5)
      (2, 2/5)
      (3, 2/5)
      (4, 1/3)
      (5, 1/3)
      (6, 0)
      (7, 0)
      (8, 0)
      (9, 0)
      (10, 0)
    };
  \end{axis}
\end{tikzpicture}
{(c) $\cR_\ebh(e^\sub_{j,i}(\tilde \alpha))$}
\end{minipage}
\caption{Illustration of the selection mechanism of e-BH with different e-values.
The x-axis corresponds to the rank of the e-value, 
and the y-axis the $k$-th largest e-value 
$e_{(k)}$. }
\label{fig:alpha_choice}
\end{figure}
\end{remark}

\begin{remark}
In Section~\ref{sec:remarks} of the Appendix, we provide further discussion on the general form of possible e-values and the choice of the e-value used in this work. We also discuss an improvement of the procedure based on the U-eBH procedure~\citep{xu2023more}, with justification that the same approach applies in our setting involving ``random hypotheses'' and aiming for conditional FDR. Detailed explanations are provided in Section~\ref{sec:uebh} of the Appendix.
\end{remark}

\subsection{Selection with hierarchical conformal e-values}\label{sec:hier}
\label{sec:hier_eval}
The procedure with the e-values~\eqref{eqn:e_j_i} leads to valid inference, and demonstrates good power, as we will illustrate through simulations. However, since it requires a subsampling step in which a sample is drawn from each group (and the rest are discarded), this may lead to unstable results when the group number is small. Specifically, although the average of the false discovery proportion (FDP) and the empirical power---i.e., 
the FDR and power---are both satisfactory across multiple trials, the FDP and empirical power in individual trials can be highly variable depending on the subsampled calibration set. 
One strategy to address this is to merge the e-values as described in Section~\ref{sec:merge} of the Appendix. 
However, it turns out that merging e-values with different thresholds/stopping times often results in a significant loss of power.

\begin{figure}[ht]
    \centering
    \includegraphics[width=0.8\linewidth]{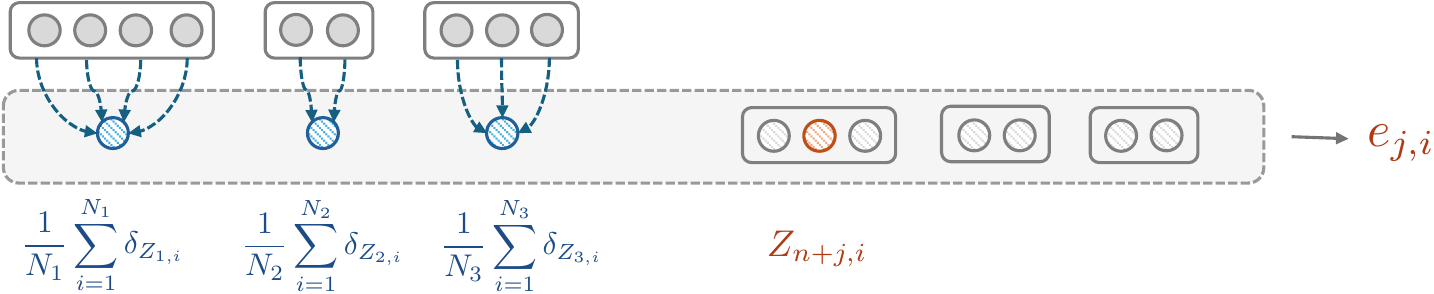}
    \caption{An illustration of the hierarchical conformal e-value. For group $k\in[K]$,
    the data points are combined with a weight $1/N_k$. The hierarchical conformal e-value 
    is constructed by contrasting the test unit with the aggregated calibration units.}
    \label{fig:hierarchical_eval}
\end{figure}

Alternatively, one can consider directly constructing a test statistic that utilizes all observations in the calibration set, yielding a non-randomized procedure with a unified threshold. The intuition behind this approach is that: instead of randomly drawing a data point from each group, we `aggregate' the data points within a group by weighting each data point inversely proportional to its group
size---such that the `aggregated unit' is still comparable to the test unit, roughly speaking. We then construct an e-value 
by contrasting the aggregated calibration units with the test unit, and show its validity using the within- and across-group 
exchangeability. Figure~\ref{fig:hierarchical_eval} is a pictorial demonstration of this scheme.

Before introducing the exact construction of such e-values, 
we first examine the following \textit{hierarchical conformal p-value} to provide intuition:
\begin{equation}\label{eqn:hier_p}
    p^\hier_{j,i} = \frac{\sum_{k=1}^K \frac{1}{N_k}\sum_{i' =1}^{N_k}\One{V_{k,i'} \leq \hat{V}_{K+j,i}}+1}{K+1}, \qquad 
    \forall j\in[M], i\in[N_{N+j}].
\end{equation}
This construction of the hierarchical conformal p-value is motivated 
by the hierarchical conformal prediction methodology introduced by~\citet{lee2023distribution}. 
With the additional assumption of score monotonicity, it can be shown that the above $p_{j,i}$ is indeed a valid p-value.
We formally state this result in the proposition below and delegate its proof to Appendix.

\begin{proposition}\label{prop:hier_p}
    Suppose that the hierarchical data $\tilde{Z}_1, \cdots, \tilde{Z}_K$, $\tilde{Z}_{K+1}, \cdots, \tilde{Z}_{K+M}$ satisfies Assumption~\ref{asm:dist}, and that $s(x,y_1) \leq s(x,y_2)$ holds for any $x \in \X$ and $y_1 \leq y_2$. Then $p_{j,i}$ defined as~\eqref{eqn:hier_p} satisfies $\PP{p_{j,i} \leq \alpha \text{ and } Y_{K+j,i} \leq C_{K+j,i}} \leq \alpha$, for any $1 \leq j \leq M$ and $1 \leq i \leq N_{K+j}$.
\end{proposition}

Note that the standard choice of score $s(x,y) = y - \hat{\mu}(x)$ satisfies the monotonicity condition. Alternatively, one can consider a direct extension of the p-value~\eqref{eq:sub_pval} as follows:
\begin{equation}\label{eqn:hier_p_2}
    p_{j,i}^\hier = \frac{\sum_{k=1}^K \frac{1}{N_k}\sum_{i'=1}^{N_k}\One{\hat{V}_{k,i'} \leq \hat{V}_{K+j,i}, Y_{K,i'} \leq C_{k,i'}}+1}{K+1}, \quad 
    \forall j\in [M], i \in [N_{K+j}].
\end{equation}
The validity of the above $p_{j,i}$ is ensured by applying
Proposition~\ref{prop:hier_p} with the score $\tilde{s}(x,y) = s(x,c(x)) 1\{y\le c(x)\} + \infty 1\{y>c(x)\}$, which is monotone with respect to $y$\footnote{Note that by construction, for any $x,x' \in \X$ and $y \in \R$, we have the equivalence $\tilde{s}(x,y) \leq \tilde{s}(x',c(x')) \iff s(x,c(x)) \leq s(x',c(x')) \text{ and } y \leq c(x)$.} Based on these observations, an intuitive option for the selection with FDR control is to apply the BH procedure to $p_{j,i}$s from~\eqref{eqn:hier_p} or~\eqref{eqn:hier_p_2}. But as before, the resulting procedure is not guaranteed to control the FDR; an example with 
FDR violation is in Section~\ref{appx:counter_sub} of the Appendix.

Instead, we now introduce the e-value counterpart to the above p-value(s) and 
propose a theoretically valid procedure. Similar to the subsampling p-value case, 
we observe that 
\[\cR_\pbh\{(p^\hier_{j,i})_{j\in[M], i\in[N_{K+j}]}\} = \cR_\ebh\{(\bar e_{j,i})_{j\in[M], i\in[N_{K+j}]}\}, \]
where
\begin{align*}
\bar e_{j,i} = \frac{\One{\hat{V}_{K+j,i} < \tilde T^\pbh}}{\sum_{k=1}^K \frac{1}{N_k} \sum_{i'=1}^{N_k} \One{\hat{V}_{k,i'} < \tilde T^\pbh, Y_{k,i'} \leq C_{k,i'}} + 1} \cdot (K+1), 
\end{align*}
and the threshold $\tilde T^\pbh$ is defined analogously to~\eqref{eq:t_pbh}
by replacing the FDP estimate with
\begin{align}\label{eqn:stopping0}
    &\overline{\FDP}^\pbh(t) = \frac{\sum_{k=1}^{K} \frac{1}{N_k}\sum_{i'=1}^{N_k} \One{\hat{V}_{k,i'} < t, Y_{k,i'} \leq C_{k,i'}}+1}{1 \vee \sum_{\ell = 1}^M \sum_{i'=1}^{N_{K+\ell}} \One{\hat{V}_{K+\ell,i'} < t}} \cdot \frac{\sum_{\ell =1}^M N_{K+\ell}}{K+1}.
\end{align}
While these $\bar e_{j,i}$'s are not valid (compound) e-values, their structure  
motivates our construction. Specifically, we construct the {\em hierarchical 
conformal e-values} by modifying the thresholds in~\eqref{eqn:stopping0},
with a crucial refinement: the thresholds in the numerator and denominator
are now modified differently. For any $j\in[M]$ and $i\in[N_{K+j}]$, let us define
\begin{equation}\label{eqn:e_hierarchical}
    e_{j,i} = \frac{\One{\hat{V}_{K+j,i} < T_j^+}}{\sum_{k=1}^K \frac{1}{N_k} \sum_{i'=1}^{N_k} \One{\hat{V}_{k,i'} < T_j^-, Y_{k,i'} \leq C_{k,i'}} + 1} \cdot (K+1), 
\end{equation}
and
\begin{equation}\label{eqn:stopping1}
\begin{split}
    &T_j^\star = \sup\left\{t \in \R : \widehat{\FDP}_j^\star(t) \leq \tilde{\alpha}\right\}, \text{ where }\\
    &\widehat{\FDP}_j^\star(t) = \frac{\sum_{k=1}^{K} \frac{1}{N_k}\sum_{i'=1}^{N_k} \One{\hat{V}_{k,i'} < t, Y_{k,i'} \leq C_{k,i'}}+\mathbbm{1}\{\star = +\}}{1 \vee \sum_{l \neq j} \sum_{i'=1}^{N_{K+l}} \One{\hat{V}_{K+l,i'} < t}} \cdot \frac{\sum_{l \neq j} N_{K+l}}{K+1},
\end{split}
\end{equation}
for $\star\in\{+,-\}$ and a predefined $\tilde{\alpha} \in (0,1)$. Intuitively, 
the use of group-specific thresholds ensures that $e_{j,i}$ coincides with 
a soft-rank e-value (as introduced in Section~\ref{rmk:softrank} of the Appendix) on the event 
$\{Y_{K+j,i}\le C_{K+j,i}, \hat V_{K+j,i} < T_j^+\}$. 
The validity of $e_{j,i}$ then follows from exchangeability. Note that the power of the e-value mostly depends on the number of groups $K$, while the within-group sample size is more closely related to its stability.
The following theorem formally shows that the above $e_{j,i}$ is a valid conditional e-value 
for the null $H_{j,i}$. As a consequence, applying the e-BH procedure to $(e_{j,i})_{1 \leq j \leq M, 1 \leq i \leq N_{K+j}}$
ensures cFDR control, and therefore FDR control.

\begin{theorem}\label{thm:hierarchical}
Suppose the hierarchical data $\tilde{Z}_1, \cdots, \tilde{Z}_K$, $\tilde{Z}_{K+1}, \cdots, \tilde{Z}_{K+M}$ satisfies Assumption~\ref{asm:dist}. Then the statistic $e_{j,i}$ defined as~\eqref{eqn:e_hierarchical} is an e-value for $H_{j,i}$, conditional on $N_{K+1:K+M}$. Consequently, the e-BH procedure, applied to $(e_{j,i})_{1 \leq j \leq M, 1 \leq i \leq N_{K+j}}$ defined by~\eqref{eqn:e_hierarchical} at level $\alpha$, controls the $\cFDR$ at level $\alpha$.
\end{theorem}

We provide the proof of Theorem~\ref{thm:hierarchical} in the Appendix and summarize the complete procedure in Algorithm~\ref{alg:hierarchical}. The hierarchical conformal e-value-based procedure utilizes all the information in the calibration set, 
and thus provides stable results. 

As in the subsampling-based method, one can boost the e-values by dividing them by a 
superuniform random variable. Again, to avoid introducing external randomness, 
we could obtain such a random variable using internal randomness as discussed in Remark 1 in the Appendix.
We will illustrate the performance of both the basic and boosted procedures through simulations 
in the next section. 

\begin{algorithm}
\caption{Selection of hierarchical data with hierarchical conformal e-values}
\label{alg:hierarchical}
\KwIn{Calibration data $(X_{k,i},Y_{k,i})_{1 \leq k \leq K, 1 \leq i \leq N_k}$, Score function $s : \X \times \Y \rightarrow \R$, Test inputs $(X_{k,i})_{K+1 \leq k \leq K+M, 1 \leq i \leq N_k}$, Cutoff function $c : \X \rightarrow \R$, Target level $\alpha$, Parameter for threshold $\tilde{\alpha}$.}

\vskip 2ex

{\bf Step 1:} Compute $C_{k,i} = c(X_{k,i})$ and $\hat{V}_{k,i} = s(X_{k,i},C_{k,i})$ for $k\in[K+M]$ and $i\in [N_k]$. 
\vskip 1ex
{\bf Step 2:} Compute $T_j^+$ and $T_j^-$ according to~\eqref{eqn:stopping1}, 
for $j\in[M]$.
\vskip 1ex
{\bf Step 3:} Compute the e-values $e_{j,i}$ according to~\eqref{eqn:e_hierarchical} 
for $j \in [M]$ and $i \in [N_{K+j}]$.
\vskip 1ex
{\bf Step 4:} Sort the e-values: $e_{(1)} \geq e_{(2)} \geq \cdots \geq e_{(n_\text{test})}$, where $n_\text{test} = \sum_{k=K+1}^{K+M} N_k$.
\vskip 1ex
{\bf Step 5:} (e-BH procedure): Compute $l^* = \max\big\{l \in [n_\text{test}] : \frac{l e_{(l)}}{n_\text{test}} \geq \frac{1}{\alpha}\big\}$. 

\vskip 2ex 

\KwOut{Selection set $\cR = \{(j,i) : e_{j,i} \geq e_{(l^*)}\}$.}
\end{algorithm}

\section{Extensions}

In Section~\ref{sec:extension} of the Appendix, we discuss possible extensions of the methods introduced in the previous sections. Here, we provide an overview and defer detailed explanations to the Appendix.

\paragraph{Procedure for joint inference on groups and individuals.}
We extend our procedure to settings where group-level hypotheses are also tested---e.g., selecting groups with large outcome means or testing group-global nulls---in addition to selecting individuals.

\paragraph{Inference under group covariate shift.}
We consider settings with group covariate shift. We introduce weighted subsampling conformal e-values---motivated by~\citet{tibshirani2019conformal}---and prove the validity of the e-BH procedure applied to these e-values.

\paragraph{Selection based on individual treatment effects.}
We consider selecting individuals based on their individual treatment effects (ITE)---e.g., testing the null hypothesis $H_{j,i}: Y_{K+j,i}(1) \leq Y_{K+j,i}(0)$ (where $Y_{K+j,i}(1)$ and $Y_{K+j,i}(0)$ denote the potential outcomes), which is equivalent to selecting individuals with positive ITE. This problem represents a nontrivial extension of the standard setting, since the threshold $Y_{K+j,i}(0)$ is unknown. We show how to construct valid e-values for the null $H_{j,i}$ and introduce subsampling- and hierarchical-version e-values tailored to this problem.

\section{Simulations}
\label{sec:experiments}
In this section, 
we illustrate the performance of the proposed procedures in a variety of 
simulation settings.\footnote{Code to reproduce the experiments is available at \url{https://github.com/yhoon31/selection_hierarchical}.}
Section~\ref{sec:sim_sub} and~\ref{sec:sim_hierarchy} 
investigate the selection of individuals, with 
subsampling conformal e-value and the 
hierarchical conformal e-values, respectively. Additional experimental results are provided in Section~\ref{sec:add_exp} of the Appendix, covering the selection of both groups and individuals, and selection based on individual treatment effects.

\paragraph{Data-generating process.}
Throughout this section, the data 
is generated from the following data-generating process:
\begin{align}\label{eq:sim_dgp}
    \begin{split}
    &G \sim \textnormal{Unif}([-5,5])^{p_G},\\
    &N \sim 2+\textnormal{Poisson}(\lambda),\\
    &X_1, X_2,\cdots,X_N \mid G,N \iidsim N_p(A G, 3 \cdot I_p),\\
    &Y_i \mid X_i, G_i, N_i \sim N(\beta_1^\top X_i + \log |\beta_2^\top G|, \sigma^2\cdot\|X\|/p), \textnormal{ for } i=1,2,\cdots,N,
    \end{split}
\end{align}
where we set the dimensions of the group-specific covariates and 
the individual-specific covariates to be $p_G = 10$ and $p=20$, respectively. 
The parameters $\beta_1 \in \R^p$ and $\beta_2 \in \R^{p_G}$ are generated by drawing each component from a uniform distribution, and $A \in \R^{p \times p_G}$ is generated by drawing each entry from a standard normal distribution (once generated, 
they are fixed throughout the repetitions). 
We consider the task of selecting individuals whose outcome values exceed $c=20$, i.e., we test $H_{j,i} : Y_{j,i} \leq 20$.

\subsection{Selecting individuals with subsampling conformal e-values}
\label{sec:sim_sub}
We first explore the performance of e-BH applied to subsampling conformal 
e-values under different group number/size settings. 
We set $\sigma=1$ and consider three scenarios of group size distributions: 
(1) $\lambda=0$, (2) $\lambda=5$, and (3) $\lambda=10$. 

\paragraph{Implementation details.}
In our simulation, $K_{\text{train}} = 100$ groups of training data are generated. 
We then use random forest regression to construct an estimator $\hat{\mu}: \R^{p_G} \times \R^p \mapsto \R$ 
that takes both the group-specific covariates and the individual-specific covariates as inputs;
based on $\hat \mu$, the score function is constructed 
as $s(g,x,y) = y - \hat{\mu}(g,x)$. We repeat the following steps 500 times, 
reporting the averaged results.

In each trial, we generate calibration data with $K=200$ groups and a test set with $M$ groups, 
where we conduct the experiment with three different choices of $M$: $20, 50$, and $200$. 
Then we run the proposed procedure (Algorithm~\ref{alg:main}), as well as the p-value-based procedure 
given by~\eqref{eq:sub_pval}, 
at levels $\alpha=0.05,0.075,\cdots,0.25$. For the e-value construction, 
we take the threshold level $\tilde{\alpha} = 0.9\alpha$. We additionally apply the U-eBH procedure with the boosted e-values as described in Section~\ref{sec:uebh} of the Appendix. We compute the false discovery proportion and the empirical power of the two procedures in each trial, and then calculate their average to obtain the estimates of the FDR and the power. 

\begin{figure}[htbp]
    \centering
    \includegraphics[width=0.85\linewidth]{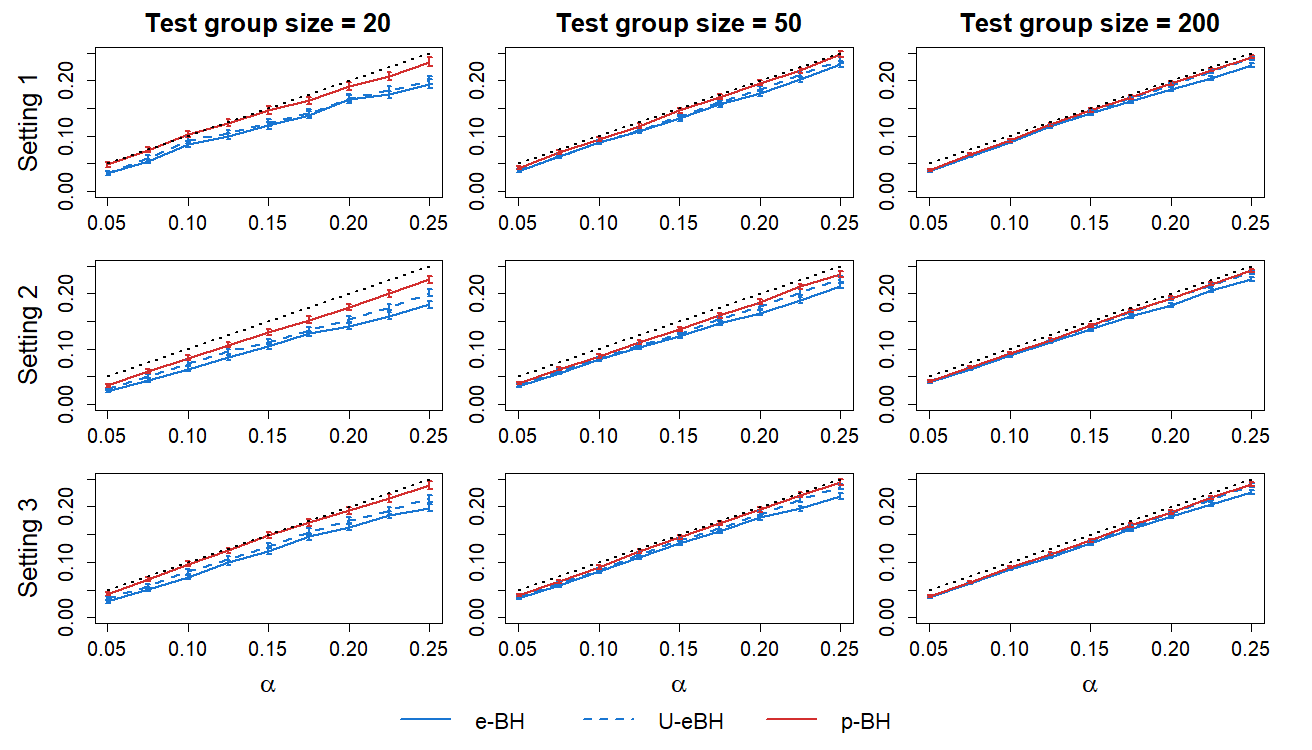}
    \caption{Estimated false discovery rates of Algorithm~\ref{alg:main} (e-BH with subsampling conformal e-values), its boosted version (U-eBH with subsampling conformal e-values), and p-BH~\eqref{eq:sub_pval}, across various group size settings and test sizes, with standard errors, at levels $\alpha=0.05,0.075,\cdots,0.25$. The dotted line corresponds to the $y=x$ line.}
    \label{fig:FDR_individual}
\end{figure}

\begin{figure}[htbp]
    \centering
    \includegraphics[width=0.85\linewidth]{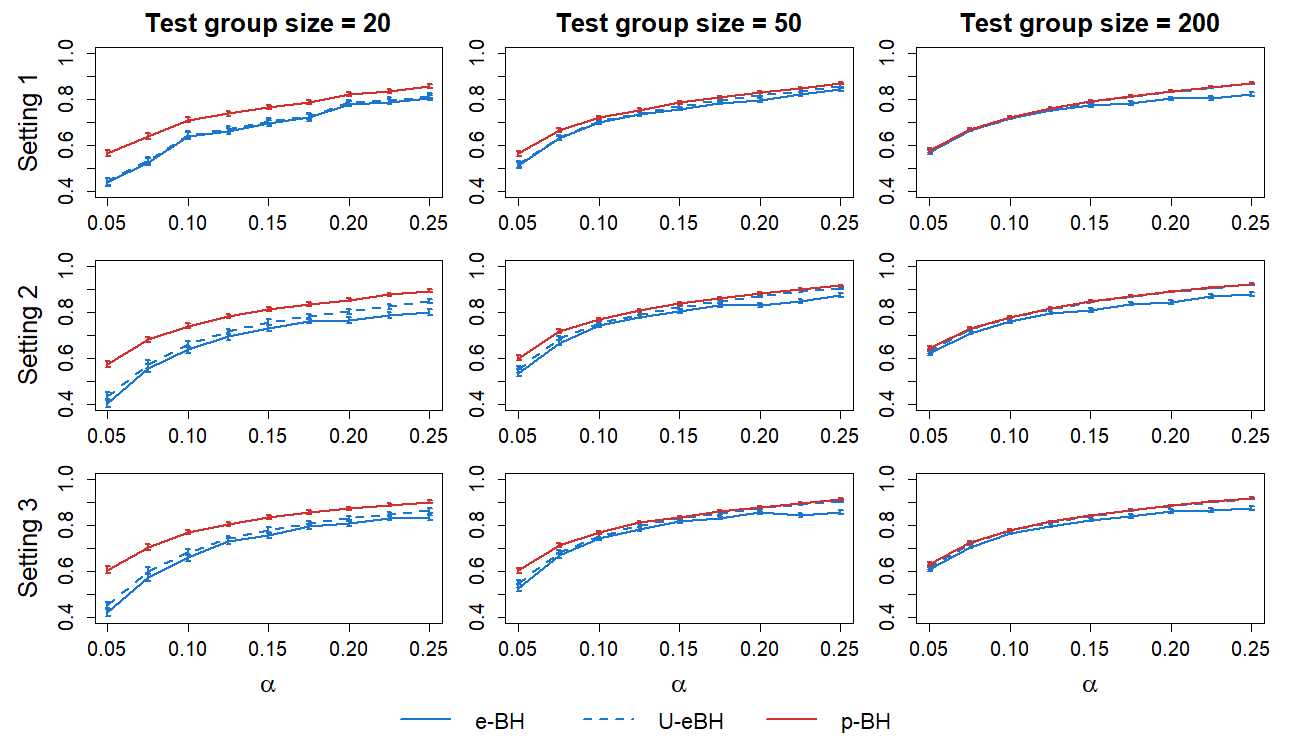}
    \caption{Estimated power of Algorithm~\ref{alg:main} (e-BH+subsampling conformal e-values), 
    its boosted version (U-eBH+subsampling conformal e-values), and p-BH~\eqref{eq:sub_pval}
    across different group size settings and test sizes, with standard errors, at levels $\alpha=0.05,0.075,\cdots,0.25$.}
    \label{fig:power_individual}
\end{figure}

\paragraph{Results.}
The estimated FDR and power, along with their standard errors, 
are shown in Figure~\ref{fig:FDR_individual} and~\ref{fig:power_individual}, respectively.
In all settings, our proposed method (Algorithm~\ref{alg:main}) controls the FDR as desired. 
The boosted version shows a slight improvement in terms of power, where the improvement 
becomes negligible for $K$ large enough. It is also worth noting that e-BH
behaves almost identically to the p-value-based method, while the latter does not 
ensure FDR control; the difference seems to diminish as $K$ increases.

\subsection{Selecting individuals with hierarchical conformal e-values}
\label{sec:sim_hierarchy}
Next, we demonstrate the experimental results for the method based on hierarchical conformal e-values. Specifically, we explore the following methods for comparison:

\begin{enumerate}
    \item [(1)] The e-BH procedure applied to the hierarchical conformal e-values (Algorithm~\ref{alg:hierarchical}).
    \item [(2)] Algorithm~\ref{alg:hierarchical} with boosting, i.e., the U-eBH procedure applied to the hierarchical conformal e-values.
    \item [(3)] BH procedure applied to the hierarchical conformal p-values~\eqref{eqn:hier_p}.
    \item [(4)] BH procedure applied to the $p_{j,i}$'s defined in~\eqref{eqn:hier_p_2}.
\end{enumerate}

Recall that the first two methods using e-values have a theoretical guarantee for FDR control, 
while the remaining two methods do not. The results of p-value-based methods are provided as a reference, 
as p-values often demonstrate strong empirical performance across many problems, despite lacking theoretical guarantees. 

We generate the data as in the previous experiments in Section~\ref{sec:sim_sub}, 
with $\lambda=5$ and test group sizes of 50, under different within-group variances $\sigma=1, 5,$ and $10$. 
We repeat the process 500 times and report the averaged results, as well as the standard errors. 

\begin{figure}[htbp]
    \centering
    \includegraphics[width=0.85\linewidth]{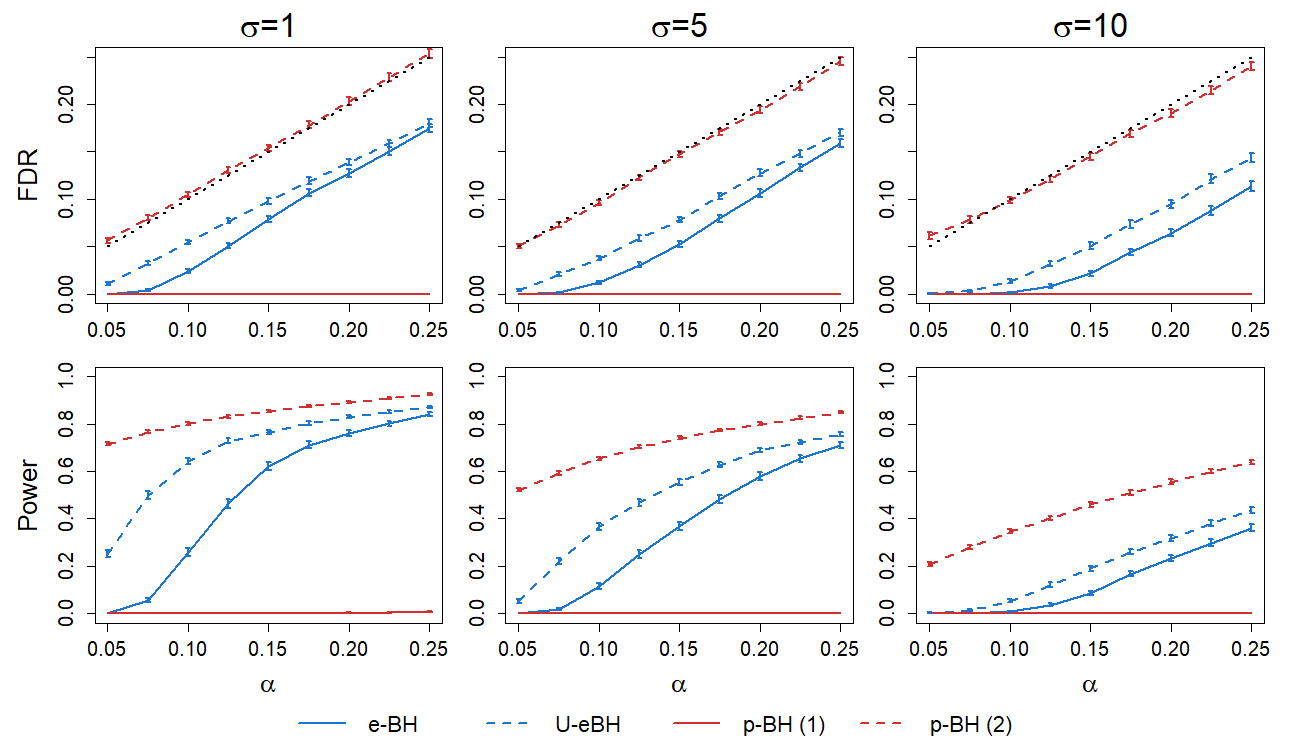}
    \caption{Estimated false discovery rate and power of Algorithm~\ref{alg:hierarchical} 
    (e-BH+hierarchical conformal e-values), its boosted version (U-eBH+hierarchical conformal e-values), 
    as well as the p-BH procedure applied to hierarchical conformal p-values~\eqref{eqn:hier_p} (p-BH (1)) 
    and~\eqref{eqn:hier_p_2} (p-BH (2)), across different within-group variances, at levels $\alpha=0.05,0.075,\cdots,0.25$.}
    \label{fig:hier_1}
\end{figure}

\paragraph{Results.}
A summary of the results is shown in Figure~\ref{fig:hier_1}, which demonstrate that selection using hierarchical conformal e-values, 
along with its boosted version, successfully controls the FDR at the desired levels. The procedure using the hierarchical conformal p-value~\eqref{eqn:hier_p} tends to be overly conservative, showing zero power in every trial. The procedure with $p_{j,i}$ values from~\eqref{eqn:hier_p_2} tends to tightly control the FDR, but its FDR occasionally exceeds the target level. 

\begin{figure}[htbp]
    \centering
    \includegraphics[width=0.85\linewidth]{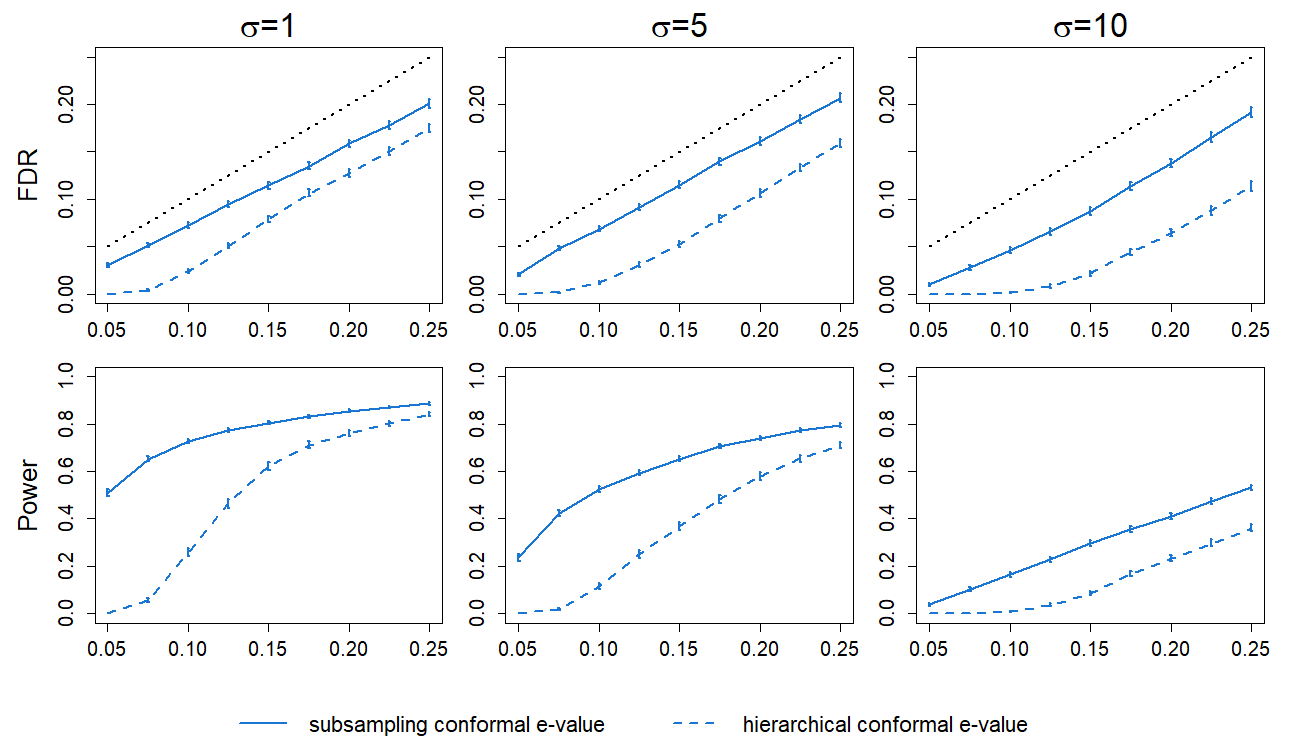}
    \caption{Estimated false discovery rate and power of Algorithm~\ref{alg:main} (e-BH+subsampling conformal e-values) and Algorithm~\ref{alg:hierarchical} 
    (e-BH+hierarchical conformal e-values), across different within-group variances, with standard errors, at levels $\alpha=0.05,0.075,\cdots,0.25$.}
    \label{fig:sub_hier}
\end{figure}

Figure~\ref{fig:sub_hier} compares the hierarchical conformal e-value-based method with the subsampling conformal e-value-based method. Overall, the procedure with hierarchical conformal e-values tends to be more conservative and less powerful compared to the procedure with subsampling conformal e-values. In other words, if we view e-value-based methods as a `correction' of p-value-based methods, the subsampling strategy provides a tighter, minimal correction that yields less conservative results. Specifically, as discussed in Section~\ref{sec:intuition}, the subsampling-based procedure closely approximates the p-value-based method, which does not provide provable FDR control in our setting but usually achieves high power in many problems. However, the procedure with hierarchical conformal e-values does not provide a tight approximation to its p-value-based counterpart.

This observation suggests a tradeoff between the randomness of the procedure and its power: the subsampling conformal e-value introduces additional randomness in the within-group selection step but achieves higher power, whereas the hierarchical conformal e-value reduces randomness at the cost of power. The choice of procedure may depend on the primary focus of the user.

\section{Real data application}
\label{sec:real_data}

\subsection{ACS income data}
We further illustrate the performance of the proposed procedure by applying it to the ACS income dataset~\citep{ding2021retiring}. 
This dataset consists of observations of U.S. adults across 50 states. 
We consider an outcome that is a binary variable indicating whether an individual's yearly income exceeds $50,000$ dollars; 
the features form a 10-dimensional vector that includes demographic and job information for each individual. 
In the experiment, we focus on the observations from California, which includes 195,665 data points 
and consider the task of identifying individuals with an income exceeding $50,000$ dollars, i.e., 
$Y=1$.

We apply the pre-trained model from~\citet{liu2023need}, which uses XGBoost, to a split of the data consisting of $50,000$ observations to construct the estimator $\hat{\mu}(\cdot)$ 
for $\PPst{Y=1}{X = \cdot}$ and define the score function $s(x,y) = y - \hat{\mu}(x)$. 
We then stratify the remaining split of $145,565$ observations based on three variables: class of worker, relationship, and occupation, retaining only the groups with at least $10$ observations. These variables correspond to the group features $G_k$ in the problem setting, but are not used to construct the estimates. This results in $858$ groups, from which we construct a calibration set consisting of $650$ groups and a test set of $200$ groups.

\begin{figure}[ht]
    \centering
    \includegraphics[width=0.75\linewidth]{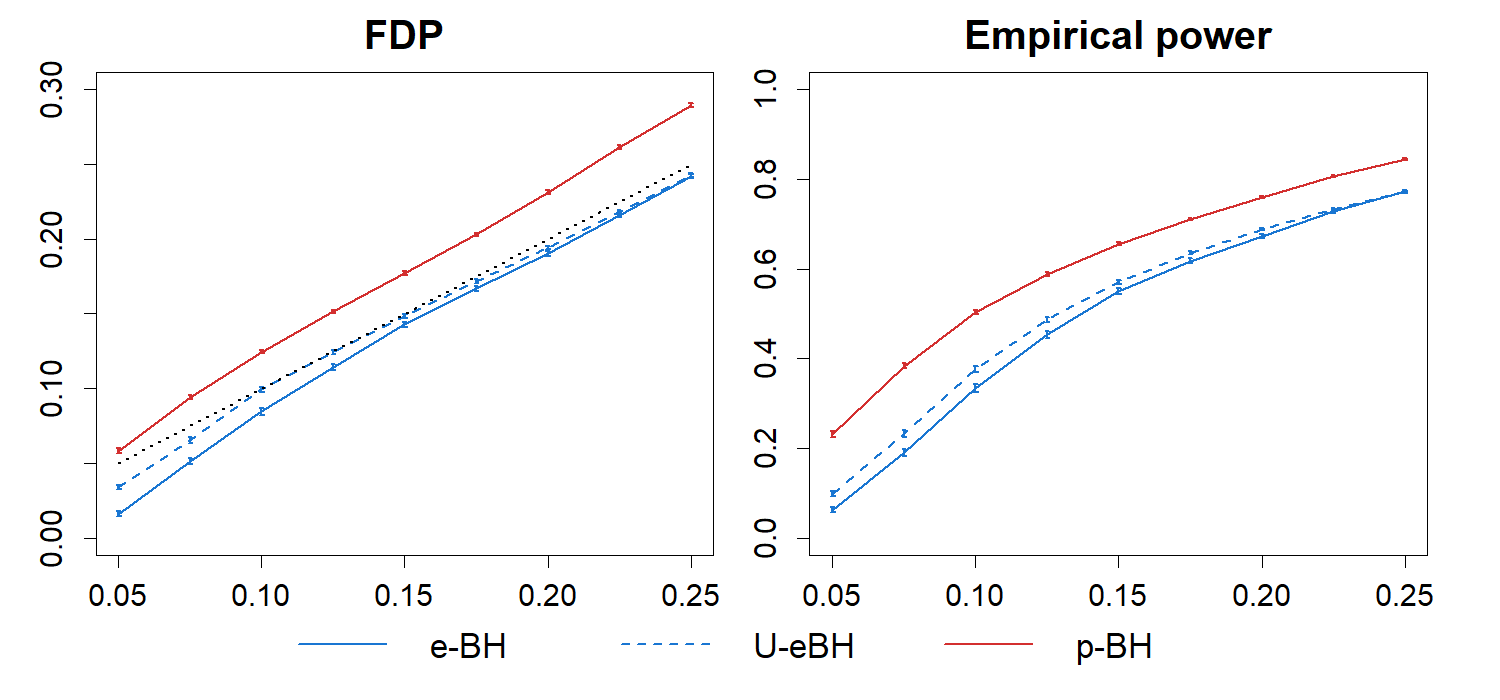}
    \caption{Results for ACS income data, with 500 trials. The dotted line corresponds to the $y=x$ line.}
    \label{fig:acs}
\end{figure}

As in the simulations, we compare the performance of three procedures: the proposed procedure with subsampling conformal e-values, its boosted procedure, and the procedure with subsampling conformal p-values. We repeat these procedures with 500 sets of samples drawn from the groups according to~\eqref{eqn:i_k_ast}, at levels $\alpha=0.05,0.075,0.1,\cdots,0.25$. The results are shown in Figure~\ref{fig:acs}. Note that we have only one set of realized data, and the plot shows the means of the FDP and empirical power---equivalently, the data-conditional FDR and power---while the procedure provides marginal FDR control. These results are provided to illustrate the overall performance of the method, though they do not exactly represent the theoretical target. Figure~\ref{fig:acs} shows that such data-conditional FDR is also well-controlled.

\subsection{Blood pressure control data}

We also explore the performance of the individual-treatment-effect-based procedure discussed in Section~\ref{sec:ite} of the Appendix, on the blood pressure control dataset~\citep{ogedegbe2018health}. 
This dataset comprises samples of patients, including their demographic information (e.g., age, gender) and clinical measurements (e.g., BMI, cardiovascular risk assessment);
the outcome variable is systolic blood pressure (SBP). 
The patients are assigned to one of two treatment groups: public health insurance coverage (HIC), or HIC combined with a nurse-led task-shifting intervention (TASSH). The task considered in this experiment is to identify individuals for whom the TASSH strategy results in a greater reduction in SBP compared to the baseline HIC strategy. In other words, letting $Y(1)$ denote the reduction under TASSH and $Y(0)$ under HIC, our aim is to test hypotheses
\begin{equation*}
    H_{j,i} : Y_{K+j,i}(1) \leq Y_{K+j,i}(0), \quad j=1,2,\cdots,M,\; i=1,2,\cdots,N_{K+j}.
\end{equation*}

\begin{figure}[ht]
    \centering
    \includegraphics[width=0.75\linewidth]{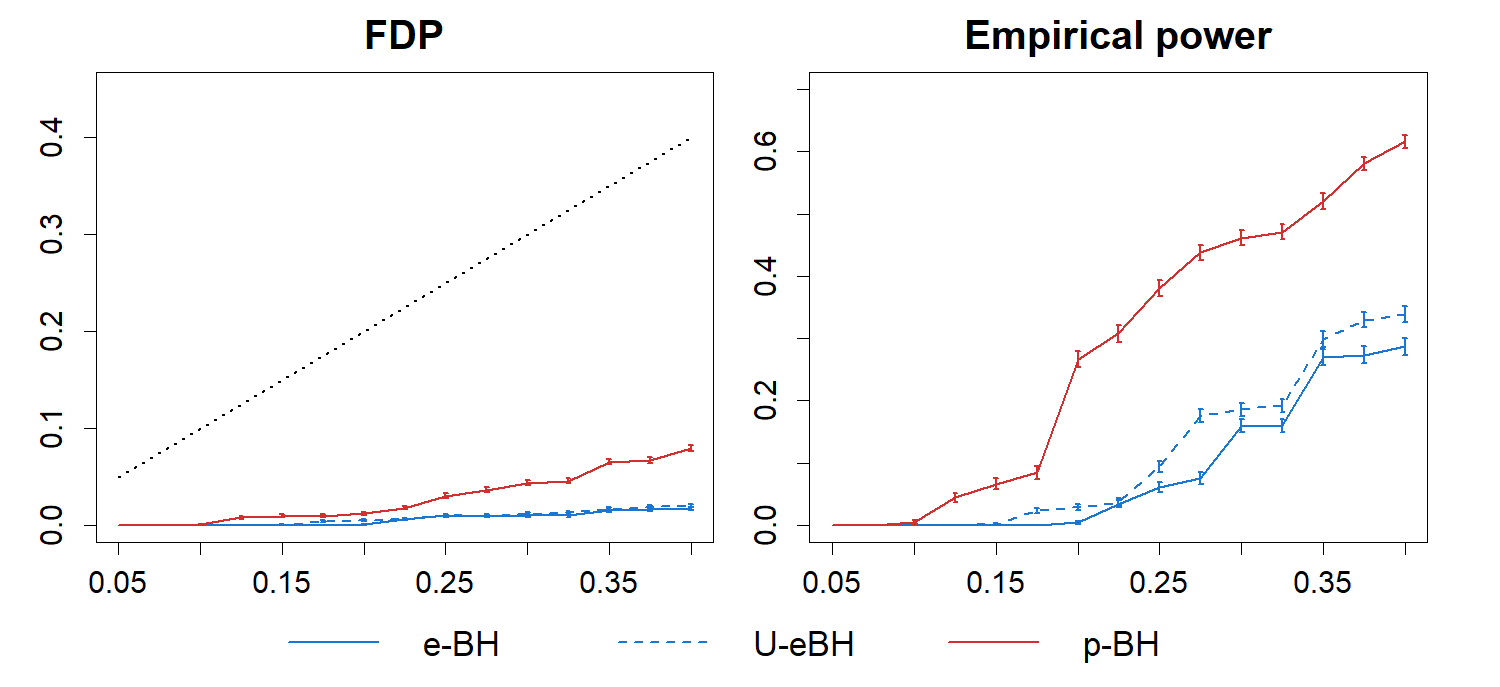}
    \caption{Results for blood pressure control data, with 500 trials. The dotted line corresponds to the $y=x$ line.}
    \label{fig:sbp}
\end{figure}

The original data were collected from 32 health centers, resulting in a natural hierarchical structure.
Patients from the same health center are likely to receive similar health care and share common environmental factors, 
leading to intra-class correlation in outcomes~\citep{mendis2010cardiovascular,ogedegbe2018health}. 
Accordingly, we assume exchangeability within the cluster 
rather than independence, which remains compatible 
with Assumption~\ref{asm:dist}.
We excluded observations with missing information. After this step, the dataset consists of 389 samples from 29 groups, with an average group size of approximately 14. For evaluation purposes, we artificially generate the counterfactual outcomes. We split the data into training, calibration, and test sets, with group sizes of 5, 20, and 4, respectively. Using the observations in the training data, we employ random forest regression with 11 covariates, comprising demographic and clinical measurements, to construct an estimator function $\hat{\mu}(\cdot)$, and a residual-score function $s(x,y) = y - \hat{\mu}(x)$. We then apply both the e-BH procedure and its boosted version to the subsampling conformal e-values described in Section~\ref{sec:ite} of the Appendix. We repeat the procedure for 500 independent trials of subsampling, at levels $\alpha = 0.05,0.075,0.1,\cdots,0.4$. 
The results are presented in Figure~\ref{fig:sbp}.

\section{Discussion}
\label{sec:discussion}
In this work, we provide predictive selection procedures with FDR control in settings where the data has a hierarchical structure. 
Our procedures are based on the e-BH method, where we construct valid e-values in two ways: subsampling conformal e-values and hierarchical conformal e-values. 
The subsampling-based method involves extra randomness but tends to more tightly control the FDR and achieve higher power empirically;
the hierarchical conformal e-value-based method is more stable, but also exhibits conservativeness empirically. 

Many open questions remain. For example, one might aim for group feature-conditional FDR control, considering that data with a hierarchical structure inherently contains information about the group-conditional distribution of the scores. How can we deal with the breakdown of between-group exchangeability after such conditioning? Can we achieve such a stronger target in a distribution-free sense? Another question is whether it would be possible to extend the procedure to settings with more complex within-group data structures, such as time series or vectors of multiple outcomes for each group—where it is hard to expect exchangeability or any kind of symmetry, so that conformal-type methods do not directly apply.  We leave these questions for future work.

\subsection*{Acknowledgment}
Z.R.~is supported by NSF grant DMS-2413135. Y.L.~is supported in part by 
NIH R01-AG065276, R01-GM139926, NSF 2210662, P01-AG041710, R01-CA222147, 
ARO W911NF-23-1-0296,
NSF 2046874, ONR N00014-21-1-2843, and the Sloan Foundation.

\newpage
\bibliographystyle{plainnat}
\bibliography{bib}

\appendix

\newpage 
\section*{Appendix}

\section{Additional remarks}\label{sec:remarks}

\subsection{Other possible choices of e-values}
\label{rmk:softrank}

Generally, a variable $e_{j,i}$ of the following form is also a valid e-value for testing $H_{j,i}$~\eqref{eqn:hypothesis}:
\begin{align}\label{eq:soft_rank_eval}
e_{j,i} = \frac{f(\hat{V}_{K+j,i})}{\sum_{k=1}^k f(\hat{V}_{k,i_k^*}) \One{Y_{k,i_k^*} \leq C_{k,i_k^*}} + f_\text{max}}\cdot(K+1),
\end{align}
where $f : \R \rightarrow \R^+$ is a nonnegative function with an upper bound $f_\text{max}$, 
and is either predetermined or dependent on the data in a manner invariant to the permutation on 
\begin{align*}
\D_{j,i} := \{(X_{1,i_1^*},Y_{1,i_1^*}), \cdots, (X_{K,i_K^*},Y_{K,i_K^*}), (X_{K+j,i},Y_{K+j,i})\}.
\end{align*}
This class of e-values generalizes the {\em soft-rank e-values} discussed by~\citet{ramdas2024hypothesis}
to the case of random hypotheses. Notably, the $e_{j,i}$ defined in~\eqref{eqn:e_j_i} 
does {\em not} belong to this class, as $T_j$ is not invariant to the permutation of $D_{j,i}$. 
However, on the event $\{Y_{K+j,i}\le C_{K+j,i},\hat V_{K+j,i} < T_j\}$, we can show that 
$e_{j,i}$ coincides with an `oracle' e-value of the form in~\eqref{eq:soft_rank_eval}---a key step in
proving its validity.


\subsection{`All-or-nothing' e-values}

The e-value defined in~\eqref{eqn:e_j_i} takes an `all-or-nothing' form: 
it is either zero or a large value. 
While such behavior is typically avoided in standard e-value applications, as it undermines desirable properties like operational continuation and post-hoc validity, 
it can be advantageous in {\em offline multiple testing} settings.
In these settings, sharp e-values discussed in Remark~\ref{rmk:alpha} 
are often of the all-or-nothing type, and they tend to achieve substantially greater 
power than the other forms; see~\citet[][Section 2.2]{lee2024boosting} for more discussion.
With this in mind, our e-value construction is designed to approximate these sharp e-values, 
aiming to achieve high statistical power comparable with that of 
(heuristic) p-value-based methods.

\subsection{Selection based on both individual and group features}

The selection procedure in Algorithm~\ref{alg:main} ensures valid $\FDR$ control 
but does not utilize all the information provided by the training/calibration data, 
in the sense that the group-feature observations $(G_k)_{1 \leq k \leq K+M}$ are not used in the inference. 
However, in settings where the group features are considered more than just side information, one might desire to construct a test statistic that depends on both the group and individual features.

In fact, the same procedure can be applied to a score that depends on both individual and group feature observations. 
Specifically, suppose we construct a score function $s:\mathcal{G} \times \X \times \Y \rightarrow \R$, 
and then define $\hat{V}_{k,i} = s(G_k,X_{k,i}, C_{k,i})$, constructing $e_{j,i}$'s 
according to~\eqref{eqn:i_k_ast} and~\eqref{eqn:e_j_i}. Under Assumption~\ref{asm:dist} with a stronger condition 
\[(Z_{k,1},\cdots,Z_{k,m}), (G_l,N_l, \tilde{Z}_l)_{l \neq k} \mid N_k = m \;\;\stackrel{d}{=}\;\; (Z_{k,1},\cdots,Z_{k,m}), (G_l,N_l, \tilde{Z}_l)_{l \neq k},\;\forall k,m\]
that includes the group features $G_l$, the same argument used in the proof of Theorem~\ref{thm:individual} shows that these 
$e_{j,i}$'s are valid e-values as in~\eqref{eqn:e_value}. Consequently, we can control the $\FDR$ by applying the e-BH procedure with such group feature-dependent test statistics. As a remark, in cases where the group features $G_k$'s are dependent of $N_k$'s and thus the above condition is not a reasonable assumption, one may apply the procedure in Algorithm~\ref{alg:main} or~\ref{alg:hierarchical} without leveraging the group features.

\subsection{Merging e-values}\label{sec:merge}

The subsampling conformal e-value-based selection procedure 
offers valid FDR control 
but relies on a random subset of the calibration data and involves 
external randomness (recall that it uses $K$ randomly drawn observations).

In settings where we have a small number of groups with large group sizes, the procedure does not efficiently utilize the information in the calibration data.
Although this does not significantly affect the validity or power of inference (since all observations in the training split can still be used for training), one might consider drawing multiple random subsets and merging the resulting e-values for more stable results.

Specifically, one option is to use the following derandomized e-values. For each $j \in [M]$ and $i\in N_{K+j}$, consider
\begin{equation}\label{eqn:e_j_i_derandomized}
    e_{j,i}^\text{derandomized} = \frac{1}{N_1 N_2\cdots N_K}\sum_{(i_1, \cdots, i_K) \in [N_1] \times \cdots [N_K]}\frac{\One{\hat{V}_{K+j,i} < T_j^*}}{\sum_{k=1}^K \One{\hat{V}_{k,i_k} < T_j^*, Y_{k,i_k} \leq C_{k,i_k}} + 1} \cdot (K+1),
\end{equation}
where for each $(i_1, \cdots, i_K)$, $T_j^*$ denotes the corresponding cutoff $T_j$, defined according to~\eqref{eqn:stopping}. Note that $e_{j,i}^{\text{derandomized}}$ is obtained by marginalizing $e_{j,i}$ in~\eqref{eqn:e_j_i} over $i_1, \dots, i_K$. It directly follows from the linearity of expectation and the argument in the proof of Theorem~\ref{thm:individual} that 
the e-BH procedure applied to $(e_{j,i}^\text{derandomized})_{1 \leq j \leq M, 1 \leq i \leq N_{K+j}}$ also controls the FDR.

Alternatively, in settings where the number of groups or the group sizes is large, 
direct computation of the derandomized e-value becomes difficult. In such cases, we can instead consider the average of e-values from multiple draws. Suppose we repeat drawing $(i_1^*, \cdots, i_K^*)$ based on~\eqref{eqn:i_k_ast}, $r$ times. Let us denote the $l$-th sample as $(i_1^l, \cdots, i_K^l)$, 
and then we consider for each $j \in [K]$ and $i \in [N_{K+j}]$ the following statistic:

\begin{equation}\label{eqn:e_j_i_average}
    e_{j,i}^\text{average} = \frac{1}{r}\sum_{l=1}^r\frac{\One{\hat{V}_{K+j,i} < T_j^l}}{\sum_{k=1}^K \One{\hat{V}_{k,i_k^l} < T_j^l, Y_{k,i_k^l} \leq C_{k,i_k^l}} + 1} \cdot (K+1),
\end{equation}
where $T_j^l$ is defined as~\eqref{eqn:stopping} with $(i_1^*, \cdots, i_K^*) = (i_1^l, \cdots, i_K^l)$. Once more, it directly follows from Theorem~\ref{thm:individual} that applying the e-BH procedure to $(e_{j,i}^\text{average})_{1 \leq j \leq M, 1 \leq i \leq N_{K+j}}$ also yields a valid FDR control.

However, it turns out that merging e-values with different stopping time thresholds, 
as described above, often leads to a significant loss of power empirically. 
For an intuitive explanation, recall that each e-value from different samples is either zero or a constant. This means the procedure strictly sacrifices power for signals deemed `ambiguous'. However, this `ambiguity' depends on the specific subsampled calibration set. During the combining steps, potentially many signals with corresponding e-values that include multiple zeroes---due to the strict sacrificing strategy---tend to have small average e-values, leading to an overall loss of power.

In Section~\ref{sec:hier}, we introduce an alternative, non-randomized method that combines information from multiple observations
and achieves reasonable power. Further discussions and comparisons will also be provided.

\subsection{Improving power with U-eBH procedure}\label{sec:uebh}

Observe that in Algorithm~\ref{alg:main}, the hypotheses for the $(K+j)$-th group share the same threshold $T_j$. Thus, the role of the e-BH procedure within each group is essentially to make a binary choice---either to reject the hypotheses whose corresponding scores $\hat{V}_{j,i}$ are below the threshold or to reject none. If the nonzero e-values have a small value, then the latter is more likely to take place, leading to a loss of power. To make the former more likely to happen, one can apply the U-eBH procedure~\citep{xu2023more}, which boosts the e-values with a uniform random variable. 
Specifically, let $U \sim \textnormal{Unif}([0,1])$ be an independent uniform random variable, and then define
\begin{equation}\label{eqn:U_e_j_i}
e_{j,i}^U = e_{j,i} / U,  \forall j \in[M], i \in [N_{K+j}],
\end{equation}
where $e_{j,i}$ follows the definition~\eqref{eqn:e_j_i}. Then the following holds.
\begin{corollary}\label{cor:uebh}
The e-BH procedure applied to $(e_{j,i}^U)_{1 \leq j \leq M, 1 \leq i \leq N_{K+j}}$ at level $\alpha$ guarantees $\cFDR \leq \alpha$.
\end{corollary}
The proof of Corollary~\ref{cor:uebh} follows directly from~\citet[Theorem 4]{xu2023more}, but requires slight modification because we consider testing ``random nulls'', and the notion of e-values differs from its standard meaning in the multiple testing literature.

\begin{remark}\label{rmk:randomness}
One might be concerned that the boosting step is introducing 
external randomness to the procedure, which  
increases the variability of the output, and can potentially be used for ``randomness hacking'', 
i.e., the user keeps generating random numbers until the desired outcome is achieved.  
As a remedy discussed in~\citet{xu2023more}, one can use the internal randomness to obtain 
such a random variable. For example, we can reserve the first group in the calibration 
set to construct a (super-)uniform random variable by permutation.

\end{remark}

\begin{remark}
Note that the boosting step~\eqref{eqn:U_e_j_i} does not change the rejection threshold $T_j$, i.e., the hypothesis $H_{j,i}$ whose corresponding score $\hat{V}_{j,i}$ is larger than $T_j$ is still never rejected regardless of the value of $U$. Thus, in our setting, the boosting step does not significantly modify the selection rule—it merely reweighs the binary decision.
\end{remark}

\begin{remark}
    Another method for boosting the power of the e-BH procedure based on the closure principle is proposed in~\citet{xu2025bringing,goeman2025partitioning}. However, this result does not directly apply to our setting, where we consider testing ``random null events'' with the notion of e-values~\eqref{eqn:e_value}, which differs from the standard formulation in multiple testing problems.
\end{remark}

\subsection{Inference with alternative selection criteria}

For simplicity, Procedures~\ref{alg:main} and~\ref{alg:hierarchical} were stated in the context where the target is to select individuals $X_{K+j,i}$ that satisfy $Y_{K+j,i} \geq c(X_{K+j,i})$. In this section, we provide a brief discussion on how the selection criterion can be generalized.

Suppose we are interested in selecting those $Y_{n+j,i}$ that fall into a set $h(X_{n+j,i})$, where $h:\mathcal{X}\to\mathcal{B}(\mathbb{R})$ is a function defining the target, and $\mathcal{B}(\mathbb{R})$ denotes the collection of measurable subsets of $\mathbb{R}$. 

The same argument can be applied to this setting by having a score that measures a distance between an estimated $y$ value and the set $h(x)$. For example, in procedures~\ref{alg:main} and~\ref{alg:hierarchical}, we used
\[s(x) = d(\hat{\mu}(x), (c(x),+\infty)) := c(x) - \hat{\mu}(x).\]
Generally, we can work with a score function $s(x) = d(\hat{\mu}(x), h(x))$ for some distance measure $d$, and the formula for the conformal e-value follows the same formula as in the paper, with
\[\hat{V}_{k,i} = s(X_{k,i}) = d(\hat{\mu}(X_{k,i}), h(X_{k,i})).\]

\subsection{Effect of group sizes}

Here we note that, regardless of the sizes of the groups, the ``influence'' from each of the groups on the inference remains the same—implying that having a large realized group size in the data does not introduce bias into the inference. The specific reasons are as follows.

\begin{enumerate}
\item \textbf{Subsampling conformal e-values.} 
This construction uses one observation from each group, 
so the group size does not play a significant role.
\item \textbf{Hierarchical conformal e-values.} 
In this case, each observation in a group is weighted
the reciprocal of the group size, ensuring that the overall ``contribution'' of each group remains comparable across groups.
\end{enumerate}

\subsection{Data-dependent choice of the hyperparameter}

Here, we discuss a data-dependent strategy for choosing the hyperparameter $\tilde{\alpha}$ used in the construction of the proposed conformal e-values. The idea is to choose $\tilde \alpha$ separately for each hypothesis:

For a test unit $(j,i)$, we could allow 
$\tilde \alpha_{j,i}$ to depend on 
the following information:
\begin{itemize}
\item $\Lbag \{X_{k,i_k^*}: k\in[K], Y_{k,i_k^*}\le c \} \cup \{X_{K+j,i}\}\Rbag$:
the multiset of the null calibration covariates and the test covariate, 
with their ordering and identities removed; 
\item $\{Z_{k,i_k^*}:k\in[K], Y_{k,i_k^*} > c\}$: 
the non-null calibration units;
\item $\{X_{K+j',i'}: j'\neq j, i'\in [N_{K+j'}]\}$:
the test units from groups other than $K+j$.
\end{itemize}
For example, consider a finite collection of 
candidate levels $(\bar \alpha_1,\dots, \bar \alpha_C$).
For each $r\in [C]$, we construct proxy e-values 
for the hypothesis $(j',i')$ with $j'\neq j$ by temporarily treating 
$(j,i)$ as an additional null calibration unit.
Concretely, we define 
\begin{align*}
T_{j,i}^{(r)} = \sup\Bigg\{t\in \mathbb{R}: \frac{1 + \One{\hat V_{j,i} < t} + 
\sum_{k\in [K]} \One{\hat V_{k,i_k^*}< t, Y_{k,i_k^*}\le C_{k,i_k^*} }}
{\sum_{\ell \neq j}\sum_{i' \in [N_{K+\ell}]} 
\One{\hat V_{K+\ell,i'} < t} } \le \bar \alpha_r\Bigg\}.
\end{align*}
The proxy e-value for $(j',i')$ can therefore be constructed as 
\begin{align*}
& e_{j',i'}^{(r)} 
= \frac{\One{\hat V_{K+j',i'} < T^{(r)}_{j,i}}}{1+ \One{\hat V_{j,i} < T^{(r)}_{j,i}} + 
\sum_{k\in [K]} \One{\hat V_{k,i_k^*}< T^{(r)}_{j,i}, Y_{k,i_k^*}\le C_{k,i_k^*} } } \cdot (K+2).
\end{align*}
Letting $\mathcal{R}^{(r)}$ denote the rejection set obtained from 
applying e-BH to the proxy e-values $(e^{(r)}_{j',i'})_{j'\neq j, i'\in[N_{K+j'}]}$, 
we choose $\tilde \alpha_{j,i} = \bar \alpha_{\hat r}$,
where $\hat r \in \arg\!\max_{r \in [C]} |\mathcal{R}^{(r)}|$.
We then construct the resampling e-value for $(j,i)$ exactly as in our 
original procedure, replacing the pre-specified $\tilde \alpha$ with 
the data-dependent value $\tilde \alpha_{j,i}$.\\

Crucially, $\tilde \alpha_{j,i}$ depends on the null 
calibration units and the test unit only through 
permutation-invariant information, which suggests that 
it is plausible to show that the resulting 
sub-sampling e-value defined with $\tilde \alpha_{j,i}$ is valid.

This data-driven approach, however, selects different values 
of $\tilde \alpha_{j,i}$ across hypotheses, and the resulting 
effect on power is unclear. In addition, it is not clear 
which selection criterion makes the optimal use of the available information. 
Developing a principled framework for data-adaptive choices of $\tilde{\alpha}_{j,i}$ and establishing their theoretical and empirical properties may be an interesting future direction.

\section{Extensions}
\label{sec:extension}
In this section, we discuss extensions of the proposed selection procedures to more general settings.
Section~\ref{sec:hybrid_null} considers the task of jointly testing hypotheses at the group and individual levels;
Section~\ref{sec:cov_shift} addresses the case where the group covariate distribution shifts 
between the calibration and test sets; and Section~\ref{sec:ite} applies the proposed methods 
to the task of selection based on individual treatment effect.

\subsection{Procedure for joint inference on the groups and the individuals}
\label{sec:hybrid_null}
We now explore the task where our objective extends beyond testing hypotheses at individual levels to 
include the selection of groups across different layers. 
For example, in drug discovery, one might be interested in selecting both individual drugs and drug classes. 
Suppose we also aim to test group-level hypotheses $\{H_j : j=1,2,\cdots,M\}$, 
in addition to the individual level hypotheses $H_{j,i}$'s. 
For example, one can consider the following types of group-level hypotheses:
\begin{enumerate}
    \item [(1)] Group-global null: $H_j : Y_{K+j,i} \leq C_{K+j,i}  \forall j \in [N_{K+j}]$,
    \item [(2)] Selecting groups with large mean: $H_j : \frac{1}{N_{K+j}}\sum_{i=1}^{N_{K+j}} Y_{K+j,i} \leq c$.
\end{enumerate}

We aim to construct a selection procedure that controls the group size-conditional FDR, which is now defined as
\begin{align}\label{eqn:cond_fdr_group}
    \begin{split}
    & \cFDR = \EEst{\FDP}{N_{K+1:K+M}}, \\
    \text{ where } & \FDP = \frac{\sum\limits_{j=1}^{M} \sum\limits_{i=1}^{N_{K+j}} \One{(j,i) \in \cR}\One{H_{j,i}} + \sum_{j=1}^M \One{j \in \cR}\One{H_j}}{|\cR| \vee 1}.
    \end{split}
\end{align}
Here, $\cR \subset \{(j,i) : j=1,\cdots,M, i=1, \cdots, N_{K+j}\} \cup \{1,2,\cdots,M\}$ denotes the rejection set.

We note that if we are interested in testing only the group-level hypotheses with marginal FDR control, 
we can simply apply the BH procedure with conformal p-values---as described in~\citet{jin2023selection}---constructed by treating each group as a single observation. Testing the individual-level hypotheses can be based on the results from the previous sections. However, jointly testing both group-level and individual-level hypotheses requires additional consideration, as we need group size-conditional e-values for the group-level hypotheses to maintain FDR control. Below, we discuss methods for the construction of valid e-values for the group level nulls.

\subsubsection{Special case: selection with group-global nulls}\label{sec:group_global}

Suppose we are interested in testing $H_{j,i}$s together with the group-global nulls

\[H_j : Y_{K+j,i} \leq C_{K+j,i} \quad \forall i \in [N_{K+j}]\]
for $j=1,2,\cdots,M$. For this goal, we can construct an e-value simply by averaging the individual level e-values in the group.
\begin{equation}\label{eqn:e_group_global}
e_j = \frac{1}{N_{K+j}}\sum_{i=1}^{N_{K+j}} e_{j,i}, \text{ where } e_{j,i} \text{ is defined as }\eqref{eqn:e_j_i}.
\end{equation}
Applying the result in the proof of Theorem~\ref{thm:individual}, 
we can demonstrate that $e_j$ is a valid e-value conditional on the group size $N_{K+j}$, 
thus ensuring that the e-BH procedure applied to both $e_{j,i}$'s and $e_j$'s controls the conditional FDR.
We rigorously state this result in the following proposition and offer the proof in Section~\ref{appx:proof_group_global}.

\begin{proposition}\label{prop:group_global}
    The e-BH procedure applied to $(e_{j,i})_{1 \leq j \leq M, 1 \leq i \leq N_{K+j}}$ from~\eqref{eqn:e_j_i} and $(e_j)_{1 \leq j \leq M}$ from~\eqref{eqn:e_group_global} at level $\alpha$ guarantees $\cFDR \leq \alpha$.
\end{proposition}

\begin{remark}[Comparison with e-filter]
The e-filter framework, studied in~\citet{gablenz2024catch}, provides a method for controlling both the individual-level and group-level false discovery rates, for group-global nulls. In contrast, the procedure we discuss above controls the overall $\FDR$ instead of the individual and group-level $\FDR$s. However, since the threshold $T_j$~\eqref{eqn:stopping} is determined based on the 
$\FDP$ estimate, the individual-level $\FDR$ is also likely controlled below the target $\alpha$. We illustrate this empirically in the next section.
    
\end{remark}

\subsubsection{Selection with general group null hypotheses}\label{sec:group_general}

Next, we consider a more general setting where the null hypothesis $H_{j}$ 
can be a condition about any function of the outcome vector of the $(K+j)$-th group $(Y_{K+j,1}, Y_{K+j,2}, \cdots, Y_{K+j,N_{K+j}})$. 
For conciseness, let us write
\begin{align*}
    \tilde{X}_k = (X_{k_1},X_{k,2},\cdots, X_{k,N_k}) \in \tilde{\X} \text{ and }
    \tilde{Y}_k = (Y_{k_1},Y_{k,2},\cdots, Y_{k,N_k}) \in \tilde{\Y},
\end{align*}
where $\tilde{\X} = \X \cup \X^2 \cup \cdots$ and $\tilde{\Y} = \Y \cup \Y^2 \cup \cdots$.
Now suppose we are interested in testing
\[H_j : h(\tilde{Y}_{K+j}) \leq \tilde{c}(\tilde{X}_{K+j}), \quad j=1,2,\cdots,M,\]
where $\tilde{c} : \tilde{\X} \mapsto \R$ is the target cutoff function and 
$h : \R \mapsto \R$ denotes the function of interest, e.g., 
$h(\tilde{Y}_{K+j}) = \frac{1}{N_{K+j}}\sum_{i=1}^{N_{K+j}} Y_{K+j,i}$, 
if we are interested in selecting groups with large mean values. We write $C_{K+j} = \tilde{c}(\tilde{X}_{K+j})$.

Now we construct the test statistic for $H_j$. 
Similarly to the construction of $e_{j,i}$ for individual level hypotheses, 
we first construct a `group score' function $s_g : \tilde{X} \times \R \mapsto \R$ 
independently of the calibration data, and write $\hat{V}_k = s_g(\tilde{X}_{k},C_{k})$, 
$\forall k \in [K+M]$. Again, we construct the score $s_g$ in the way that a smaller value of $\hat{V}_{K+j}$ can be viewed as stronger evidence against $H_j$. 

Next, let
\[I_{\geq r} = \{k \in [K] : N_k \geq r\},\; r=1,2,\cdots\]
and
\[\hat{V}_k^r = s_g(\tilde{X}_k^r,\tilde{c}(\tilde{X}_k^r)), \text{ for } r \leq N_k, \text{ where } \tilde{X}_k^r = (X_{k,1}, X_{k,2}, \cdots, X_{k,r}).\]
Observe that $I_\geq r$ denotes the set of all groups with size at least $r$, and that for such groups, the $\hat{V}_k^r$s are exchangeable. As a remark, $\hat{V}_k^r$ can be constructed using a randomly chosen set of $r$ data points from the group, rather than the first $r$ points as described above.

Then we define
\begin{equation}\label{eqn:e_group_general}
    e_j = \frac{\One{\hat{V}_{K+j} < T_j}}{\sum_{k \in I_{\geq N_{K+j}}} \One{\hat{V}_{k}^{N_{K+j}} < T_j, h(\tilde{Y}_k^{N_{K+j}}) \leq \tilde{c}(\tilde{X}_k^{N_{K+j}})} +1} \cdot |I_{\geq N_{K+j}} + 1|,
\end{equation}
where $T_j$ is defined as
\begin{equation}\label{eqn:stopping_group}
\begin{split}
&T_j= \sup\left\{t \in \R : \widehat{\FDP}^{N_{K+j}}(t) \leq \tilde{\alpha}\right\}, \text{ where }\\
&\widehat{\FDP}^r(t) = \frac{\sum_{k \in I_{\geq r}} \One{\hat{V}_{k}^r < t, h(\tilde{Y}_k^r) \leq \tilde{c}(\tilde{X}_k^r)} + 1}{1 \vee \sum_{l=1}^M \One{\hat{V}_{K+l} \leq t}}  \cdot \frac{M}{|I_{\geq r}| + 1},  \text{ for } r \geq 1.
\end{split}
\end{equation}
The parameter $\tilde{\alpha}$ above may differ from the one used to construct the individual level e-values.
Intuitively, the e-value $e_j$ is constructed by comparing the group score $\hat{V}_{K+j}$ 
with the scores of groups with the same size, and the group null $H_j$ is rejected if the 
score is significantly smaller than the scores of the groups with similar size. 
We prove that $e_j$ in~\eqref{eqn:e_group_general} is a valid conditional e-value for $H_j$ in the following theorem,
with the proof deferred to Section~\ref{appx:proof_group_general}.

\begin{theorem}\label{thm:group_general}
    The statistic $e_j$ defined as~\eqref{eqn:e_group_general} is an e-value for $H_j$ conditional on $N_{K+1:K+M}$. Consequently, the e-BH procedure applied to $(e_{j,i})_{1 \leq j \leq M, 1 \leq i \leq N_{K+j}}$ from~\eqref{eqn:e_j_i} and $(e_j)_{1 \leq j \leq M}$ from~\eqref{eqn:e_group_general} at level $\alpha$ controls the $\cFDR$~\eqref{eqn:cond_fdr_group} at level $\alpha$.
\end{theorem}

By Theorem~\ref{thm:group_general}, the procedure controls the overall $\FDR$ that accounts for both the group-level and the individual-level hypotheses, while the individual-level $\FDR$ and the group-level $\FDR$ are also likely controlled at the target level $\alpha$. This is because the thresholds $T_j$ in~\eqref{eqn:stopping} and~\eqref{eqn:stopping_group} are determined using estimates of the individual-level and group-level $\FDP$, respectively. We illustrate this with experiments in the next section.

\subsection{Inference under group-covariate shift}
\label{sec:cov_shift}
In this section, we discuss the setting where we have a group-covariate shift, meaning that the test group features $G_{K+1}, \cdots, G_{K+M}$ are drawn from a distribution $\tilde{P}_G$, 
potentially distinct from $P_G$. For example, the test groups may be drawn with selection procedure 
depending on the group features, leading to a group-covariate distribution that is different from the calibration set. 

For simplicity, we explicitly assume the following model.
\begin{equation}\label{eqn:cov_shift}
\begin{split}
    &G_1, G_2, \cdots, G_K \iidsim P_G,\quad G_{K+1}, \cdots, G_{K+M} \iidsim \tilde{P}_G\\
    &N_1, N_2, \cdots, N_{K+M} \iidsim P_{N},\\
    &Z_{k,1}, Z_{k,2}, \cdots, Z_{k,N_k} \mid G_k, N_k \iidsim P_{Z \mid G} \quad\text{ for } k=1,2,\cdots,K+M,
    \end{split}
\end{equation}
where we do not observe $Y_{k,i}$ for $K+1 \leq k \leq K+M$ and $1 \leq i \leq N_k$.
Let $w(x) = \frac{d\tilde{P}_G(x)}{dP_G(x)}$ represent the likelihood ratio between $\tilde{P}_G$ 
and $P_G$, and we assume it is known. Leveraging the weighted conformal inference scheme proposed by~\citet{tibshirani2019conformal}, 
we construct the e-values by properly weighting the calibration data.

Here, we only discuss the extension of subsampling conformal e-values to the weighted case; 
the hierarchical conformal e-values can be extended in a similar fashion.
The weighted subsampling conformal e-values are constructed as follows. 
For any $j\in [M]$ and $i\in [N_{K+j}]$,  
\[
e_{j,i}^w = \frac{\One{\hat{V}_{K+j,i} < T_j^w}}{\sum_{k=1}^K p_k^j \cdot \One{\hat{V}_{k,i_k^*} < T_j^w, Y_{k,i_k^*} \leq C_{k,i_k^*}} + p_{K+j}^j}, 
\text{ where } p_k^j = \frac{w(G_k)}{\sum_{l=1}^K w(G_l) + w(G_{K+j})},
\]
and $i_k^*$ is the index of the randomly drawn observation in the $k$-th group as 
in~\eqref{eqn:i_k_ast}.
Here, the threshold $T_j^w$ is defined as
\begin{equation}\label{eqn:stopping_cov_shift}
\begin{split}
&T_j^w = \sup\left\{t \in \R : \widehat{\FDP}_j^w(t) \leq \tilde{\alpha}\right\}, \text{ where }\\
&\widehat{\FDP}_j^w(t) = \frac{\sum_{k=1}^K p_k^j\cdot \One{\hat{V}_{k,i_k^*} < t, Y_{k,i_k^*} \leq C_{k,i_k^*}} + p_{K+j}^j}{1 \vee \sum_{l \neq j} \sum_{i'=1}^{N_{K+l}} \One{\hat{V}_{K+l,i'} < t}} \cdot \frac{\sum_{l=1}^{M} N_{K+l}}{K+1}.
\end{split}
\end{equation}
We prove that applying the e-BH procedure with the above likelihood ratio-weighted test statistics controls the FDR.
The proof can be found in Section~\ref{appx:proof_cov_shift}.

\begin{theorem}\label{thm:cov_shift}
Under model~\eqref{eqn:cov_shift}, 
the e-BH procedure applied to $(e_{j,i}^w)_{1 \leq j \leq M, 1 \leq i \leq N_{K+j}}$ from~\eqref{eqn:e_j_i} at level $\alpha$ guarantees $\cFDR \leq \alpha$.
\end{theorem}

\subsection{Selection based on individual treatment effects}\label{sec:ite}

In this section, we consider a concrete application arising from causal inference:
we are to extend the selection procedure to the problem of selecting test points 
based on their individual treatment effects (ITEs)~\citep{lei2021conformal,jin2023sensitivity}. 
To set the stage, we follow the potential outcomes framework~\citep{imbens2015causal}, assuming 
that each unit is associated with two potential outcomes, one under treatment and one without. 
Specifically, we consider the setting where samples are drawn as in~\eqref{eqn:model}, 
with $Z_{k,i} = (X_{k,i}, Y_{k,i}(1), Y_{k,i}(0)) \in \X \times \Y \times \Y$, 
where $Y_{k,i}(1)$ and $Y_{k,i}(0)$ denote the counterfactual outcomes with and without treatment. We assume that we only observe $(X_{k,i}, A_k, Y_{k,i})$ for each individual, where $Y_{k,i} = (1 - A_k)Y_{k,i}(0) + A_k Y_{k,i}(1)$, and that treatment is assigned groupwise independently of the data, i.e.,
\[A_1,\cdots,A_{K} \mid (\tilde{Z}_k)_{1 \leq k \leq K} \iidsim \textnormal{Bernoulli}(p_A),\]
for some $p_A \in (0,1)$.

Suppose we have $K$ treatment groups and $M$ control groups---we are conditioning everything on the treatment assignments. Without loss of generality, we may assume $A_1 =\cdots = A_K = 1$ and $A_{K+1} = \cdots = A_{K+M} = 0$. Now, we consider the task of selecting individuals in the control groups whose individual treatment effect $Y_{K+j,i}(1) - Y_{K+j,i}(0)$ exceeds certain threshold $c$. For conciseness, we describe the procedure for the case when $c = 0$; the extension to a general $c \in \mathbb{R}$ is straightforward. 

For this task, the corresponding hypotheses can be written as
\begin{equation}\label{eqn:hyp_ite}
    H_{j,i} : Y_{K+j,i}(1) \leq Y_{K+j,i}(0), \quad j=1,2,\cdots,M, i=1,2,\cdots,N_{K+j}.
\end{equation}
Compared to the original problem~\eqref{eqn:hypothesis}, 
this can be viewed as a setting where the threshold 
$C_{k,i} = Y_{k,i}(0)$ is unobserved for each calibration point. 
Therefore, the procedure from the previous section cannot be directly applied to this setting. However, a similar approach can still be adopted to achieve FDR control, as we describe below.

Given a score function $s : \X \times \Y \rightarrow \R$, let us write $\hat{V}_{k,i}^0 = s(X_{k,i}, Y_{k,i}(0))$ and $\hat{V}_{k,i}^1 = s(X_{k,i}, Y_{k,i}(1))$, for $k \in [K+M]$ and $i \in [N_{K}]$. For example, one can construct a function $\hat{\mu} : \X \rightarrow \Y$ using the training set such that $\hat{\mu}(X_{k,i})$ estimates $Y_{k,i}(1)$, and then define $s(x,y) = y - \hat{\mu}(x)$. Note that we only have access to $\hat{V}_{k,i}^1$ for $1 \leq k \leq K$ and $\hat{V}_{k,i}^0$ for $K+1 \leq k \leq K+M$. Throughout this section, we assume that the score function satisfies the following monotonicity condition.

\begin{assumption}\label{asm:mono}
     For any $x \in \X$ and $y_1 \leq y_2$, it holds that $s(x,y_1) \leq s(x,y_2)$.
\end{assumption}

\subsubsection{Subsampling conformal e-values for ITEs.}
We first consider generalizing the subsampling approach. 
Suppose we draw $i_1^*, \cdots, i_K^*$ as in~\eqref{eqn:i_k_ast}. The subsampling conformal p-value for $H_{j,i}$~\eqref{eqn:hyp_ite} can be constructed as
\begin{equation}\label{eqn:p_ite}
    p_{j,i} = \frac{\sum_{k=1}^K \One{\hat{V}_{K+j,i}^0 > \hat{V}_{k,i_k^*}^1}+1}{K+1}.
\end{equation}

\begin{proposition}\label{prop:p_ite}
        Suppose Assumptions~\ref{asm:dist} and~\ref{asm:mono} hold. Then for any $\alpha \in (0,1)$, $p_{j,i}$ defined as~\eqref{eqn:p_ite} satisfies
        \[\PP{p_{j,i} \leq \alpha \text{ and } Y_{K+j,i}(1) \leq Y_{K+j,i}(0)} \leq \alpha,\]
        for any $1 \leq j \leq M$ and $1 \leq i \leq N_{K+j}$.
\end{proposition}
For completeness, we provide the proof of Proposition~\ref{prop:p_ite} in Section~\ref{appx:proof_p_ite}.
Again, it is not guaranteed that the BH procedure applied to the p-values above controls the FDR. 
For a theoretically valid procedure, we construct the subsampling conformal e-value as follows.
\begin{equation}\label{eqn:e_ite}
e_{j,i} = \frac{\One{\hat{V}_{K+j,i}^0 < T_j}}{\sum_{k=1}^K \One{\hat{V}_{k,i_k^*}^1 < T_j}+ 1}\cdot(K+1),\quad 
\forall j \in [M], i \in [N_{K+j}],  
\end{equation}
where
\begin{equation}\label{eqn:stopping_ite}
\begin{split}
   &T_j = \sup\left\{t \in \R : \widehat{\FDP}_j(t) \leq \tilde{\alpha}\right\}, \text{ where }\\
    &\widehat{\FDP}_j(t) = \frac{\sum_{k=1}^K \One{\hat{V}_{k,i_k^*}^1 < t} + 1}{1 \vee \sum_{l \neq j} \sum_{i'=1}^{N_{K+l}} \One{\hat{V}_{K+l,i'}^0 < t}} \cdot \frac{\sum_{l=1}^M N_{K+l}}{K+1},
\end{split}
\end{equation}
for a predefined $\tilde{\alpha} \in (0,1)$. Under the monotonicity condition of the score function, it can be shown that the $e_{j,i}$ above is a valid e-value for $H_{j,i}$, 
as stated in the following theorem. The proof of Theorem~\ref{thm:e_ite} is 
delegated to Section~\ref{appx:proof_ite}.

\begin{theorem}\label{thm:e_ite}
Suppose Assumptions~\ref{asm:dist} and~\ref{asm:mono} hold. 
Then $e_{j,i}$ from~\eqref{eqn:e_ite} is an e-value for $H_{j,i}$~\eqref{eqn:hyp_ite}, 
conditional on $N_{K+1:K+M}$. Consequently, the e-BH procedure, 
applied to $(e_{j,i})_{1 \leq j \leq M, 1 \leq i \leq N_{K+j}}$ 
defined by~\eqref{eqn:e_ite} at level $\alpha$, controls the $\cFDR$ at level $\alpha$.
\end{theorem}

\subsubsection{Hierarchical conformal e-values for ITEs.}
Next, we generalize the hierarchical conformal e-value. 
Similar to the previous case, we construct the e-value based on two stopping times:

\begin{equation}\label{eqn:e_ite_hier}
e_{j,i} = \frac{\One{\hat{V}_{K+j,i}^0 < T_j^+}}{\sum_{k=1}^K \frac{1}{N_k} \sum_{i'=1}^{N_k} \One{\hat{V}_{k,i'}^1 < T_j^-}+ 1}\cdot(K+1),\quad 
\forall j \in [M], i \in [N_{K+j}],  
\end{equation}
where
\begin{equation*}\label{eqn:stopping_ite_1}
\begin{split}
   &T_j^+ = \sup\left\{t \in \R : \widehat{\FDP}_j^+(t) \leq \tilde{\alpha}\right\}, \text{ where }\\
    &\widehat{\FDP}_j^+(t) = \frac{\sum_{k=1}^K \frac{1}{N_k} \sum_{i'=1}^{N_k} \One{\hat{V}_{k,i'}^1 < t} + 1}{1 \vee \sum_{l \neq j} \sum_{i'=1}^{N_{K+l}} \One{\hat{V}_{K+l,i'}^0 < t}} \cdot \frac{\sum_{l \neq j} N_{K+l}}{K+1},
\end{split}
\end{equation*}
and
\begin{equation*}\label{eqn:stopping_ite_2}
\begin{split}
   &T_j^- = \sup\left\{t \in \R : \widehat{\FDP}_j^-(t) \leq \tilde{\alpha}\right\}, \text{ where }\\
    &\widehat{\FDP}_j^-(t) = \frac{\sum_{k=1}^K \frac{1}{N_k} \sum_{i'=1}^{N_k} \One{\hat{V}_{k,i'}^1 < t}}{1 \vee \sum_{l \neq j} \sum_{i'=1}^{N_{K+l}} \One{\hat{V}_{K+l,i'}^0 < t}} \cdot \frac{\sum_{l \neq j} N_{K+l}}{K+1},
\end{split}
\end{equation*}
for a predefined $\tilde{\alpha} \in (0,1)$. The following theorem 
shows that when the score function is nondecreasing in $y$, the hierarchical conformal e-value 
defined in~\eqref{eqn:e_ite_hier} is a valid conditional e-value for $H_{j,i}$.
The proof of Theorem~\ref{thm:ite_hier} is presented in Section~\ref{appx:proof_ite_hier}. 
\begin{theorem}\label{thm:ite_hier}
    Suppose Assumptions~\ref{asm:dist} and~\ref{asm:mono} hold. Then $e_{j,i}$ from~\eqref{eqn:e_ite_hier} is an e-value for $H_{j,i}$~\eqref{eqn:hyp_ite}, conditional on $N_{K+1:K+M}$. Consequently, the e-BH procedure, applied to $(e_{j,i})_{1 \leq j \leq M, 1 \leq i \leq N_{K+j}}$ defined by~\eqref{eqn:e_ite_hier} at level $\alpha$, controls the $\cFDR$ at level $\alpha$.
\end{theorem}
We conclude this section by noting that the proposed procedures can be extended to the setting where the treatment assignment is dependent on the covariates, 
using a weighting strategy as discussed in Section~\ref{sec:cov_shift}.

\section{FDR violation of p-BH under hierarchical exchangeability}
\label{appx:counter_sub}
Suppose the target FDR level is $\alpha$.
Consider a setting where the group size $N_k\equiv 2$, 
$\forall k\in[K+M]$. In group $k$, the covariates are generated independently of other groups 
via 
\begin{align*}
    X_{k,1} \sim \text{Unif}([0,1]),~X_{k,2} = 
    \begin{cases}
        X_{k,1}, & \text{if } X_{k,1} \in [\alpha/2,\alpha] \cup [1-\alpha,1-\alpha/2],\\
        1 - X_{k,1}, & \text{Otherwise.} 
    \end{cases}
\end{align*}
The response $Y_{k,i} = X_{k,i}$, for $i\in\{1,2\}$. It can be checked that 
the data-generating process satisfies the within- and across-group exchangeability.
We then sample $K = 100$ calibration groups and $M = 1$ test group from the distribution,
constructing $p^{\text{sub}}_j$'s and $p^\hier_j$'s according to~\eqref{eq:sub_pval}
and~\eqref{eqn:hier_p_2}, respectively; the score function is $s(x,c) = c-x$. 
We then run the p-BH with the threshold 
$c \equiv 0.9$ and target FDR level $\alpha \in \{0.1,0.2,0.3\}$.
Table~\ref{tab:sub_pval_fdr} lists the averaged FDR across $N=10,000$ repetitions
along with the $99\%$ confidence intervals for both p-values. 
We can clearly see that the FDR is violated at all levels.
\begin{table}[h!]
    \centering
    \begin{tabular}{c|c|c}
    \toprule
    $\alpha$ & subsampling p-value & hierarchical p-value \\
    \midrule 
    $0.1$ & $0.121$ $[0.119,0.124]$ & $0.120$ $[0.118,0.123]$\\ 
    $0.2$ &  $0.260$ $[0.256, 0.263]$ & $0.262$ $[0.259,0.266]$\\ 
    $0.3$ &  $0.401$ $[0.397, 0.405]$ & $0.408$ $[0.404,0.412]$\\ 
    \bottomrule
    \end{tabular}
    \caption{Averaged FDR (99\% confidence interval) resulting from running p-BH with 
    the $p_j^{\text{sub}}$'s and $p_j^\hier$'s across $N=10,000$ independent trials.}
    \label{tab:sub_pval_fdr}
\end{table}

\section{Additional experimental results}\label{sec:add_exp}

\subsection{Selecting both groups and individuals}
\label{sec:sec_joint}
\begin{figure}[htbp]
    \centering
    \includegraphics[width=0.9\linewidth]{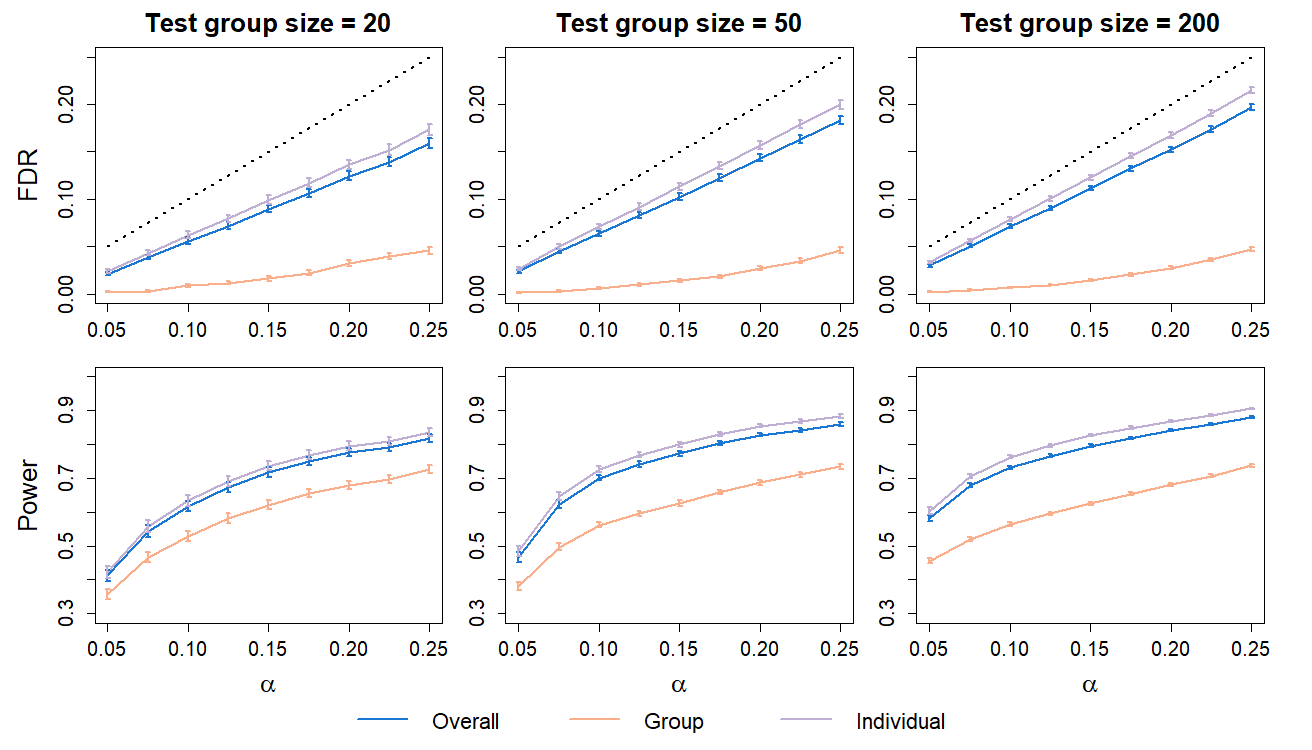}
    \caption{The false discovery rate and power of the joint selection procedure for individual nulls and group-global nulls at different test group sizes (20, 50, and 200) and levels.}
    \label{fig:group_global}
\end{figure}

Next, we investigate the performance of the procedure for selecting both the groups and the individuals. 
We first examine the selection with group-global nulls. 
The data is generated as in~\eqref{eq:sim_dgp}, with $\lambda=5$, and we run the procedure as described in Section~\ref{sec:group_global}. 
We present the results for test group sizes of 20, 50, and 200. 
Figure~\ref{fig:group_global} illustrates the overall FDR that the procedure aims to control, 
as well as the group-FDR and individual-FDR, which are computed using only the group level nulls and the individual level nulls, respectively. 
The results indicate that the overall FDR is controlled as proved in Proposition~\ref{prop:group_global};
moreover, the FDR within each level (group and individual) is also controlled fairly well.

\begin{figure}[htbp]
    \centering
    \includegraphics[width=0.9\linewidth]{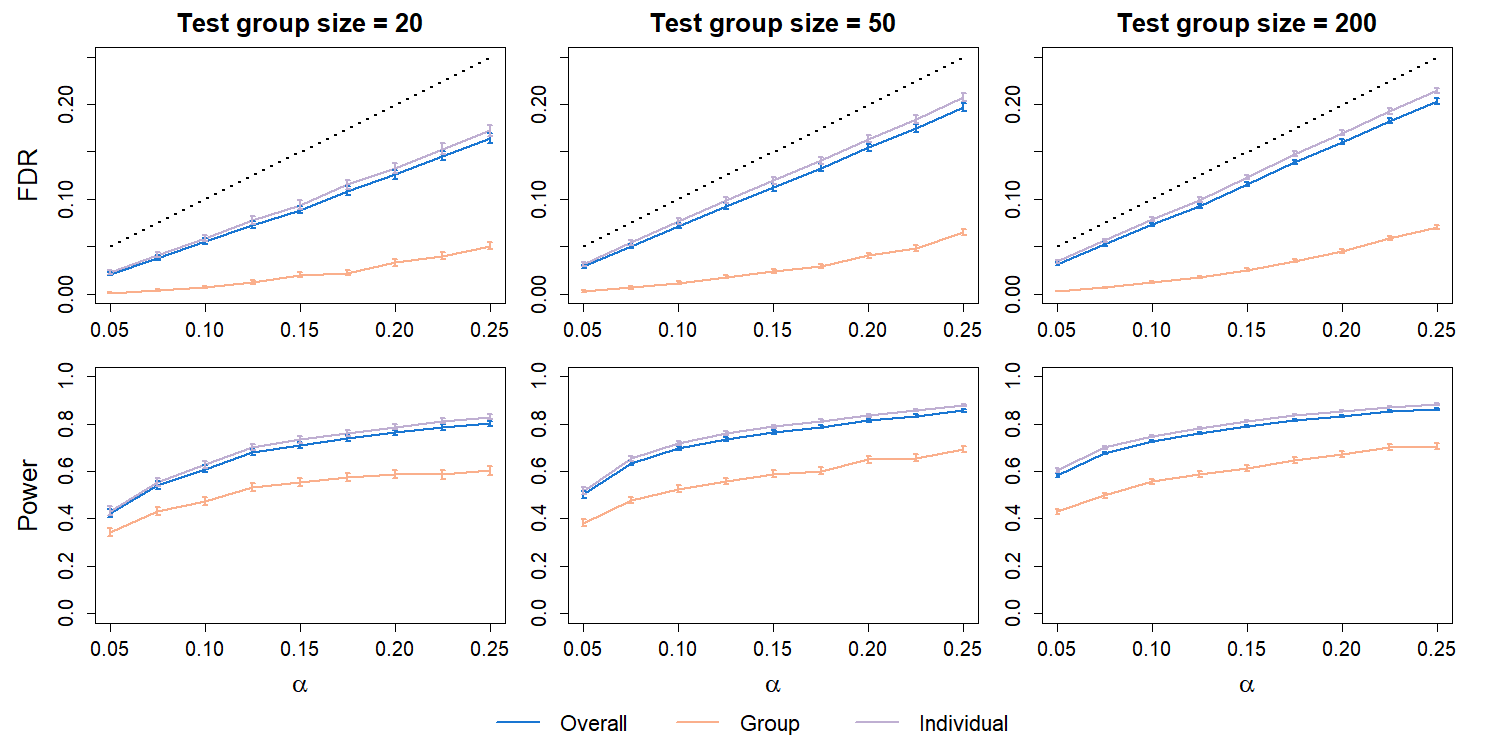}
    \caption{The false discovery rate and power of the joint selection procedure for individual nulls and group nulls~\eqref{eqn:group_null_mean} in Setting 1, at different test group sizes (20, 50, and 200) and levels.}
    \label{fig:group_mean_1}
\end{figure}

\begin{figure}[htbp]
    \centering
    \includegraphics[width=0.9\linewidth]{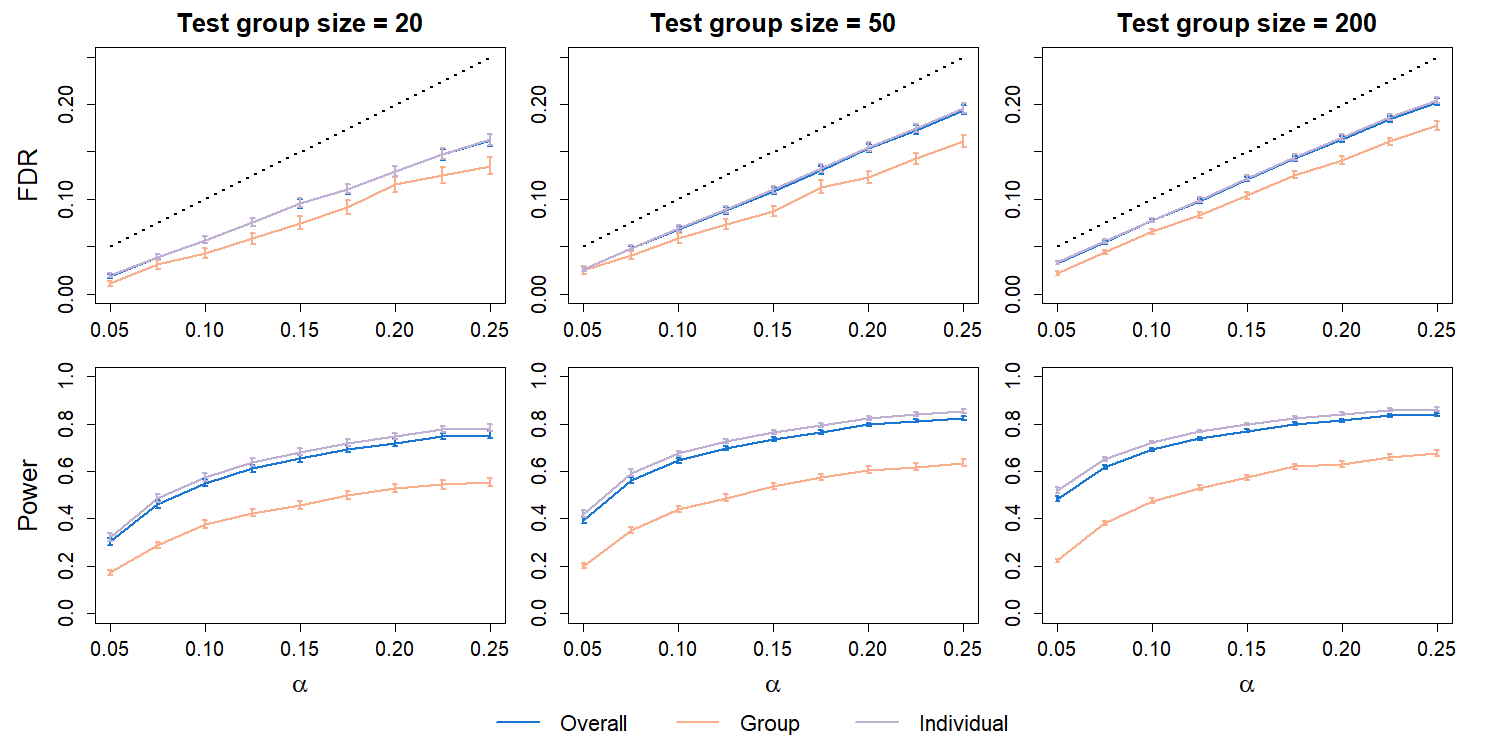}
    \caption{The false discovery rate and power of the joint selection procedure for individual nulls and group nulls~\eqref{eqn:group_null_mean} in Setting 2, at different test group sizes (20, 50, and 200) and levels.}
    \label{fig:group_mean_2}
\end{figure}

Next, we examine the selection procedure for the case where the group nulls are given in the form of
\begin{equation}\label{eqn:group_null_mean}
    H_{j} : \frac{1}{N_{K+j}}\sum_{i=1}^{N_{K+j}} Y_{K+j,i} \leq c,
\end{equation}
which means we select groups with sufficiently large average outcomes. We run the procedure as described in Section~\ref{sec:group_general}, in two settings:
\begin{enumerate}
    \item [(1)] Setting 1 (constant group size) : $N = 10$ almost surely,
    \item [(2)] Setting 2 (heterogeneous group size) : $N \sim 2+\textnormal{Poisson}(5)$.
\end{enumerate}
Figure~\ref{fig:group_mean_1} and~\ref{fig:group_mean_2} show the results for Setting 1 and Setting 2, respectively.
Here, we see a similar story: the overall FDR is controlled as desired, and the group-FDR and individual-FDR are also well-controlled.

\subsection{Simulations with an alternative score}

\begin{figure}[htbp]
    \centering
    \includegraphics[width=0.9\linewidth]{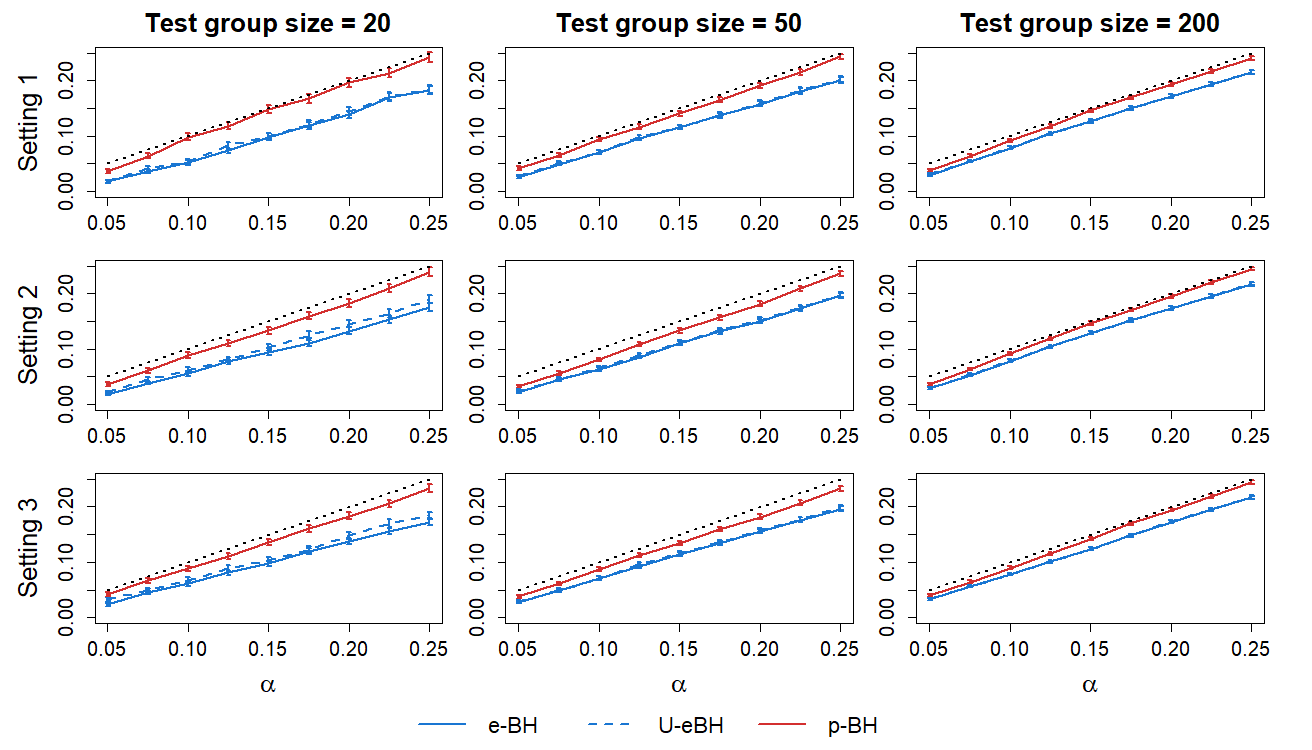}
    \caption{False discovery rate of Procedure~\ref{alg:main} under an alternative score.}
    \label{fig:alt_score_fdr}
\end{figure}

\begin{figure}[htbp]
    \centering
    \includegraphics[width=0.9\linewidth]{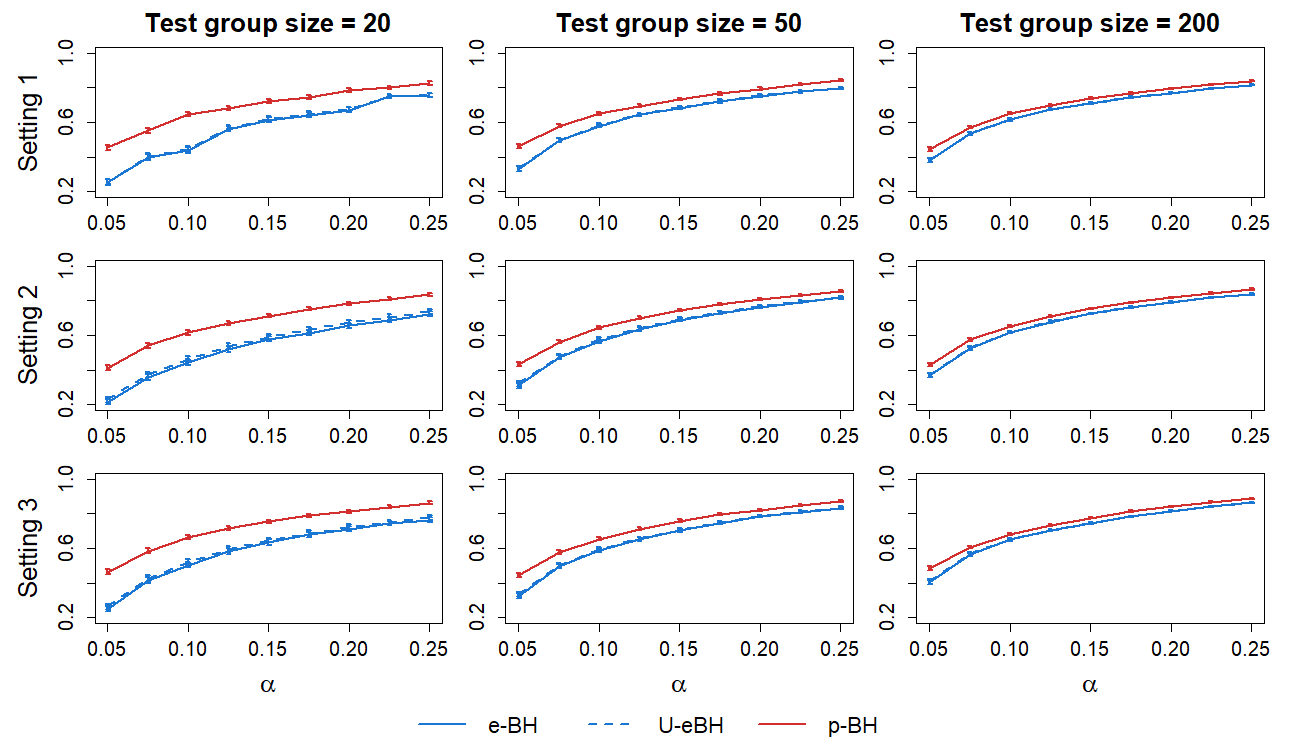}
    \caption{Power of Procedure~\ref{alg:main} under an alternative score.}
    \label{fig:alt_score_power}
\end{figure}

In this section, we further explore the performance of the subsampling-based procedure~\ref{alg:main}—which showed higher performance—under a different choice of the score function. Here, we construct a score $s: \X \times \R \rightarrow [0,1]$ such that $s(g,x,c)$ estimates the probability $\PPst{Y \leq c}{G=g,X=x}$. We follow the same experimental setting as in Section~\ref{sec:experiments}, with three different settings of the group sizes and the test group numbers as described in that section, and fit a random forest classifier on the training data with the transformed outcomes $\One{Y_i \leq c}$ to construct the score. The results are shown in Figures~\ref{fig:alt_score_fdr} and~\ref{fig:alt_score_power}, and show comparable performance to the residual-based score.

\subsection{Simulations with different group size distributions}

\begin{figure}[htbp]
    \centering
    \includegraphics[width=0.96\linewidth]{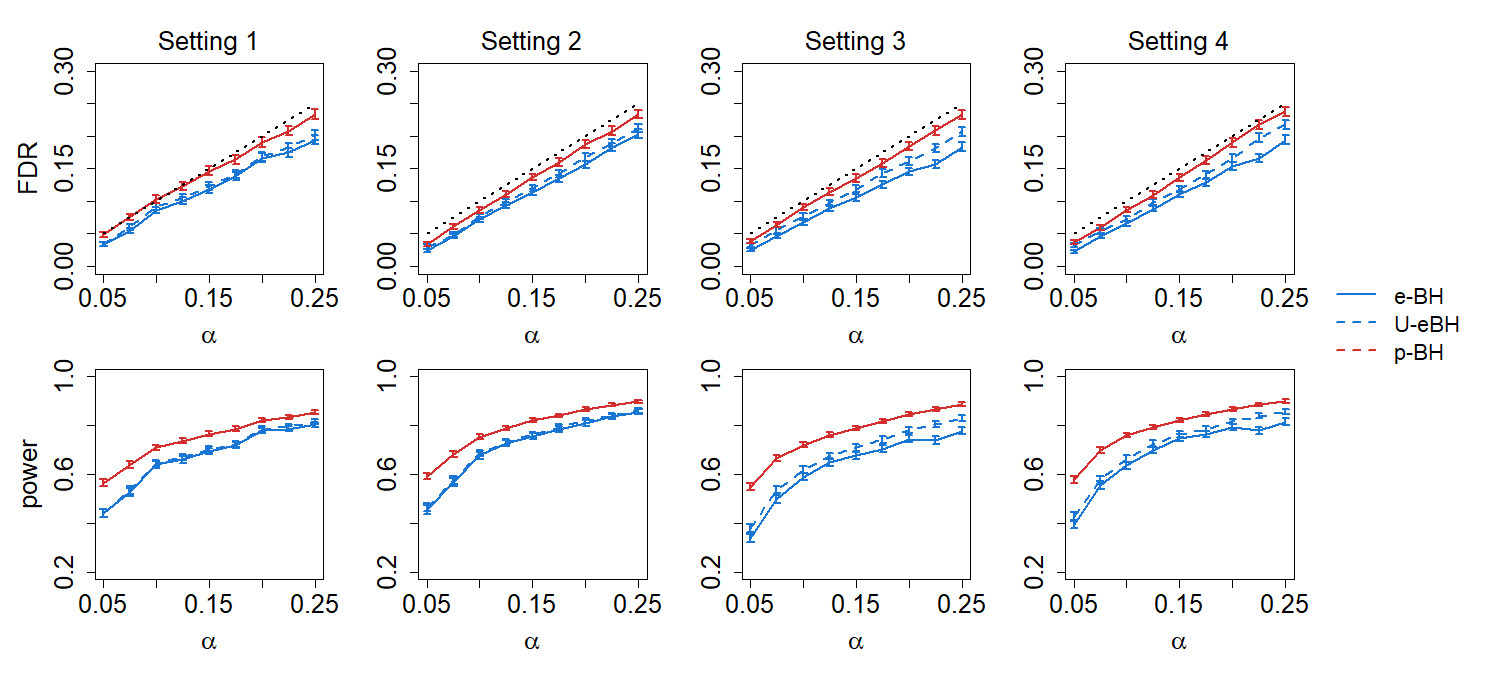}
    \caption{False discovery rate and power of Procedure~\ref{alg:main} under different group size distributions.}
    \label{fig:diff_group_size}
\end{figure}

In this section, we repeat the simulations in Section~\ref{sec:experiments} with varying group size distributions. We follow the experimental setting in Section~\ref{sec:experiments} with test group size $M = 20$ and consider four different group size distributions:
\begin{enumerate}
\item Setting 1: $N \equiv 2$ (i.e., $N = 2$ almost surely),
\item Setting 2: $N \equiv 10$,
\item Setting 3: $N \sim \text{Unif}(\{2,3,\ldots,10\})$,
\item Setting 4: $N \sim 2 + \text{Poisson}(5)$.
\end{enumerate}
The false discovery rate and power of Procedure~\ref{alg:main} under the above settings are shown in Figure~\ref{fig:diff_group_size}.

\subsection{Simulations under non-i.i.d. sampling}

\begin{figure}[htbp]
    \centering
    \includegraphics[width=0.96\linewidth]{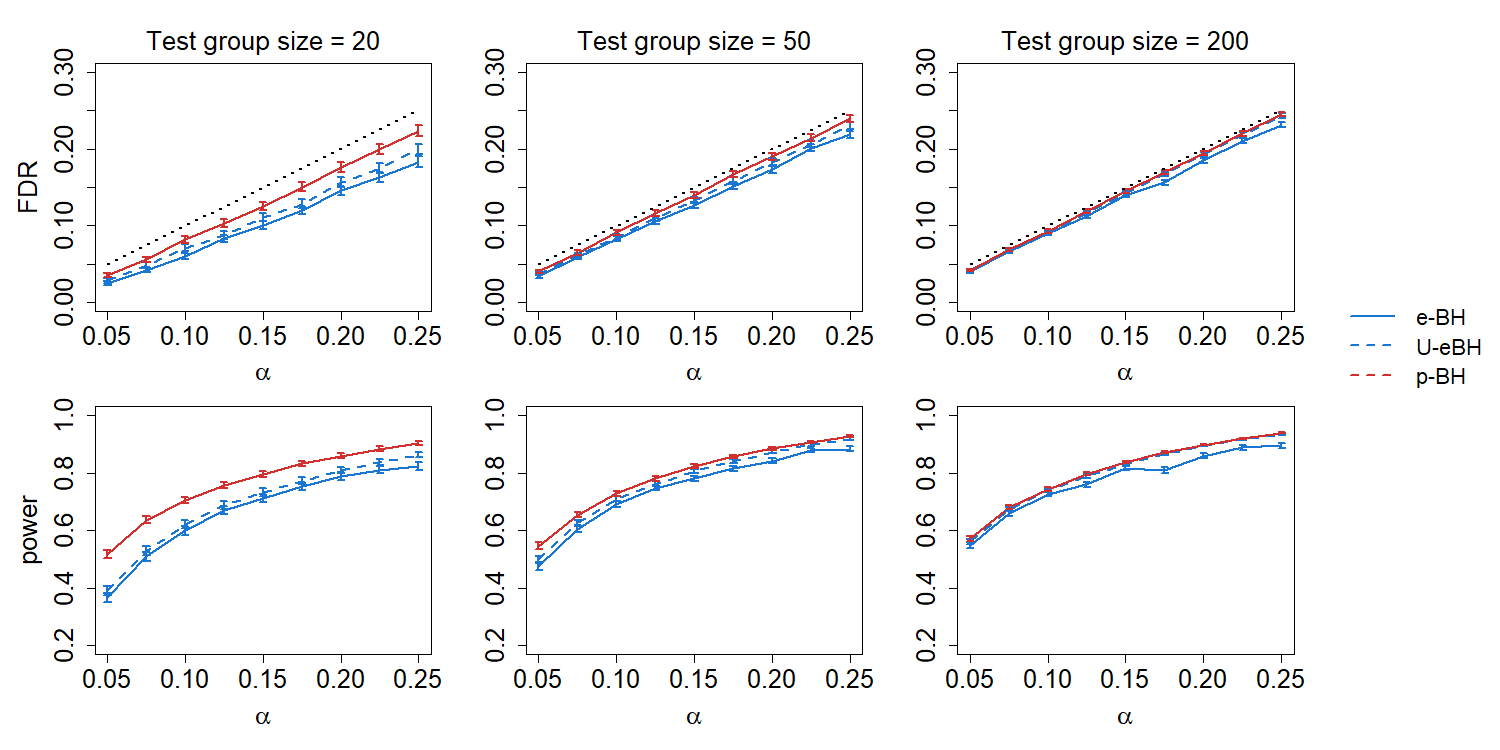}
    \caption{False discovery rate and power of Procedure~\ref{alg:main} under the setting~\ref{eq:sim_dgp_exch}.}
    \label{fig:exch}
\end{figure}

Here, we provide simulation results under the setting where the groups, as well as the within-group individual samples, satisfy hierarchical exchangeability but are not independent. We generate the data as follows.
\begin{align}\label{eq:sim_dgp_exch}
    \begin{split}
    &Z \sim N(0,1),\\
    &G_k = G_k' + Z, \textnormal{ for } k=1,\ldots,K, \textnormal{ where } G_1',\ldots,G_{K+M}' \iidsim \textnormal{Unif}([-5,5])^{p_G}\\
    &N_k \iidsim 2+\textnormal{Poisson}(\lambda), \textnormal{ for } k=1,\ldots,K,\\
    &X_{k,1}, X_{k,2},\cdots,X_{k,N_k} \mid G_k,N_k \iidsim N_p(A G_k, 3 \cdot I_p),\\
    &Z_1,\ldots,Z_K \iidsim N(0,5^2),\\
    &Y_{k,i} = Y_{k,i}' + Z_k, \text{ where } Y_{k,i}' \mid X_{k,i}, G_k, N_k \sim N(\beta_1^\top X_i + \log |\beta_2^\top G_k|, \sigma^2\cdot\|X\|/p) + Z_k
    \end{split}
\end{align}
Note that the variable $Z$ induces the between-group dependence, while the variables $Z_k$ induce the within-group dependence in group $k$. The dimensions $p_G$ and $p$, the parameters $\beta_1$ and $\beta_2$, and the matrix $A$ are defined as in the experiments in Section~\ref{sec:experiments}. We set $\lambda = 5$ and conduct experiments for test group sizes 20, 50, and 200. The results are shown in Figure~\ref{fig:exch}.

\subsection{Analysis on the resampling risk}

The experimental results from the previous sections show that Procedure~\ref{alg:main} achieves good performance across different settings, tightly approximating the p-value-based procedure. In this section, we explore whether such performance can be affected by the randomization component of the procedure. Specifically, we investigate the potential risk of low performance due to having a ``bad subsample.”

\begin{figure}[htbp]
    \centering
    \includegraphics[width=0.96\linewidth]{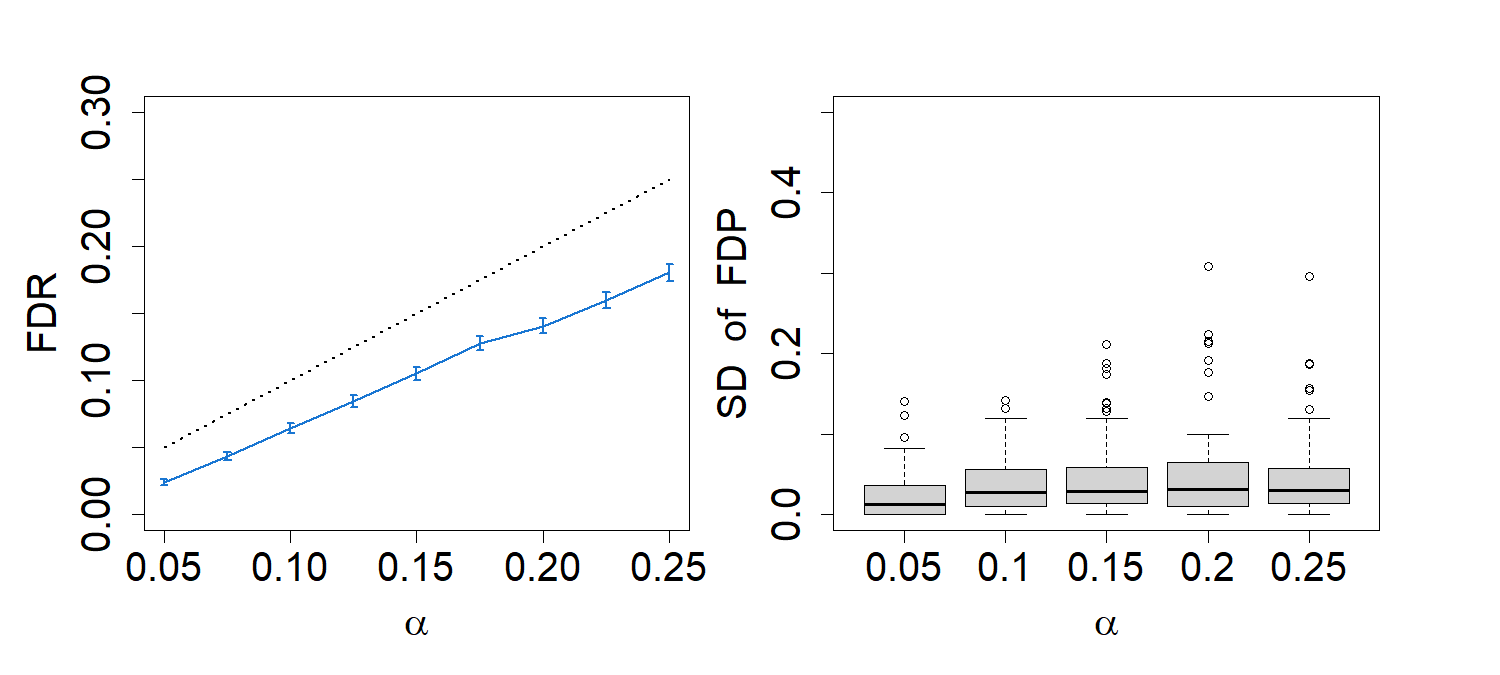}
    \caption{False discovery rate and standard deviation of the FDP of Procedure~\ref{alg:main}.}
    \label{fig:subsampling}
\end{figure}

We generate the data as in Section~\ref{sec:experiments}, but in each trial---i.e., for each realization of the data---we repeat the process of constructing a subsample and running the BH procedure 50 times. We compute the standard deviation of the false discovery proportion of Procedure~\ref{alg:main} across multiple subsamples. The results are shown in Figure~\ref{fig:subsampling}, illustrating that the variance across different subsamples is relatively small and thus it is unlikely that we face a high false discovery proportion due to a bad subsample.

\subsection{Selection based on individual treatment effects}
\label{appx:sim_ite}

\begin{figure}[htbp]
    \centering
    \includegraphics[width=0.9\linewidth]{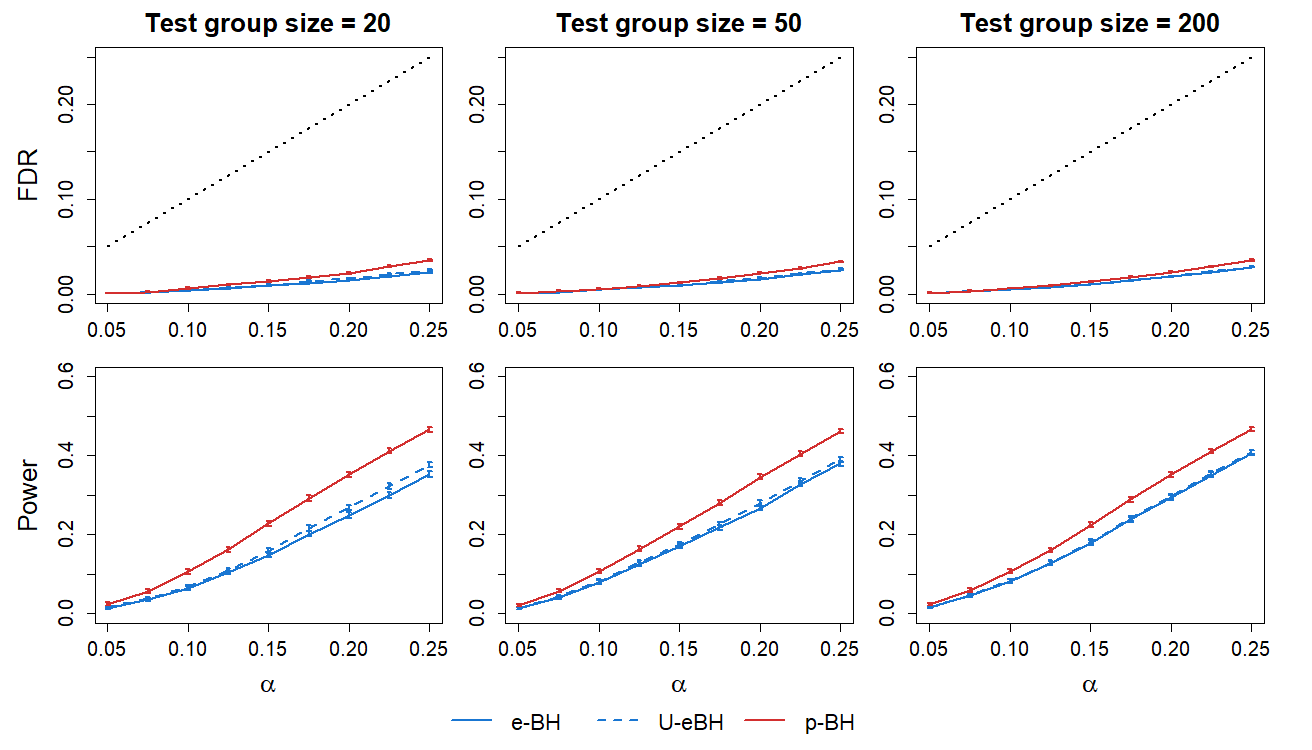}
    \caption{False discovery rate and power of the selection procedure for the nulls~\eqref{eqn:hyp_ite} using conformal e-values~\eqref{eqn:e_ite} (e-BH) and its boosted version (U-eBH), along with the procedure using conformal p-values~\eqref{eqn:p_ite}, across different test group sizes at levels $\alpha=0.05,0.075,\cdots,0.25$.}
    \label{fig:ite}
\end{figure}

In this section, we illustrate the performance of the selection procedure based on individual treatment effects, as discussed in Section~\ref{sec:ite}. We generate the data as previously, with $\lambda=10$ and test group sizes $20, 50$, and $200$. We compare the performance of procedures using e-values and p-values obtained from subsampling. The results shown in Figure~\ref{fig:ite}, demonstrating that the proposed procedure controls the FDR while closely approximating the p-value-based method.

\subsection{Choice of the hyperparameter}

\begin{figure}[ht]
\centering
\includegraphics[width=0.8\linewidth]{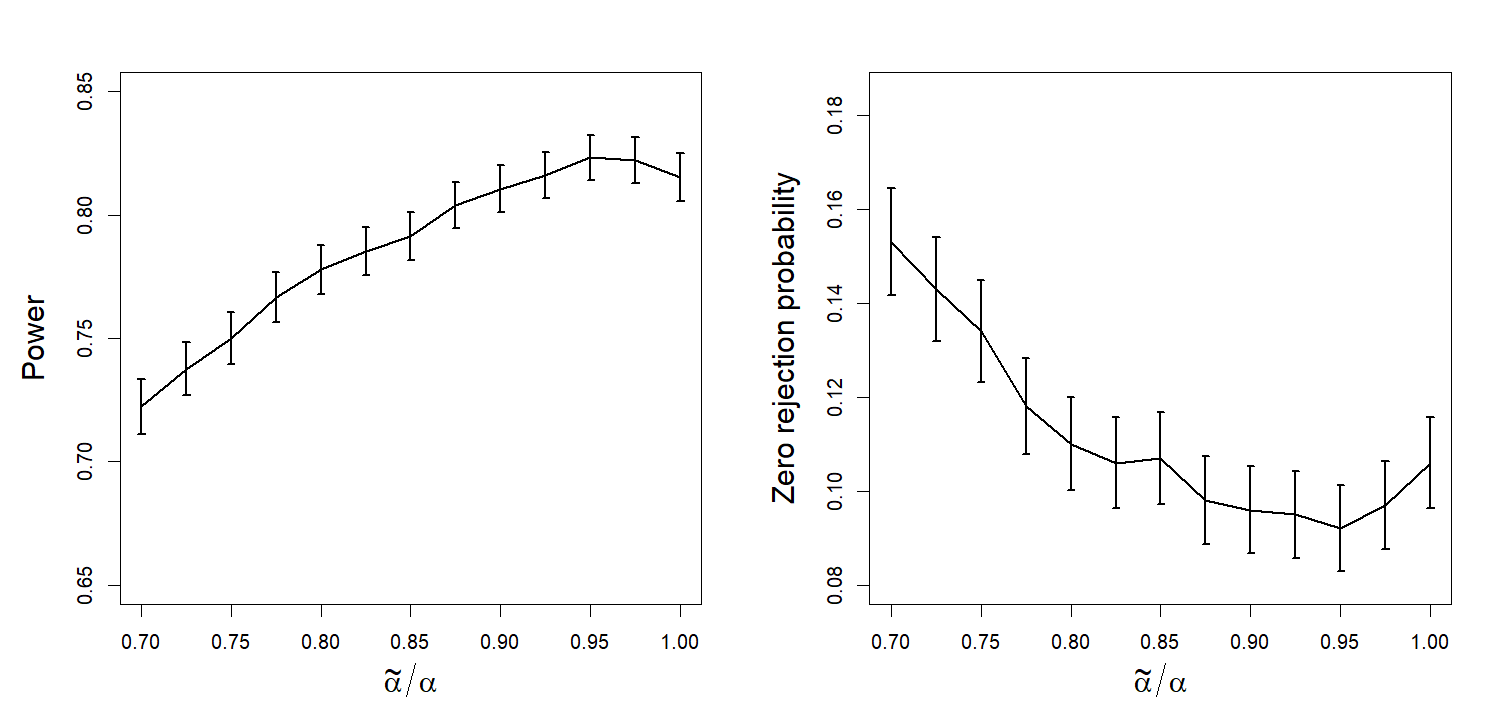}
\caption{Power and Probability of making zero rejections as a function of 
$\tilde \alpha /\alpha$.}
\label{fig:tilalpha}
\end{figure}

In this section, we provide an empirical illustration for the choice $\tilde{\alpha} = 0.9\alpha$ of the hyperparameter, which is discussed in Remark~\ref{rmk:alpha}. We follow the simulation setting in Section~\ref{sec:sim_sub}, with parameters $p_G=30, p=15, \sigma=1$, and group sizes $K_{\text{train}} = 50, K=100$, and $M=20$. We repeat the simulation under different values of $\tilde{\alpha}$ and examine the power of the procedure and the proportion of trials with zero rejections. The results are shown in Figure~\ref{fig:tilalpha}.

\section{Technical proofs}\label{sec:proofs}

\subsection{Proof of Proposition~\ref{prop:equiv}}\label{sec:prop_proof}
Throughout the proof, we use $\cR_{\pbh}$ as a shorthand 
for $\cR_{\pbh}\{(p_{j,i}^\sub)_{j\in[M],i\in [N_{K+j}]}\}$
and $\cR_\ebh$ as a shorthand for 
$\cR_{\ebh}\{(\tilde e_{j,i})_{j\in[M],i\in [N_{K+j]}}\}$.
Since both p-BH and e-BH construct their rejection sets by 
thresholding the $\hat V_{K+j}$'s. It suffices
to show that $|\cR_\pbh| = |\cR_\ebh|$.
We consider two cases.

\paragraph{Case 1.}
When $|\cR_\pbh| = 0$, we prove by contradiction that 
$|\cR_\ebh| = 0$. Assuming $\cR_\ebh \neq \varnothing$, we 
let $(i_0,j_0)$ denote the index of the unit with the 
largest $\hat V_{K+j,i}$ in $\cR_\ebh$, with ties broken arbitrarily. 
By the property of e-BH, there is 
$\hat V_{K+j_0,i_0} < T^\pbh$ and that
\begin{align}\label{eq:eval_thres}
\tilde e_{j_0,i_0} = 
\frac{K+1}{1 + \sum^K_{k=1} \One{\hat V_{k,i_k^*} < T^\pbh,Y_{k,i_k^*}\le C_{k,i_k^*}}}
\ge \frac{\sum^M_{j=1}N_{K+j}}{\alpha |\cR_\ebh|}.
\end{align}
On the other hand, the p-value corresponding to $(j_0,i_0)$ is 
\begin{align*}
p_{j_0,i_0}^\sub & = \frac{1 + \sum^K_{k=1} \One{Y_{k,i_k^*} \le C_{k,i_k^*}, \hat V_{k,i_k^*}\le \hat V_{K+j_0,i_0} }}{K+1} \\
& \le \frac{1 + \sum^K_{k=1} \One{Y_{k,i_k^*} \le C_{k,i_k^*}, \hat V_{k,i_k^*} < T^\pbh}}{K+1}
\le \frac{\alpha |\cR_\ebh|}{\sum^M_{j=1}N_{K+j}},
\end{align*}
where the last step follows from~\eqref{eq:eval_thres}. 
Now using the definition of p-BH, we have $(j_0,i_0)\in \cR_\pbh$, which is a contradiction.
We therefore conclude that $|\cR_\ebh|=0$.

\paragraph{Case 2.}
We now consider the case that $|\cR_\pbh|>0$. 
Let $(j_1,i_1)$ denote the index corresponding to the largest $\hat V_{K+j,i}$
in $\cR_{\pbh}$, with the ties broken arbitrarily. 
There is 
\begin{align} \label{eq:pval_st}
p_{j_1,i_1}^\sub \le \frac{\alpha |\cR_\pbh|}{\sum_{j=1}^M N_{K+j}} 
& \Rightarrow 
\frac{\sum^M_{j=1} N_{K+j}}{K+1} \frac{1+\sum_{k=1}^K
\One{Y_{k,i_k^*} \le C_{k,i_k^*}, \hat V_{k,i_k^*} \le \hat V_{K+j_1,i_1} }}{|\cR_\pbh|} \le \alpha\notag\\
& \Rightarrow 
\frac{\sum^M_{j=1} N_{K+j}}{K+1} \frac{1+\sum_{k=1}^K
\One{Y_{k,i_k^*} \le C_{k,i_k^*}, \hat V_{k,i_k^*} \le \hat V_{K+j_1,i_1} }}
{\sum^M_{j=1}\sum^{N_{K+j}}_{i=1} \One{\hat V_{K+j,i} \le \hat V_{K+j_1,i_1}}} \le \alpha.
\end{align}
Choose $t_1 > \hat V_{K+j_1,i_1}$ such that 
$t_1 < \min\{\hat V_{j,i}: \hat V_{j,i} > \hat V_{K+j_1,i_1}, j\in[K+M], i\in N_j\}$. Then 
\begin{align*}
\widehat{\FDP}^\pbh(t_1) & =  
\frac{\sum^M_{j=1} N_{K+j}}{K+1} \frac{1+\sum_{k=1}^K
\One{Y_{k,i_k^*} \le C_{k,i_k^*}, \hat V_{k,i_k^*} < t_1}}{\sum^M_{j=1}\sum^{N_{K+j}}_{i=1} \One{\hat V_{K+j,i} < t_1}}\\
& = \frac{\sum^M_{j=1} N_{K+j}}{K+1} \frac{1+\sum_{k=1}^K
\One{Y_{k,i_k^*} \le C_{k,i_k^*}, \hat V_{k,i_k^*} \le \hat V_{K+j_1,i_1} }}{\sum^M_{j=1} \sum^{N_{K+j}}_{i=1}\One{\hat V_{K+j,i} \le \hat V_{K+j_1,i_1}}} \le \alpha,
\end{align*}
where the last step follows from Equation~\eqref{eq:pval_st}.
Recalling the definition of $T^\pbh$, we have that $t_1 \le T^\pbh$.
For any $(j,i) \in \cR_\pbh$, there is 
$\hat V_{K+j,i} \le \hat V_{K+j_1,i_1} < t_1 \le T^\pbh$.
Now let $(j_2,i_2)$ denote the index corresponding to 
the largest $\hat V_{K+j,i}< T^\pbh$. By definition, 
$\hat V_{K+j_1,i_1} \le \hat V_{K+j_2,i_2} $. Suppose that 
$\hat V_{K+j_1,i_1} < \hat V_{K+j_2,i_2} $. Then,
\begin{align*} 
p_{j_2,i_2}^\sub & =  
\frac{1+\sum_{k=1}^K\One{Y_{k,i_k^*} \le C_{k,i_k^*}, \hat V_{k,i_k^*} \le \hat V_{K+j_2,i_2} }}{K+1} \\
& \le \widehat{\FDP}^\pbh(T^\pbh) 
\frac{\sum^M_{j=1}\sum_{i=1}^{N_{K+j}} \One{\hat V_{K+j,i}  \le \hat V_{K+j_2,i_2}}}{\sum^M_{j=1}N_{K+j}}\\
&\le \alpha \cdot \frac{\sum^M_{j=1}\sum_{i=1}^{N_{K+j}} \One{p_{j,i}^\sub  \le \hat p^\sub_{j_2,i_2}}}{\sum^M_{j=1}N_{K+j}},
\end{align*}
where the last step uses the definition of $T^\pbh$.
The above implies that $(j_2,i_2)$ should be selected by $\cR_\pbh$, which is a
contradiction. We therefore conclude that $\hat V_{K+j_1,i_1} = \hat V_{K+j_2,i_2}$.

For any $(j,i)\in \cR_\pbh$, there is 
\begin{align*}
\tilde e_{j,i} & = \frac{K+1}{1+\sum^K_{k=1}\One{\hat V_{k,i_k^*}< T^\pbh, Y_{k,i_k^*} \le C_{k,i_k^*}}} \\
& \ge \frac{1}{\alpha} \frac{\sum^M_{j=1}N_{K+j}}{\sum^M_{j=1}\sum^{N_{K+j}}_{i=1} 
\One{\hat V_{K+j,i} < T^\pbh}} \\
& = \frac{1}{\alpha} \frac{\sum^M_{j=1}N_{K+j}}{\sum^M_{j=1}\sum^{N_{K+j}}_{i=1} 
\One{\hat V_{K+j,i} \le \hat V_{K+j_1,i_1}}}  = \frac{\sum^M_{j=1}N_{K+j}}{\alpha |\cR_\pbh|}.
\end{align*}
The above implies that $(j,i)\in \cR_\ebh$. Moreover, for any $(j,i)\notin \cR_\pbh$, 
there must be $\hat V_{K+j,i} > \hat V_{K+j_1,i_1} = \hat V_{K+j_2,i_2}$, which also implies 
$\hat V_{k+j} \ge T^\pbh$. As a result, $\tilde e_{j,i} = 0$ and will never be included in 
$\cR_\ebh$. The proof is completed.

\subsection{Proof of Lemma~\ref{lem:ebh}}
\label{appx:proof_ebh}
Let $\cR_{\ebh}$ denote the rejection set obtained from 
applying the e-BH procedure to $(e_{j,i})_{1 \leq j \leq M, 1 \leq i \leq N_{K+j}}$.
Following the proof steps of~\citet{wang2022false}, we express the FDP 
of $\cR_\ebh$ as
\begin{align} \label{eqn:FDP_bound}
\FDP & = \frac{\sum_{j=1}^M \sum_{i=1}^{N_{K+j}} \mathbbm{1}\{H_{j,i}\} \cdot \One{e_{j,i} \ge 
\frac{\sum_{\ell = 1}^M N_{K+ \ell}}{\alpha |\cR_\ebh|}}}
{|\cR_\ebh|} \notag\\
& \leq \alpha \cdot \frac{\sum_{j=1}^M \sum_{i=1}^{N_{K+j}} \One{H_{j,i}} \cdot e_{j,i} }{\sum_{j=1}^M N_{K+j}},
\end{align}
where the inequality follows from $\mathbbm{1}\{x \ge t\} \le x/t$ for any $t>0$.
Taking the expectation of both sides of~\eqref{eqn:FDP_bound} conditional on $N_{K+1:K+M}$, we have
\begin{align*}
        \cFDR = \EEst{\FDP}{N_{K+1:K+M}}&  \leq \alpha \cdot \EEst{\frac{\sum_{j=1}^M \sum_{i=1}^{N_{K+j}} \One{H_{j,i}} \cdot e_{j,i} }{\sum_{j=1}^M N_{K+j}}}{N_{K+1:K+M}}\\
        & = \frac{\alpha}{\sum_{j=1}^M N_{K+j}} \cdot \sum_{j=1}^M \sum_{i=1}^{N_{K+j}} \EEst{\One{H_{j,i}} \cdot e_{j,i}}{N_{K+1:K+M}}\\
        & \leq \frac{\alpha}{\sum_{j=1}^M N_{K+j}} \cdot \sum_{j=1}^M \sum_{i=1}^{N_{K+j}} 1 = \alpha,
    \end{align*}
    where the inequality uses the definition of conditional e-values~\eqref{eqn:e_value}.

\subsection{Proof of Theorem~\ref{thm:individual}}
\label{appx:proof_individual}
Our proof strategy involves introducing an `oracle' threshold and showing that,  
on the event $\{H_{j,i}, \hat V_{j,i} < T_j\}$, we can replace 
our `working' threshold $T_j$ with the oracle one; we then leverage 
the properties of the oracle thresholds to establish the validity of $e_{j,i}$.

We begin by defining the FDP estimate corresponding to the oracle threshold:
\[\widetilde{\FDP}_{j,i}(t) = \frac{\sum_{k=1}^K \One{\hat{V}_{k,i_k^*} < t, Y_{k,i_k^*} \leq C_{k,i_k^*}} + \One{\hat{V}_{K+j,i} < t, Y_{K+j,i} \leq C_{K+j,i}}}{1 \vee \sum_{\substack{l \in [M], l \neq j}} \sum_{i'=1}^{N_{K+l}} \One{\hat{V}_{K+l, i'} < t}} \cdot \frac{\sum_{l=1}^M N_{K+l}}{K+1},\]
where we replace the `$+1$' in the numerator of $\widehat{\FDP}_j$
with $\mathbbm{1}\{\hat{V}_{K+j,i} < t, Y_{K+j,i} \leq C_{K+j,i}\}$ 
(this is oracle since we do not have access to $Y_{K+j,i}$).
The oracle threshold is correspondingly
\[ \tilde{T}_{j,i} = \sup\big\{t \in \R : \widetilde{\FDP}_{j,i}(t) \leq \tilde{\alpha}\big\}. \]
Next, we show that 
$T_j = \tilde{T}_{j,i}$ holds if $\hat{V}_{K+j,i} < T_j$ and $Y_{K+j,i} \leq C_{K+j,i}$. Observe that $\widehat{\FDP}_j(t) = \widetilde{\FDP}_j(t)$ for any $t > \hat{V}_{K+j,i}$, provided that $Y_{K+j,i} \leq C_{K+j,i}$ holds. Fix any $\eps > 0$. By definition of $T_j$, if $\hat{V}_{K+j,i} < T_j$, then there exists $\max\{T_j-\eps, \hat{V}_{K+j,i}\} < t < T_j$ such that $\widehat{\FDP}_j(t) \leq \alpha$. Since $t > \hat{V}_{K+j,i}$, this implies $\widetilde{\FDP}_j(t) \leq \alpha$, and consequently $\tilde{T}_{j,i} \geq t > T_j-\eps$. As this holds for any $\eps > 0$, we have $\tilde{T}_{j,i} \geq T_j$. Therefore, the condition $\hat{V}_{K+j,i} < T_j$ implies $\hat{V}_{K+j,i} < \tilde{T}_{j,i}$, allowing us to apply analogous arguments to derive $T_j \geq \tilde{T}_{j,i}$. Consequently, we have $T_j = \tilde{T}_{j,i}$.

Based on the above observation, we have
\begin{align*}
    &\EEst{\One{H_{j,i}} \cdot e_{j,i}}{N_{K+1:K+M}}\\
    &= \EEst{\frac{\One{\hat{V}_{K+j,i} < T_j}}{\sum_{k=1}^K \One{\hat{V}_{k,i_k^*} < T_j, Y_{k,i_k^*} \leq C_{k,i_k^*}} + 1} \cdot (K+1) \One{Y_{K+j,i} \leq C_{K+j,i}}}{N_{K+1:K+M}}\\
    &= \mathbb{E}\left[\rule{0cm}{1cm}\frac{\One{\hat{V}_{K+j,i} < T_j}}{\sum_{k=1}^K \One{\hat{V}_{k,i_k^*} < T_j, Y_{k,i_k^*} \leq C_{k,i_k^*}} + 1} \cdot (K+1)\cdot \One{\hat{V}_{K+j,i} < T_j}\One{Y_{K+j,i} \leq C_{K+j,i}}\right.\\
    &\left.\hspace{14cm} \left|\rule{0cm}{1cm}\right. N_{K+1:K+M}\rule{0cm}{1cm}\right]\\
    &\stackrel{\textnormal{(a)}}{=}\mathbb{E}\left[\rule{0cm}{1cm}\frac{\One{\hat{V}_{K+j,i} < \tilde{T}_{j,i}}}{\sum_{k=1}^K \One{\hat{V}_{k,i_k^*} < \tilde{T}_{j,i}, Y_{k,i_k^*} \leq C_{k,i_k^*}} + 1} \cdot (K+1)\cdot \One{\hat{V}_{K+j,i} < T_j}\One{Y_{K+j,i} \leq C_{K+j,i}}\right.\\
     &\left.\hspace{14cm} \left|\rule{0cm}{1cm}\right.N_{K+1:K+M}\right]\\
    &\leq \EEst{\frac{\One{\hat{V}_{K+j,i} < \tilde{T}_{j,i}, Y_{K+j,i} \leq C_{K+j,i}}}{\sum_{k=1}^K \One{\hat{V}_{k,i_k^*} < \tilde{T}_{j,i}, Y_{k,i_k^*} \leq C_{k,i_k^*}} + 1} \cdot (K+1)}{N_{K+1:K+M}}.
\end{align*}
Above, step (a) uses the fact that $T_j = \tilde T_{j,i}$ on the event 
$\{\hat V_{K+j,i} < T_j, Y_{K+j,i} \le C_{K+j,i}\}$.
Due to the within-group and between-group exchangeability, we have that the subsamples
\[
(X_{1,i_1^*}, Y_{1,i_1^*}), \cdots, (X_{K,i_K^*}, Y_{K,i_K^*}), (X_{K+j,i},Y_{K+j,i})
\] 
are exchangeable conditional on the group sizes.
Also note that $\tilde{T}_{j,i}$ is invariant with respect to arbitrary permutations 
on these subsamples, so for any $k \in [K]$, we have
\begin{align*}
&\EEst{\frac{\One{\hat{V}_{K+j,i} < \tilde{T}_{j,i}, Y_{K+j,i} \leq C_{K+j,i}}}{\sum_{k=1}^K \One{\hat{V}_{k,i_k^*} < \tilde{T}_{j,i}, Y_{k,i_k^*} \leq C_{k,i_k^*}} + 1}}{N_{K+1:K+M}}\\
& = \EEst{\frac{\One{\hat{V}_{K+j,i} < \tilde{T}_{j,i}, Y_{K+j,i} \leq C_{K+j,i}}}{\sum_{k=1}^K \One{\hat{V}_{k,i_k^*} < \tilde{T}_{j,i}, Y_{k,i_k^*} \leq C_{k,i_k^*}} + \One{\hat{V}_{K+j,i} < \tilde{T}_{j,i}, Y_{K+j,i} \leq C_{K+j,i}}}}{N_{K+1:K+M}}\\
&= \EEst{\frac{\One{\hat{V}_{k,i^*_k} < \tilde{T}_{j,i}, Y_{k,i_k^*} \leq C_{k,i_k^*}}}{\sum_{k'=1}^K \One{\hat{V}_{k',i_{k'}^*} < \tilde{T}_{j,i}, 
Y_{k',i_{k'}^*} \leq C_{k',i_{k'}^*}} + \One{\hat{V}_{K+j,i} < \tilde{T}_{j,i}, Y_{K+j,i} \leq C_{K+j,i}}}}{N_{K+1:K+M}},
\end{align*}
where the first step replaces the $1$ in the denominator 
with $\One{\hat V_{K+j,i} < \tilde{T}_{j,i}, Y_{K+j,i} \leq C_{K+j,i}}$---this 
does not change the value of the ratio when the latter equals to one;
the last step follows from the fact that the joint distribution of 
$(X_{1,i_1^*}, Y_{1,i_1^*}), \cdots, (X_{K,i_K^*}, Y_{K,i_K^*}), (X_{K+j,i},Y_{K+j,i})$
remains invariant upon the swap of $(k,i_k^*)$ and $(K+j,i)$.
As a result, 
\begin{align*}
&\EEst{\One{H_{j,i}} \cdot e_{j,i}}{N_{K+1:K+M}}\\
& \le \sum_{k =1}^K
\EEst{\frac{\One{\hat{V}_{k,i_k^*} < \tilde{T}_{j,i}, Y_{k,i_k^*} \leq C_{k,i_k^*}}}{\sum_{k'=1}^K \One{\hat{V}_{k',i_{k'}^*} < \tilde{T}_{j,i}, Y_{k',i_{k'}^*} \leq C_{k',i_{k'}^*}} + \One{\hat{V}_{K+j,i} < \tilde{T}_{j,i}, Y_{K+j,i} \leq C_{K+j,i}}}}{N_{K+1:K+M}}\\
& \qquad + \EEst{\frac{\One{\hat{V}_{K+j,i} < \tilde{T}_{j,i}, Y_{K+j,i} \leq C_{K+j,i}}}{\sum_{k=1}^K \One{\hat{V}_{k,i_k^*} < \tilde{T}_{j,i}, Y_{k,i_k^*} \leq C_{k,i_k^*}} + \One{\hat{V}_{K+j,i} < \tilde{T}_{j,i}, Y_{K+j,i} \leq C_{K+j,i}} }}{N_{K+1:K+M}}\\
& \le 1.
\end{align*}
We have so far established that $\EEst{e_{j,i}}{N_{K+1:K+M}} \le 1$, i.e., 
$e_{j,i}$ is a conditional e-value.
Finally, by Lemma~\ref{lem:ebh}, the e-BH procedure applied to $(e_{j,i})_{1 \leq j \leq M, q \leq i \leq N_{K+j}}$ controls the cFDR.

\subsection{Proof of Corollary~\ref{cor:uebh}}
Following the arguments in~\citet{xu2023more}, we first show that, given an e-value $e$ and a grid $\mathcal{G}$, the stochastically rounded e-value $S_\mathcal{G}(e)$ is also an e-value. Here, $S_\mathcal{G}(e)$ is defined as  
\[S_\mathcal{G}(e) = e_- \One{U > \frac{e - e_-}{e_+ - e_-}} + e_+ \One{U \leq \frac{e - e_-}{e_+ - e_-}},\]
where $U \sim \text{Unif}([0,1])$, and $e_-$ and $e_+$ denote the lower and upper grid points bracketing $e$ with respect to $\mathcal{G}$.

The variable $e$ is assumed to be an e-value in the sense that $\EE{e \cdot \One{H_0}} \leq 1$ for some null event $H_0$, and we aim to show that this implies $\EE{S_\mathcal{G}(e) \One{H_0}} \leq 1$. We compute
\begin{align*}
    &\EE{S_\mathcal{G}(e) \One{H_0}} = \EE{\left(e_- \One{U > \frac{e - e_-}{e_+ - e_-}} + e_+ \One{U \leq \frac{e - e_-}{e_+ - e_-}}\right)\One{H_0}}\\
    &= \EE{\EEst{\left(e_- \One{U > \frac{e - e_-}{e_+ - e_-}} + e_+ \One{U \leq \frac{e - e_-}{e_+ - e_-}}\right)\One{H_0}}{e}}\\
    &= \EE{e_- \cdot \left(1-\frac{e - e_-}{e_+ - e_-}\right)\cdot \One{H_0} + e_+ \cdot \left(\frac{e - e_-}{e_+ - e_-}\right) \cdot \One{H_0}}\\
    &= \EE{\frac{e_-\cdot e_+ - e_- \cdot e + e_+ \cdot e - e_+ \cdot e_-}{e_+ - e_-}\cdot \One{H_0}} = \EE{e \cdot \One{H_0}} \leq 1.
\end{align*}
Therefore, the e-values $e_{j,i}$ in the definition of~\eqref{eqn:e_value} remain valid e-values after stochastic rounding. Hence, by Lemma~\ref{lem:ebh}, applying the e-BH procedure to the rounded e-values provides valid $\cFDR$ control. Moreover, by the result of~\citet{xu2023more}, applying the e-BH procedure to the rounded e-values is deterministically equivalent to applying the e-BH procedure to the original e-values divided by $U$. This completes the proof.

\subsection{Proof of Proposition~\ref{prop:hier_p}}
\label{appx:proof_hier_p}
The proof follows the arguments used in the proof for hierarchical conformal prediction~\citep{lee2023distribution}. Fix any $1 \leq j \leq m$ and $1 \leq i \leq N_{K+j}$. We compute
\begin{align*}
    &\PP{p_{j,i} \leq \alpha \text{ and } Y_{K+j,i} \leq C_{K+j,i}}\\
    &=\PP{\frac{\sum_{k=1}^K \frac{1}{N_k}\sum_{i'=1}^{N_k}\One{V_{k,i'} \leq \hat{V}_{K+j,i}}+1}{K+1} \leq \alpha \text{ and } Y_{K+j,i} \leq C_{K+j,i}}\\
    &\leq \PP{\frac{\sum_{k=1}^K \frac{1}{N_k} \sum_{i' = 1}^{N_k} 
    \One{V_{k,i'} \leq V_{K+j,i}}+1}{K+1} \leq \alpha} \; \textnormal{ since $\hat{V}_{K+j,i} \geq V_{K+j,i}$ if $Y_{K+j,i} \leq C_{K+j,i}$ holds}\\
    &\leq\PP{\frac{\sum_{k \in [K] \cup \{K+j\}} \frac{1}{N_k}\sum_{i'=1}^{N_k}
    \One{V_{k,i} \leq V_{K+j,i}}}{K+1} \leq \alpha}\\
    &\leq \PP{V_{K+j,i} \leq 
    Q_\alpha'\left(\sum_{k \in [K] \cup \{K+j\}} \sum_{i'=1}^{N_k} \frac{1}{(K+1)N_k} \delta_{V_{k,i'}}\right)},
\end{align*}
where we define $Q_\alpha'(P) = \sup\{x : \Pp{X \sim P}{X \leq x} \leq \alpha\}$ for a distribution $P$. 
Now let us write $q = Q_{\alpha}'\left(\sum_{k \in [K]\cup\{K+j\}} \sum_{i'=1}^{N_k} \frac{1}{(K+1)N_k} \delta_{V_{k,i'}}\right)$, 
and observe that $q$ is invariant to the permutation of $(V_{K+j,1},\cdots,$ $V_{K+j,N_{K+j}})$. 
Therefore, by the exchangeability of $(V_{K+j,1}, \cdots, V_{K+j,N_{K+j}})$ 
(conditional on $N_{K+j}$), we have
\begin{multline*}
    \PPst{V_{K+j,i} \leq q}{N_{K+j}} = \EEst{\One{V_{K+j,i} \leq q}}{N_{K+1}} 
    = \frac{1}{N_{K+j}}\sum_{i'=1}^{N_{K+j}} \EEst{\One{V_{K+j,i'} \leq q}}{N_{K+j}}\\
    = \EEst{\frac{1}{N_{K+j}}\sum_{i'=1}^{N_{K+j}}\One{V_{K+j,i'} \leq q}}{N_{K+j}}.
\end{multline*}
Thus, by marginalizing with respect to $N_{K+1}$, we have
\[\PP{V_{K+j,i} \leq q} = \EE{\frac{1}{N_{K+j}}\sum_{i'=1}^{N_{K+j}}\One{V_{K+j,i'} \leq q}}.\]
Next, observe that $q$ also invariant with respect to group-level permutations, 
i.e., any permutation of $(\tilde{V}_1, \cdots, \tilde{V}_K, \tilde{V}_{K+j})$, 
where $\tilde{V}_k = (V_{k,1}, \cdots, V_{k,N_k})$. Therefore, 
by the between-group exchangeability of the data, we have
\begin{align*}
    \EE{\frac{1}{N_{K+j}}\sum_{i'=1}^{N_{K+j}}\One{V_{K+1,i'} \leq q}} 
    & = \frac{1}{K+1}\sum_{k \in [K]\cup\{K+j\}} \EE{\frac{1}{N_k}\sum_{i'=1}^{N_k}\One{V_{k,i'} \leq q}}\\
    & = \EE{\sum_{k\in[K]\cup \{K+j\}} \sum_{i'=1}^{N_k} \frac{1}{(K+1) N_k} \One{V_{k,i'} \leq q}}.
\end{align*}
Note that by the definition of $Q_\alpha'$, 
we have $\sum_{k \in [K]\cup \{K+j\}} \sum_{i'=1}^{N_k} \frac{1}{(K+1) N_k} \One{V_{k,i'} \leq q} \leq \alpha$ deterministically ---observe that the supremum in the definition of $Q_\alpha'$ is equivalent to maximum for discrete distributions.

Therefore, putting everything together, we have
\begin{multline*}
    \PP{p_{j,i} \leq \alpha \text{ and } Y_{K+j,i} \leq C_{K+j,i}} \leq \PP{V_{K+j,i} \leq q}\\
=  \EE{\sum_{k \in [K]\cup\{K+j\}} \sum_{i'=1}^{N_k} \frac{1}{(K+1) N_k} \One{V_{k,i'} \leq q}} \leq \alpha.
\end{multline*}

\subsection{Proof of Theorem~\ref{thm:hierarchical}}
\label{appx:proof_hierarchical}
As in the proof of Theorem~\ref{thm:individual}, we shall define an ``oracle'' stopping time that is invariant to 
permutations within and across groups, and then connect $T_j^+$ and $T_j^-$ to it.
To start, we define for each $j\in[M]$ that 
\begin{equation*}
\begin{split}
    &\tilde{T}_j = \sup\left\{t \in \R : \widetilde{\FDP}_j(t) \leq \tilde{\alpha}\right\}, \text{ where }\\
    &\widetilde{\FDP}_j(t) = \frac{\frac{1}{N_{K+j}}\sum\limits_{i'=1}^{N_{K+j}} \One{\hat{V}_{K+j,i'} < t, Y_{K+j,i'} \leq C_{K+j,i'}}+\sum\limits_{k=1}^{K} \frac{1}{N_k}\sum\limits_{i'=1}^{N_k} \One{\hat{V}_{k,i'} < t, Y_{k,i'} \leq C_{k,i'}}}{1 \vee \sum\limits_{l \neq j} \sum\limits_{i'=1}^{N_{K+l}} \One{\hat{V}_{K+l,i'} < t}}\\
    &\hspace{140mm}\cdot \frac{\sum\limits_{l \neq j} N_{K+l}}{K+1}.
\end{split}
\end{equation*}
It is straightforward to see that $\tilde T_j$ is invariant to the permutations of 
$\tilde{Z}_1,\ldots, \tilde{Z}_K$ and $\tilde{Z}_{K+j}$.
Next, observe that $\widehat{\FDP}_j^-(t) \leq \widetilde{\FDP}_j(t) \leq \widehat{\FDP}_j^+(t)$ for any $t$, implying $T_j^+ \leq \tilde{T}_j \leq T_j^-$. Now, let $N_{-(K+j)} = (N_{K+l})_{l \neq j}$. Note also that $e_{j,i}$ is independent of $N_{K+j}$ conditional on $N_{-(K+j)}$, by the construction of $T_j^+$ and $T_j^-$. Therefore, we have
\begin{multline*}
    \EEst{e_{j,i} \One{H_{j,i}}}{N_{-(K+j)}} = \EEst{e_{j,i}\One{H_{j,i}}}{N_{K+1:K+M}}\\
    = \frac{1}{N_{K+j}}\sum_{i'=1}^{N_{K+j}}\EEst{e_{j,i'}\One{H_{j,i'}}}{N_{K+1:K+M}}
    = \EEst{\frac{1}{N_{K+j}}\sum_{i'=1}^{N_{K+j}} e_{j,i'}\One{H_{j,i'}}}{N_{K+1:K+M}},
\end{multline*}
where the second inequality holds since the $e_{j,i}$'s have the same distribution for $1 \leq i \leq N_{K+j}$. 
Since all terms in the above equation are constants---which are equal to the first term---after conditioning on $N_{-(K+j)}$, this also implies
\begin{equation}\label{eqn:cond_exp}
    \EEst{e_{j,i}\One{H_{j,i}}}{N_{K+1:K+M}} =  \EEst{\frac{1}{N_{K+j}}\sum_{i'=1}^{N_{K+j}} e_{j,i'}\One{H_{j,i'}}}{N_{-(K+j)}}.
\end{equation}
Therefore, by applying $T_j^+ \leq \tilde{T}_j \leq T_j^-$, we have
\begin{align*}
     &\frac{1}{K+1}\cdot \EEst{e_{j,i}\One{H_{j,i}}}{N_{K+1:K+M}}\\
     &= \EEst{\frac{\frac{1}{N_{K+j}}\sum\limits_{i'=1}^{N_{K+j}}\One{\hat{V}_{K+j,i'} < T_j^+, Y_{K+j,i'} \leq C_{K+j,i'}}}{\sum\limits_{k=1}^K \frac{1}{N_k} \sum\limits_{i'=1}^{N_k} \One{\hat{V}_{k,i'} < T_j^-, Y_{k,i'} \leq C_{k,i'}} + 1} }{N_{-(K+j)}}\\
     &\leq \EEst{\frac{\frac{1}{N_{K+j}}\sum\limits_{i'=1}^{N_{K+j}}\One{\hat{V}_{K+j,i'} < \tilde{T}_j, Y_{K+j,i'} \leq C_{K+j,i'}}}{\sum\limits_{k=1}^K \frac{1}{N_k} \sum\limits_{i'=1}^{N_k} \One{\hat{V}_{k,i'} < \tilde{T}_j, Y_{k,i'} \leq C_{k,i'}} + 1}}{N_{-(K+j)}}\\
     &\leq \mathbb{E}\left[\rule{0cm}{1.4cm}\frac{\frac{1}{N_{K+j}}\sum\limits_{i'=1}^{N_{K+j}}\One{\hat{V}_{K+j,i'} < \tilde{T}_j, Y_{K+j,i'} \leq C_{K+j,i'}}}{\sum\limits_{k=1}^K \frac{1}{N_k} \sum\limits_{i'=1}^{N_k} \One{\hat{V}_{k,i'} < \tilde{T}_j, Y_{k,i'} \leq C_{k,i'}} + \frac{1}{N_{K+j}}\sum\limits_{i'=1}^{N_{K+j}}\One{\hat{V}_{K+j,i'} < \tilde{T}_j, Y_{K+j,i'} \leq C_{K+j,i'}}} \left|\rule{0cm}{1.4cm}\right.N_{-(K+j)}\right]\\
     & = \frac{1}{K+1},
\end{align*}
where the last equality holds since $\tilde{T}_j$ is invariant with respect to permutations of $\tilde{Z}_1, \cdots, \tilde{Z}_K$ and $\tilde{Z}_{K+j}$, which are exchangeable and are independent of $N_{-(K+j)}$. Therefore, we have
\[\EEst{e_{j,i}\One{H_{j,i}}}{N_{K+1:K+M}} \leq 1.\]

The second claim follows directly from Lemma~\ref{lem:ebh}.

\subsection{Proof of Proposition~\ref{prop:group_global}}
\label{appx:proof_group_global}
By the result of Theorem~\ref{thm:individual}, we have
\begin{multline*}
    \EEst{e_j \One{H_j}}{N_{K+1:K+M}} = \EEst{\frac{1}{N_{K+j}}\sum_{i=1}^{N_{K+j}} e_{j,i} \One{H_j}}{N_{K+1:K+M}}\\ = \frac{1}{N_{K+j}} \cdot \EEst{\sum_{i=1}^{N_{K+j}} e_{j,i} \One{\cap_{i=1}^{N_{K+j}} H_{j,i}}}{N_{K+1:K+M}}\\
    \leq \frac{1}{N_{K+j}} \EEst{\sum_{i'=1}^{N_{K+j}} e_{j,i} \One{H_{j,i'}}}{N_{K+1:K+M}} \leq 1,
\end{multline*}
for each $j =1,2,\cdots,M$. Therefore, $e_1,\cdots,e_M$ are e-values conditionally on the group sizes $N_{K+1:K+M}$, and thus the e-BH procedure that includes $e_1,\cdots,e_M$ controls the conditional FDR.

\subsection{Proof of Theorem~\ref{thm:group_general}}
\label{appx:proof_group_general}
It is sufficient to show that $e_j$ defined as~\eqref{eqn:e_group_general} is an e-value for $H_j$, 
conditional on $N_{K+j}$. Let
\begin{equation*}
\begin{split}
&\tilde{T}_j= \sup\left\{t \in \R : \widetilde{\FDP}_j^{N_{K+j}}(t) \leq \tilde{\alpha}\right\}, \text{ where }\\
&\widetilde{\FDP}_j^r(t) = \frac{\sum_{k \in I_{\geq r}} \One{\hat{V}_{k}^r < t, h(\tilde{Y}_k^r) \leq \tilde{c}(\tilde{X}_k^r)} + \One{\hat{V}_{K+j}^r < t, h(\tilde{Y}_{K+j}^r) \leq \tilde{c}(\tilde{X}_{K+j}^r)}}
{1 + \sum_{l \neq j} \One{\hat{V}_{K+l} < t} }  \cdot \frac{M}{|I_{\geq r}| + 1}
\end{split}
\end{equation*}
Applying arguments similar to the proof of Theorem~\ref{thm:individual}, 
we have $T_j = \tilde{T}_j$ on the event $\{\hat{V}_{K+j}^{N_{K+j}} < T_j, h(\tilde{Y}_{K+j}) \leq \tilde{c}(\tilde{X}_{K+j})\}$. Therefore, for any $r \geq 1$, letting $N_{-(K+j)} = (N_{K+l})_{l \neq j}$, we have
\begin{align*}
    &\EEst{e_j \One{H_j}}{N_{K+j}=r, N_{-(K+j)}}\\
    &=\mathbb{E}\left[\rule{0cm}{0.8cm}\frac{\One{\hat{V}_{K+j} < T_j}}{\sum_{k \in I_{\geq r}} \One{\hat{V}_{k}^r < T_j, h(\tilde{Y}_k^r) \leq \tilde{c}(\tilde{X}_k^r)} +1} \times 
    (|I_{\geq r}| + 1) \times \One{h(\tilde{Y}_{K+j}) \leq \tilde{c}(\tilde{X}_{K+j})}\right.\\
    &\left.\hspace{12.1cm} \bigg|\; N_{K+j}=r, N_{-(K+j)}\rule{0cm}{0.8cm}\right]\\
    &=\mathbb{E}\left[\rule{0cm}{0.8cm}\frac{\One{\hat{V}_{K+j} < \tilde{T}_j} \cdot \One{\hat{V}_{K+j} < T_j}}
    {\sum_{k \in I_{\geq r}} \One{\hat{V}_{k}^r < \tilde{T}_j, h(\tilde{Y}_k^r) \leq \tilde{c}(\tilde{X}_k^r)} + \One{\hat{V}_{K+j} < \tilde{T}_j, h(\tilde{Y}_{K+j}) \leq \tilde{c}(\tilde{X}_{K+j})}} \right.\\
    &\left.\hspace{5.2cm}\times 
    (|I_{\geq r}| + 1) \times \One{h(\tilde{Y}_{K+j}) \leq \tilde{c}(\tilde{X}_{K+j})} \left\vert\; N_{K+j}=r, N_{-(K+j)} \rule{0cm}{0.8cm}\right.\right]\\
    &\leq \mathbb{E}\left[\rule{0cm}{1.2cm}\frac{\One{\hat{V}_{K+j} < \tilde{T}_j}}{\sum\limits_{k \in I_{\geq r}} \One{\hat{V}_{k}^r < \tilde{T}_j, h(\tilde{Y}_k^r) \leq \tilde{c}(\tilde{X}_k^r)} +\One{\hat{V}_{K+j} < \tilde{T}_j, h(\tilde{Y}_{K+j}) \leq \tilde{c}(\tilde{X}_{K+j})}} 
    \cdot (|I_{\geq r}| + 1)\right.\\
    &\left.\hspace{12.2cm} \left|\rule{0cm}{1.2cm}\right.\;N_{K+j}=r, N_{-(K+j)}\rule{0cm}{1.2cm}\right]\\
    &\le 1,
\end{align*}
where the last equality holds since $\{(\tilde{X}_k^r, \tilde{Y}_k^r) : k \in I_{\geq N_{K+j}}\} \cup \{(\tilde{X}_{K+j}, \tilde{Y}_{K+j})\}$ 
are conditionally exchangeable given $N_{K+j} = r$ and $N_{-{K+j}}$, and $\tilde{T}_j$ is invariant with respect to their permutations. Therefore, $e_j$ is a conditional e-value given $N_{K+j}$, as desired.

\subsection{Proof of Theorem~\ref{thm:cov_shift}}
\label{appx:proof_cov_shift}
Similar to the proof of Theorem~\ref{thm:individual}, we define the oracle stopping time that is 
invariant to permutations within and across groups. For each $j \in [M]$, we define
\begin{equation*}
\begin{split}
&\tilde{T}_{j,i}^w = \sup\left\{t \in \R : \widetilde{\FDP}_{j,i}^w(t) \leq \tilde{\alpha}\right\}, \text{ where }\\
&\widetilde{\FDP}_{j,i}^w(t) = \frac{\sum_{k=1}^K p_k^j\cdot \One{\hat{V}_{k,i_k^*} < t, Y_{k,i_k^*} \leq C_{k,i_k^*}} + p_{K+j}^j\cdot \One{\hat{V}_{K+j,i} < t, Y_{K+j,i} \leq C_{K+j,i}}}{1 \vee \sum_{l \neq j} \sum_{i'=1}^{N_{K+l}} \One{\hat{V}_{K+l,i'} < t}} \cdot \frac{\sum_{l=1}^{M} N_{K+l}}{K+1}.
\end{split}
\end{equation*}
By similar arguments as in the proof of Theorem~\ref{thm:individual}, we have $T_j^w = \tilde{T}_{j,i}^w$ on the event $\{\hat{V}_{K+j,i} < T_j^w, Y_{K+j,i} \leq C_{K+j,i}\}$. 
Now fix any $(g_1,z_1),(g_2,z_2),\cdots,(g_K,z_K), (g_{K+j},z_{K+j}) \in \mathcal{G} \times (\X \times \Y)$ and let $\mathcal{E}_g$ 
denote the event that 
\[[(G_1,Z_1),\cdots,(G_K,Z_K),(G_{K+j},Z_{K+j})] = [(g_1,z_1),\cdots,(g_K,z_K), (g_{K+j},z_{K+j})],\]
where $[\cdot]$ denotes a multiset---a set allowing for repeated elements.

Then we have
\begin{align*}
    &\EEst{e_{j,i}^w \One{H_{j,i}}}{N_{K+1:K+M}}\\
    &= \EEst{\frac{\One{\hat{V}_{K+j,i} < T_j^w, Y_{K+j,i} \leq C_{K+j,i}}}{\sum_{k=1}^K p_k^j \cdot \One{\hat{V}_{k,i_k^*} < T_j^w, Y_{k,i_k^*} \leq C_{k,i_k^*}} + p_{K+j}^j}}{N_{K+1:K+M}}\\
    &\le \EEst{\frac{\One{\hat{V}_{K+j,i} < \tilde{T}_{j,i}^w, Y_{K+j,i} \leq C_{K+j,i}}}{\sum_{k=1}^K p_k^j \cdot \One{\hat{V}_{k,i_k^*} < \tilde{T}_{j,i}^w, Y_{k,i_k^*} \leq C_{k,i_k^*}} + p_{K+j}^j}}{N_{K+1:K+M}}\\
    &=\EEst{\EEst{\frac{\One{\hat{V}_{K+j,i} < \tilde{T}_{j,i}^w, Y_{K+j,i} \leq C_{K+j,i}}}{\sum_{k=1}^K p_k^j \cdot \One{\hat{V}_{k,i_k^*} < \tilde{T}_{j,i}^w, Y_{k,i_k^*} \leq C_{k,i_k^*}} + p_{K+j}^j}}{\mathcal{E}_g, N_{K+j}}}{N_{K+1:K+M}}\\
    &= \EEst{\EEst{\frac{\One{\hat{V}_{K+j,i} < \tilde{T}_{j,i}^w, Y_{K+j,i} \leq C_{K+j,i}}}{\sum_{k \in [K]\cup\{K+j\}} p_k^j \cdot \One{\hat{V}_{k,i_k^*} < \tilde{T}_{j,i}^w, Y_{k,i_k^*} \leq C_{k,i_k^*}}}}{\mathcal{E}_g, N_{K+j}}}{N_{K+1:K+M}}\\
    &\le 1.
\end{align*}
Above, the last inequality holds since
\[(G_{K+j},Z_{K+j}) \mid \mathcal{E}_g, N_{K+1:K+M} \sim \sum_{k \in [K] \cup \{K+j\}} p_k^j \cdot \delta_{(g_k,z_k)}\]
and $\tilde{T}_{j,i}^w$ depends only on the set $\{(G_1,Z_1),\cdots,(G_K,Z_K),(G_{K+j},Z_{K+j})\}$ (and $(\hat{V}_{K+l,i})_{l \neq j, 1 \leq i \leq N_{K+l}}$). 

Therefore, we have shown that $e_{j,i}^w$ is a valid group size-conditional e-value for each $(j,i)$, and the claim follows directly.

\subsection{Proof of Proposition~\ref{prop:p_ite}}
\label{appx:proof_p_ite}
Observe that by the monotonicity condition, the null event $Y_{K+j,i}(1) \leq Y_{K+j,i}(0)$ implies $\hat{V}_{K+j,i}^1 \leq \hat{V}_{K+j,i}^0$. Therefore,
\begin{align*}
    &\PP{p_{j,i} \leq \alpha \text{ and } Y_{K+j,i}(1) \leq Y_{K+j,i}(0)}\\
    &\leq \PP{\frac{\sum_{k=1}^K \One{\hat{V}_{K+j,i}^0 > \hat{V}_{k,i_k^*}^1}+1}{K+1} \leq \alpha \text{ and } \hat{V}_{K+j,i}^1 \leq \hat{V}_{K+j,i}^0}\\
    &\leq \PP{\frac{\sum_{k=1}^K \One{\hat{V}_{K+j,i}^1 > \hat{V}_{k,i_k^*}^1}+1}{K+1} \leq \alpha} \leq \alpha,
\end{align*}
where the last step applies the exchangeability of $\hat{V}_{1,i_1^*}^1, \cdots, \hat{V}_{K,i_K^*}^1, \hat{V}_{K+j,i}^1$.

\subsection{Proof of Theorem~\ref{thm:e_ite}}
\label{appx:proof_ite}
In this setting, we define an oracle permutation-invariant stopping time as 
\begin{equation*}
\begin{split}
   &\tilde{T}_{j,i} = \sup\left\{t \in \R : \widetilde{\FDP}_{j,i}(t) \leq \tilde{\alpha}\right\}, \text{ where }\\
    &\widetilde{\FDP}_{j,i}(t) = \frac{\sum_{k=1}^K \One{\hat{V}_{k,i_k^*}^1 < t} + \One{\hat{V}_{K+j,i}^1 < t}}{1 \vee \sum_{l \neq j} \sum_{i'=1}^{N_{K+l}} \One{\hat{V}_{K+l,i'}^0 < t}} \cdot \frac{\sum_{l=1}^M N_{K+l}}{K+1}.
\end{split}
\end{equation*}
By applying argument similar to the proof of Theorem~\ref{thm:individual}, we have $T_j = \tilde{T}_{j,i}$ under the event $\{\hat{V}_{K+j,i}^1 \leq T_j\}$
(but in this case, we do not directly observe the event $\{\hat V^1_{K+j,i} \le T_j\}$ since $Y_{K+j,i}(1)$ is 
not accessible). Therefore,

\begin{align*}
    &\EEst{e_{j,i} H_{j,i}}{N_{K+1:K+M}}\\
    &= \EEst{\frac{\One{\hat{V}_{K+j}^0 < T_j}\cdot\One{Y_{K+j,i}(1) \leq Y_{K+j,i}(0)}}{\sum\limits_{k=1}^K \One{\hat{V}_{k,i_k^*}^1 < T_j}+ 1}\cdot(K+1)}{N_{K+1:K+M}}\\
    &= \EEst{\frac{\One{\hat{V}_{K+j,i}^0 < T_j}\cdot\One{Y_{K+j,i}(1) \leq Y_{K+j,i}(0)}}{\sum\limits_{k=1}^K \One{\hat{V}_{k,i_k^*}^1 < T_j}+ \One{\hat{V}_{K+j,i}^1 < T_j}}\cdot \One{\hat{V}_{K+j,i}^1 < T_j} \cdot(K+1)}{N_{K+1:K+M}}\\
    &= \mathbb{E}\left[\rule{0cm}{1.1cm}\frac{\One{\hat{V}_{K+j,i}^0 < \tilde{T}_{j,i}}\cdot\One{Y_{K+j,i}(1) \leq Y_{K+j,i}(0)}}{\sum\limits_{k=1}^K \One{\hat{V}_{k,i_k^*}^1 < \tilde{T}_{j,i}}+ \One{\hat{V}_{K+j,i}^1 < \tilde{T}_{j,i}}}\cdot \One{\hat{V}_{K+j,i}^1 < \tilde{T}_{j,i}, \hat{V}_{K+j,i}^1 < T_j} \cdot(K+1) \left|\rule{0cm}{1.1cm}\right. N_{K+1:K+M}\rule{0cm}{1.1cm}\right]\\
    &\leq \EEst{\frac{\One{\hat{V}_{K+j,i}^1 < \tilde{T}_{j,i}}}{\sum\limits_{k=1}^K \One{\hat{V}_{k,i_k^*}^1 < \tilde{T}_{j,i}}+ \One{\hat{V}_{K+j,i}^1 < \tilde{T}_{j,i}}} \cdot(K+1)}{N_{K+1:K+M}} \le 1,
\end{align*}
where the second equality holds since $\hat{V}_{K+j,i}^0 < T_j$ and $Y_{K+j,i}(1) \leq Y_{K+j,i}(0)$,
implying that $\hat{V}_{K+j,i}^1 < T_j$ by the monotonicity condition on $s$; 
the third equality holds by the observation above. 
The last inequality holds due to the exchangeability of $(X_{1,i_1^*}, Y_{1,i_1^*}), \cdots, (X_{K,i_K^*}, Y_{K,i_K^*})$, $(X_{K+j,i}, Y_{K+j,i})$ and the invariance of $\tilde{T}_j$ under arbitrary permutations of these pairs. The next claim follows directly from Lemma~\ref{lem:ebh}.

\subsection{Proof of Theorem~\ref{thm:ite_hier}}
\label{appx:proof_ite_hier} 
The proof applies similar ideas as those used in the proofs of  Theorem~\ref{thm:hierarchical} and Theorem~\ref{thm:e_ite},
where we let
\begin{equation*}
\begin{split}
   &\tilde{T}_j = \sup\left\{t \in \R : \widetilde{\FDP}_j(t) \leq \tilde{\alpha}\right\}, \text{ where }\\
    &\widetilde{\FDP}_j(t) = \frac{\sum_{k=1}^K \frac{1}{N_k} \sum_{i'=1}^{N_k} \One{\hat{V}_{k,i'}^1 < t} + \frac{1}{N_{K+j}} \sum_{i'=1}^{N_{K+j}} \One{\hat{V}_{k,i'}^1 < t}}{1 \vee \sum_{l \neq j} \sum_{i'=1}^{N_{K+l}} \One{\hat{V}_{K+l,i'}^0 < t}} \cdot \frac{\sum_{l \neq j} N_{K+l}}{K+1},
\end{split}
\end{equation*}
for each $j \in [M]$. Then we have $T_j^+ \leq \tilde{T}_j \leq T_j^-$, and we also have the equality~\eqref{eqn:cond_exp} for the e-values $(e_{j,i})_{1 \leq j \leq M, 1 \leq i \leq N_{K+j}}$, by applying the same steps as in the proof of Theorem~\ref{thm:hierarchical}.

Therefore,
\begin{align*}
    &\frac{1}{K+1}\cdot \EEst{e_{j,i}\One{H_{j,i}}}{N_{K+1:K+M}}\\
    &= \EEst{\frac{\frac{1}{N_{K+j}}\sum_{i'=1}^{N_{K+j}}\One{\hat{V}_{K+j,i'}^0 < T_j^+} \cdot \One{Y_{K+j,i'}(1) \leq Y_{K+j,i'}(0)}}{\sum_{k=1}^K \frac{1}{N_k} \sum_{i'=1}^{N_k} \One{\hat{V}_{k,i'}^1 < T_j^-} + 1} }{N_{-(K+j)}}\\
    &\leq \EEst{\frac{\frac{1}{N_{K+j}}\sum_{i'=1}^{N_{K+j}}\One{\hat{V}_{K+j,i'}^0 < \tilde{T}_j}\cdot \One{Y_{K+j,i'}(1) \leq Y_{K+j,i'}(0)}}{\sum_{k=1}^K \frac{1}{N_k} \sum_{i'=1}^{N_k} \One{\hat{V}_{k,i'}^1 < \tilde{T}_j} + 1} }{N_{-(K+j)}}\\
    &\leq \EEst{\frac{\frac{1}{N_{K+j}}\sum_{i'=1}^{N_{K+j}}\One{\hat{V}_{K+j,i'}^1 < \tilde{T}_j}}{\sum_{k=1}^K \frac{1}{N_k} \sum_{i'=1}^{N_k} \One{\hat{V}_{k,i'}^1 < \tilde{T}_j} + \frac{1}{N_{K+j}}\sum_{i'=1}^{N_{K+j}}\One{\hat{V}_{K+j,i'}^1 < \tilde{T}_j}} }{N_{-(K+j)}} 
    \le \frac{1}{K+1},
\end{align*}
where the last inequality holds since $Y_{K+j,i'}(1) \leq Y_{K+j,i'}(0)$ implies $\hat{V}_{K+j,i'}^1 \leq \hat{V}_{K+j,i'}^0$. Therefore, $e_{j,i}$ is a valid group size-conditional e-value for $H_{j,i}$ and applying Lemma~\ref{lem:ebh} completes the proof.

\end{document}